\documentclass[fleqn,usenatbib]{mnras}

\usepackage{newtxtext,newtxmath}

\usepackage[T1]{fontenc}
\newcommand{\angstrom}{\mbox{\normalfont\AA}}
\DeclareRobustCommand{\VAN}[3]{#2}
\let\VANthebibliography\thebibliography
\def\thebibliography{\DeclareRobustCommand{\VAN}[3]{##3}\VANthebibliography}

\usepackage{graphicx}	
\usepackage{amsmath}	
\usepackage{float}	

\usepackage{ulem}

\title[Introducing SPHINX-MHD]{Introducing SPHINX-MHD: The Impact of Primordial Magnetic Fields on the First Galaxies, Reionization, and the Global 21cm Signal}

\author[H. Katz \& S. Martin-Alvarez et al.] {Harley Katz$^{1}$\thanks{E-mail:
  \href{mailto:harley.katz@physics.ox.ac.uk}{harley.katz@physics.ox.ac.uk}},
  Sergio Martin-Alvarez$^2$\thanks{Co-first author, E-mail: \href{mailto:smartin@ast.cam.ac.uk}{smartin@ast.cam.ac.uk}},
  Joakim Rosdahl$^{3}$,
  Taysun Kimm$^{4}$,
  J\'er\'emy Blaizot$^{3}$,
  \newauthor 
  Martin G. Haehnelt$^{2}$,
  L\'eo Michel-Dansac$^{3}$,
  Thibault Garel$^{5}$,
  Jose Oñorbe$^6$,
  \newauthor
  Julien Devriendt$^{1}$,
  Adrianne Slyz$^{1}$,
  Omar Attia$^{7,8}$,
  and Romain Teyssier$^{8}$\\ \\
  $^1$Sub-department of Astrophysics, University of Oxford,
  Keble Road, Oxford OX1 3RH, UK \\
  $^2$Kavli Institute for Cosmology and Institute of Astronomy,
  Madingley Road, Cambridge CB3 0HA, UK \\ 
  $^2$Univ Lyon, Univ Lyon1, Ens de Lyon, CNRS, Centre de Recherche
  Astrophysique de Lyon UMR5574, F-69230, Saint-Genis-Laval, France \\
  $^3$Department of Astronomy, Yonsei University, 50 Yonsei-ro,
  Seodaemun-gu, Seoul 03722, Republic of Korea \\  
  $^5$Observatoire de Geneve, Universite de Geneve, 51 Ch. des Maillettes, 1290 Versoix, Switzerland\\
  $^6$Facultad de Fisicas, Universidad de Sevilla, Avda. Reina Mercedes s/n, Campus de Reina Mercedes, 41012 Sevilla, Spain \\
  $^{7}$Department of Physics, ETH Zurich, Wolfgang-Pauli-Strasse 27, CH-8093 Zurich, Switzerland\\
  $^{8}$Institute for Computational Science, University of Zurich, Winterthurerstrasse 190, CH-8057 Zurich, Switzerland
  }

\date{Accepted XXX. Received YYY; in original form ZZZ}

\pubyear{2020}

\begin{document}
\label{firstpage}
\pagerange{\pageref{firstpage}--\pageref{lastpage}}
\maketitle

\definecolor{lblue}{rgb}{0,0.65,0.9}
\definecolor{dblue}{rgb}{0,0,0.66667}
\definecolor{dred}{rgb}{0.66667,0.3333,0}

\begin{abstract}
We present the first results from SPHINX-MHD, a suite of cosmological radiation-magnetohydrodynamics simulations designed to study the impact of primordial magnetic fields (PMFs) on galaxy formation and the evolution of the intergalactic medium (IGM) during the epoch of reionization. The simulations are among the first to employ multi-frequency, on-the-fly radiation transfer and constrained transport ideal MHD in a cosmological context to simultaneously model the inhomogeneous process of reionization as well as the growth of primordial magnetic fields. We run a series of $(5\,\text{cMpc})^3$ cosmological volumes, varying both the strength of the seed magnetic field and its spectral index. We find that PMFs with a spectral index ($n_B$) and a comoving amplitude ($B_0$) that have $n_B > -0.562\log_{10}\left(\frac{B_0}{1{\rm n}G}\right) - 3.35$ produce electron optical depths ($\tau_e$) that are inconsistent with CMB constraints due to the unrealistically early collapse of low-mass dwarf galaxies. For $n_B\geq-2.9$, our constraints are considerably tighter than the $\sim{\rm n}G$ constraints from Planck. PMFs that do not satisfy our constraints have little impact on the reionization history or the shape of the UV luminosity function. Likewise, detecting changes in the Ly$\alpha$ forest due to PMFs will be challenging because photoionisation and photoheating efficiently smooth the density field. However, we find that the first absorption feature in the global 21cm signal is a particularly sensitive indicator of the properties of the PMFs, even for those that satisfy our $\tau_e$ constraint. Furthermore, strong PMFs can marginally increase the escape of LyC photons by up to 25\% and shrink the effective radii of galaxies by $\sim44\%$ which could increase the completeness fraction of galaxy surveys.  Finally, our simulations show that surveys with a magnitude limit of ${\rm M_{UV,1500\angstrom}=-13}$ can probe the sources that provide the majority of photons for reionization out to $z=12$.
\end{abstract}



\section{Introduction}
\label{intro}
Understanding the formation of the first generation of galaxies and how the Universe emerged from the Dark Ages remains one of the most interesting frontiers in modern cosmology. During the first billion years, the Universe evolved from a nearly neutral state after recombination to being almost completely ionised. This process of reionization likely began at $z\gtrsim30$ with the formation of the first metal-free Population~III stars \citep{Wise2012} and ended somewhere in the redshift range of $z\sim5-6$ \citep{Fan2006,Kulkarni2019}. 

Currently, the most commonly discussed scenario is that reionization was primarily driven by photons emitted by dwarf galaxies \citep[e.g.,][]{Finkelstein2019}. The number density of these objects combined with their predicted high LyC\footnote{Throughout the paper, we refer to hydrogen-ionising photons with energy above 13.6 eV as Lyman continuum photons, or LyC for short.} escape fractions ($f_{\rm esc}$, e.g., \citealt{Kimm2017}) makes them a prime candidate to be the dominant sources of reionization.  The majority of these galaxies are too dim to be directly observed, even with our most powerful space telescopes; however, deep observations of lensed objects behind massive galaxy clusters have hinted at a steep faint-end slope to the UV luminosity function \citep{Livermore2017}. The upcoming launch of the James Webb Space Telescope (JWST) is expected to shed a significant amount of light on the sources of reionization \citep{Gardner2006}.

Much of our detailed knowledge of the physics during the epoch of reionization stems from high-resolution cosmological radiation hydrodynamics simulations. These simulations generally fall into one of two categories: those that model reionization on large (i.e. $\sim100$ cMpc) scales, which are required for an accurate determination of the 21cm signal \citep[e.g.,][]{Iliev2014} and to capture the inhomogeneities of reionization relevant for Ly$\alpha$ forest measurements \citep{Kulkarni2019}, and those that simulate much smaller volumes, focused on galaxy formation and the escape of LyC radiation from a multi-phase interstellar medium (ISM) as well as the back-reaction of the formation of the first galaxies and reionization on subsequent galaxy formation \citep[e.g.,][]{Oshea2015,Rosdahl2018,Katz2020}. Due to computational limitations, simulations that model the large-scale inhomogeneous process of reionization on $\sim100$ cMpc scales while simultaneously resolving the escape of ionizing radiation from dwarf galaxies are beyond current capabilities.  However, despite the differences between the simulations, they tend to agree that reionization is a highly inhomogeneous process that evolved in an inside-out manner \citep[e.g.,][]{Iliev2006,Lee2008,Gnedin2014,Katz2017}.

Nearly all modern cosmological simulations model their volumes assuming a $\Lambda$CDM Universe and a concordance cosmology consistent with that measured from either the Planck \citep{Planck2018} or WMAP \citep{Hinshaw2013} satellites. The simulations are evolved using the Friedmann equations \citep{Friedmann1922}, and generally explosive feedback is input into the simulation, for example from different types of supernova (SN) or accreting black holes (AGN), in order to explain the Schechter function shape of the $z=0$ stellar mass function \citep[e.g.,][]{Vogelsberger2013,Crain2015}. On-the-fly radiation hydrodynamics has become more common in cosmological simulations \citep{Gnedin2001,Pawlik2008,Wise2011,Rosdahl2013,Kannan2019,Hopkins2020}, but very few include magnetic fields. Magnetic fields are dynamically important in numerous astrophysical contexts such as the ISM \citep{Beck2007} and the intracluster medium \cite[ICM,][]{Feretti2012}, yet only recently have their effects been self-consistently modelled in cosmological simulations \citep[e.g.,][]{Dubois2008,Dolag2009,Doumler2010,Rieder2017,Marinacci2018,Vazza2018,Garaldi2020}. 

The origin of cosmological magnetic fields is currently unknown and various theories have been proposed to explain their existence. Depending on e.g. the inflationary scenario, primordial magnetic fields (PMFs) could be generated before recombination and magnetic fields of this type will be the primary focus of our work. Current constraints from Planck have placed an upper limit of $\sim10^{-9}G$, dependent on the exact structure of the PMF, based on a number of effects PMFs have on CMB anisotropies \citep{PlanckPMF2016}. Additional constraints on PMFs can be derived from their impact on Big-Bang nucleosynthesis \citep[BBN;][]{Grasso1995,Caprini2002} and structure formation \citep{Wasserman1978,Blasi1999}. 

During reionization, magnetic fields can also be generated at the edges of ionisation fronts, when electron density and pressure gradients are misaligned \citep{Biermann1950} or because of charge segregation \citep{Durrive2015,Durrive2017}. They can also be generated during galaxy formation, in or around compact objects such as stars \citep{Beck2013,Butsky2017,Sergio2020b} and black holes \citep{Vazza2017} and expelled into the low density regions of the Universe via magnetised winds. We reserve studying these scenarios for future work. Regardless of their origin, any acceptable magnetogenesis theory must be able to simultaneously explain the order $10^{-6}G$ level magnetic fields observed in galaxies \citep{Davis1951,Basu2013} as well as the weak ($\gtrsim10^{-16}\ G$) magnetic fields in the intergalactic medium \citep[IGM, e.g.][]{Neronov2010,Dolag2011,Tavecchio2011}. 

Magnetic fields are particularly interesting in the context of reionization for a number of reasons:
\begin{enumerate}
\item Magnetic fields can have a significant impact on the structure and pressure support of the multi-phase ISM \citep{Kortgen2019}. The magnetic energy in the ISM is observed to be in equipartition with the turbulent, thermal and cosmic ray energy densities \citep{Tabatabaei2008,Beck2015}. \cite{Marinacci2016} demonstrated that star formation histories are not significantly affected, unless the strength of the PMF is $\gtrsim10^{-9}G$ at which point the additional pressure can suppress gas accretion onto low mass galaxies. However, \cite{Sergio2020} showed that, in the context of ideal MHD, the structure of the ISM and the size of galaxies is noticeably different in the case where seed fields have a strength $<10^{-9}G$, even if the star formation rates remain unchanged as physics associated with the magnetic field can drain angular momentum from the gas and deposit it further away from the centre of the galaxy. As magnetic fields modify the structure of the ISM, they may change $f_{\rm esc}$, affecting both the history of reionization as well as the sources responsible. If the effective radii of galaxies decrease significantly due to strong PMFs, our interpretations of the high-redshift UV luminosity function may also change due to a systematic difference in the size-luminosity relation \citep{Kawamata2018}.

\item After recombination, magnetic fields can deposit their energy into the IGM via ambipolar diffusion and decaying MHD turbulence \citep[e.g.,][]{Jedamzik1998,Subramanian1998,Sethi2005}. These processes can impact the thermal and ionisation history of the IGM in the early Universe which subsequently translates to a change in the optical depth to the CMB. For sufficiently strong PMFs, these two effects alone can generate electron temperatures of $T_e>10^4$K and ionise the universe to levels of $\sim10\%$ \citep{Sethi2005,Chluba2015}. These effects become particularly important for seed fields with $B_0\gtrsim10^{-9}G$ \citep{Sethi2005} although this value is approximately equal to the current upper limits on the strength of the PMF \citep[e.g][]{PlanckPMF2016}.

\item Magnetic fields can generate density perturbations, which depending on the strength and spectral slope of the PMF will impact structure formation on small scales with $k$-modes $\gtrsim1{\rm cMpc^{-1}}$ \citep{Wasserman1978,Gopal2003,Shaw2012}. This modification to the matter power spectrum is coincidentally in the region of $k$-space that predominantly affects the formation of dwarf galaxies which, as discussed earlier, are the primary candidates to be the dominant source of reionization. \cite{Pandey2015,Sanati2020} have shown that the reionization history can significantly change depending on the assumptions regarding PMFs. Hence, the observed reionization history itself provides a constraint on the properties of the PMFs \citep{Sanati2020}. Similarly, the Ly$\alpha$ forest is currently our best constraint on the tail-end of reionization \citep{Fan2006,Kulkarni2019} and the effective optical depth is sensitive to the presence of sub-n$G$ PMFs \citep[e.g.,][]{Pandey2013,Meiksin2014}.

\item Although we will not consider them in this work, magnetic fields themselves can be generated at the interfaces of ionisation fronts and density irregularities in the neutral IGM via the Biermann battery \citep{Subramanian1994} and other mechanisms \citep[e.g.][]{Durrive2015}. The Biermann battery will generate magnetic fields as long as there is a gradient in the electron density that is perpendicular to a temperature gradient, for example, in an ionisation front sweeping over a gas filament. \cite{Gnedin2000} post-processed cosmological simulations and found that this mechanism can generate seed magnetic fields on the order of $B\lesssim10^{-21}G$ in the IGM (see also \citealt{Attia2021}) and the simulations of \cite{Garaldi2020} show that the Durrive battery generates magnetic fields slightly weaker than those produced by the Biermann battery.
\end{enumerate}

\begin{table*}
    \centering
    \caption{List of the simulations in the SPHINX-MHD suite along with details of each run. From left to right, columns indicate the name of the simulation, the comoving strength of the initial magnetic field ($B_0$; see text for details), the spectral index of the magnetic field, the dark matter ($M_{\rm DM}$) and stellar ($M_{*}$) particle mass resolution, the cell sizes at $z = 6$ ($\Delta x_{z=6}$) and $z = 20$ ($\Delta x_{z=20}$), the final redshift of the simulation, and the initial structure of the magnetic field. In each simulation, we only change the initial conditions for magnetic fields varying both their strength and spectral index $n_B$. }
    \begin{tabular}{lcccccccc}
    \hline
    Simulation Name & $B_0\ [G]$ & $n_B$ & $M_{\rm DM}\ {\rm [M_{\odot}]}$ &  $M_{*}\ {\rm [M_{\odot}]}$ & $\Delta x_{z=6}$ [pc] & $\Delta x_{z=20}$ [pc] & $z_{\rm final}$& Seed Field Structure \\
    \hline
        B21 & $10^{-21}$ & - & $3.1\times10^4$ & $10^3$ & 7.31$h^{-1}$ & 2.44$h^{-1}$ & 6.0 & Uniform, $z$-direction  \\
        B14 & $10^{-14}$ & - & $3.1\times10^4$ & $10^3$ & 7.31$h^{-1}$ & 2.44$h^{-1}$ &  6.0 &Uniform, $z$-direction  \\
        B11 & $5\times10^{-11}$ & - & $3.1\times10^4$ & $10^3$ & 7.31$h^{-1}$ & 2.44$h^{-1}$ & 6.0 &Uniform, $z$-direction  \\
        B11\_29 & $5\times10^{-11}$ & $-2.9$ & $3.1\times10^4$ & $10^3$ & 7.31$h^{-1}$ & 2.44$h^{-1}$ & 6.0 &Random  \\
        B11\_27 & $5\times10^{-11}$ & $-2.7$ & $3.1\times10^4$ & $10^3$ & 7.31$h^{-1}$ & 2.44$h^{-1}$ & 6.0 &Random  \\
        B11\_24 & $5\times10^{-11}$ & $-2.4$ & $3.1\times10^4$ & $10^3$ & 7.31$h^{-1}$ & 2.44$h^{-1}$ & 54.6 & Random  \\
    \hline
    \end{tabular}
    \label{sims_tab}
\end{table*} 

While it is clear that PMFs can significantly impact galaxy formation during the first billion years as well as the reionization history, there are currently no cosmological simulations that systematically study these effects using coupled radiation-magnetohydrodynamics. In this work, we introduce SPHINX-MHD, a suite of simulations that self-consistently model
\begin{enumerate}
\item the formation and evolution of galaxies during the reionization epoch, 
\item the growth and amplification of primordial magnetic seed fields, 
\item the escape of ionising radiation from a multi-phase ISM, and
\item an inhomogeneous reionization process, using multi-frequency radiation transfer and ideal magnetohydrodynamics. 
\end{enumerate}
These simulations are an extension of the SPHINX project \citep{Rosdahl2018,Katz2020} which aims to address numerous goals including: understanding the primary sources of reionization, the statistical behaviour of $f_{\rm esc}$, the back-reaction of radiation feedback on the formation of dwarf galaxies, and the observational signatures of EoR galaxies. SPHINX-MHD goes beyond the goals of the original SPHINX, with the additional aspiration of constraining the impact of magnetic fields on reionization and the formation of the first galaxies. We vary the strength and spectral slope of the PMFs, taking into account their effects on the matter power spectrum \citep{Shaw2012}. When exploring different spectral slopes, our PMF models emulate inflationary magnetogenesis scenarios capable of generating strong magnetic fields \citep[e.g., the models of][]{Turner1988,Ratra1992}. We depart from an approximately scale-invariant red spectrum case, frequently associated with inflationary magnetogenesis and commonly addressed in the literature \citep[][]{Bonvin2013,Tasinato2015,PlanckPMF2016}. Many of these inflation models allow for a range of spectral indices \citep{Subramanian2010,Subramanian2016} such that there is no back reaction on the expansion during inflation. Furthermore the amplitude of the magnetic field is extremely sensitive to parameters chosen for the inflation model and depends as well on coupling function between the inflaton and the electromagnetic field.

This paper is organised as follows.  In Section~\ref{methods}, we describe the numerical methods including initial condition generation, simulation physics, and halo finding.  In Section~\ref{results}, we present our results on the impact of magnetic fields on reionization and galaxy formation during the first billion years. Finally, in Section~\ref{dc}, we present our discussion and conclusions.

\section{Numerical Methods}
\label{methods}
In total, we run six simulations as listed in Table~\ref{sims_tab}, varying the strength of the primordial seed magnetic field and its spectral slope, accounting for its effect on the matter power spectrum \citep{Shaw2012}. These simulations are designed to explore a representative subset of the parameter space for which magnetic fields may be interesting in the context of reionization. The lowest magnetic field strength is approximately that of a Biermann battery-generated magnetic field. The B14 simulation is used for comparison to other simulations in the literature that explored primordial seed fields of this value \citep[e.g.,][]{IllustrisTNG}.  Finally the B11 simulations represent scenarios where the magnetic field is likely to have a dynamic impact on galaxy structure \citep[e.g.,][]{Sergio2020}, and where the impact of the primordial seed field on the matter power spectrum can modify structure formation and potentially the reionization history \citep{Shaw2012,Pandey2015,Sanati2020}. We describe below in detail the numerical methods for each of these simulations.

\subsection{Initial Conditions}
Initial conditions for the simulations are generated with {\small MUSIC} \citep{Hahn2011} at $z=150$ assuming a $\Lambda$CDM Universe with the following cosmological parameters: $\Omega_{\Lambda}=0.6825$, $\Omega_{\rm m}=0.3175$, $\Omega_{\rm b}=0.049$, $h=0.6711$, $\sigma_8=0.83$, and $n_s=0.962$, consistent with the Planck~2013 results \citep{Planck2014}.  We set the initial gas composition of the simulation to be 76\% H and 24\% He by mass and assume a metallicity floor of $3.2\times10^{-4}Z_{\odot}$ to account for the lack of cooling due to molecular hydrogen and associated Pop. III star formation in our simulation \citep{Wise2012}.  

The initial conditions represent a comoving cubic volume with a side length of 5~cMpc\footnote{Note that all length units that are prefaced with a "c" represent comoving units while all others are physical.}, initially populated with $512^3$ dark matter particles of mass $3.1\times10^4{\rm M_{\odot}}$ and the same number of gas cells.  Assuming that we require 300 dark matter particles to resolve a halo, the minimum halo mass resolved by our simulation is $9.3\times10^6{\rm M_{\odot}}$, which is well below the atomic cooling threshold. Because the simulated volume is considerably smaller than the cosmological homogeneity scale of $\sim$100~cMpc, the volume was selected among 60 different dark matter-only simulations for having the most average halo mass function at $z=6$, $z=8$, and $z=10$ (see \citealt{Rosdahl2018}).

The magnetic fields in the simulations are initialised in two different ways. The B21, B14, and B11 simulations exhibit uniform seed fields that are aligned with the $z$-axis of the simulation. The comoving strengths of the seed fields for these simulations are listed in Table~\ref{sims_tab}. The transfer function used to compute the initial conditions for these simulations is calculated from the fitting function presented in \cite{Eisenstein1999} and the initial density field is identical between these different models. The seed fields in the B11\_29, B11\_27, and B11\_24, are initialised to have a power spectrum that is described as:
\begin{equation}
P(k)=Ak^{n_B},
\end{equation}
where $n_B$ is the spectral slope and 
\begin{equation}
A=\frac{(2\pi)^{n_B+5}B_{1{\rm cMpc}}^2}{2\Gamma\left(\frac{n_B+3}{2}\right)k_{1{\rm cMpc}}^{n_B+3}}.
\label{eq:normB}
\end{equation}
In this initial configuration, we follow the convention to normalise the strength of the magnetic field at a scale of 1~cMpc ($B_0=B_{\lambda= 1 \text{cMpc}}$, e.g. \citealt{Shaw2012,PlanckPMF2016}). Our method for generating the magnetic component of the initial conditions for the B11\_29, B11\_27, and B11\_24 simulations will be further described in Martin-Alvarez et al. {\it in prep}). In short, to generate these ICs, we initialise a random Gaussian vector potential field in Fourier space over a uniform grid. We modulate this spectrum at each wavelength $k$ so that the magnetic field resulting from the curl of the vector potential has a spectral slope $n_B$. The computed magnetic field is divergenceless by construction and spatially-displaced in Fourier space so that it is defined at cell interfaces. Finally, the norm of the magnetic power spectrum is set using Equation \ref{eq:normB}. The resulting magnetic field perturbations deviate from the average field with a magnetic field rms $\sigma_{{\rm rms},B} / \langle B \rangle$ of $\sim 1.25$ for each set of ICs. 

For the B11\_29, B11\_27, and B11\_24 simulations, we also account for the impact that the PMFs have on the matter power spectrum. In order to account for the impact of the PMF on the matter power spectrum, we use a modified version of {\small CAMB} \citep{CAMB,Shaw2012} to compute the transfer function used by {\small MUSIC}. The general idea is that magnetic fields generate density perturbations in the baryons via the Lorentz force \citep{Wasserman1978,Kim1996,Subramanian1998,Gopal2003,Sethi2005,Tashiro2006}. These magnetically-induced perturbations grow at the same rate as the primordial density fluctuations and couple to the dark matter via gravity. For strong enough PMFs, the additional density perturbations can dominate over primordial fluctuations, especially at high $k$. Hence, the initial density field is different in the simulations with non-uniform magnetic fields compared to the B21, B14, and B11 simulations. 

\begin{figure}
\centerline{\includegraphics[scale=1,trim={0 0.6cm 0 1.4cm},clip]{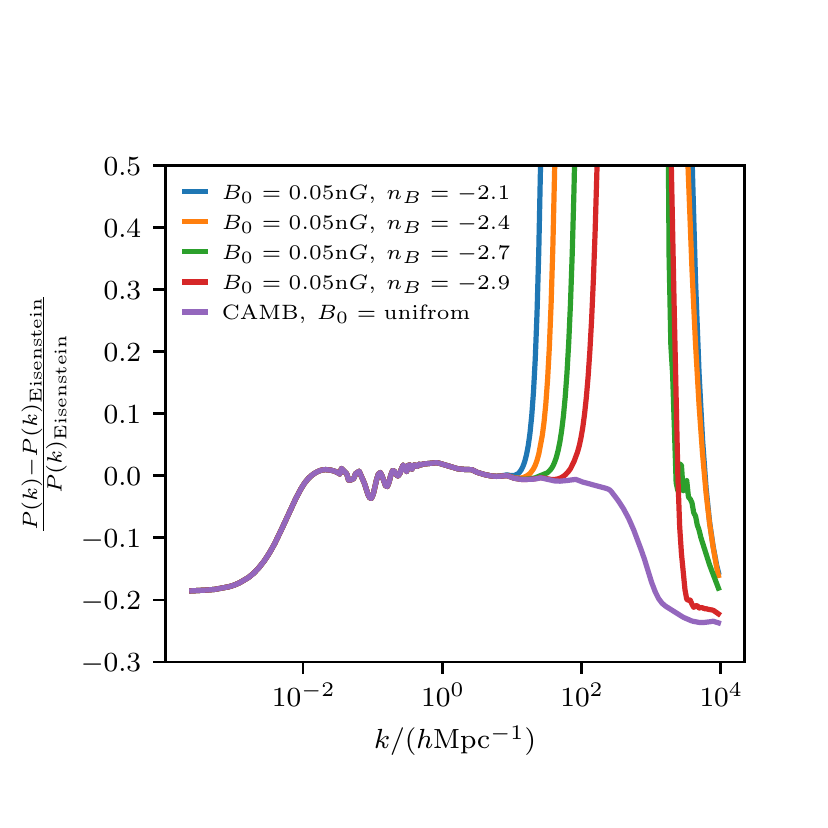}}
\caption{Fractional difference between the input matter power spectra for simulations that use {\small CAMB} (with no effects due to $B_0$) or the modified version of {\small CAMB} that accounts for PMFs when compared with those of the fiducial model that employ the fitting function from \protect\cite{Eisenstein1999}. In the standard $\Lambda$CDM case, the results differ by up to $\sim20\%$ at high and low $k$ while the differences due to the PMFs completely dominate any systematic error due to differences in transfer function.}
\label{ICmag}
\end{figure}

\begin{figure}
\centerline{\includegraphics[scale=0.99,trim={0 0.6cm 0 1.4cm},clip]{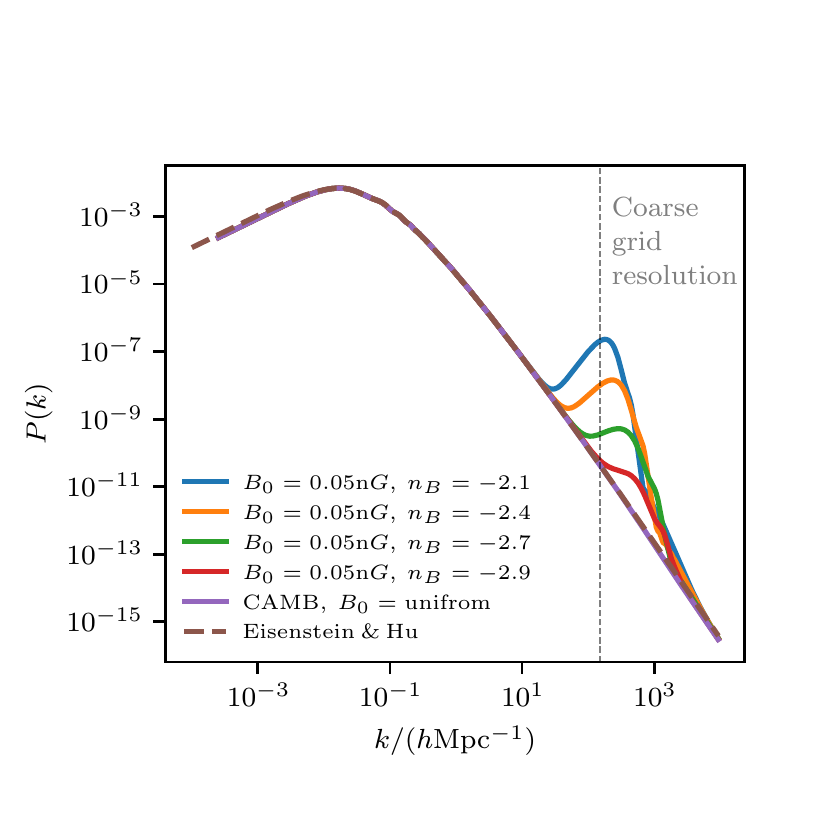}}
\centerline{\includegraphics[scale=0.99,trim={0 0.6cm 0 1.4cm},clip]{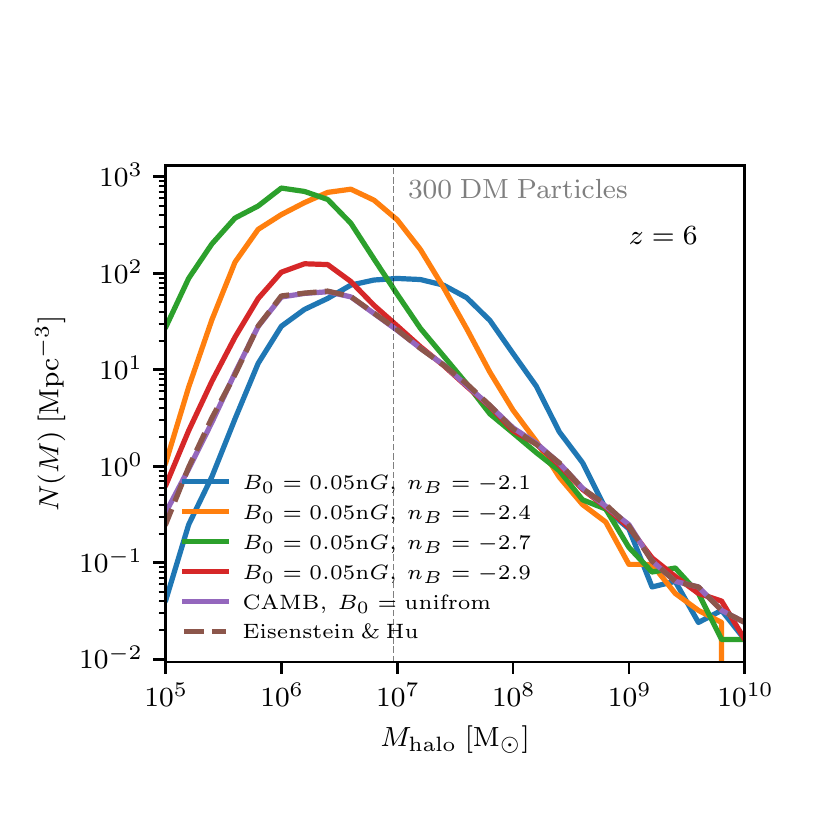}}
\caption{Input matter power spectra (top) and dark matter halo mass function at $z=6$ for the various sets of initial conditions computed as the number of halos in mass bins of width 0.2~dex (bottom) for dark matter-only simulations. The {\small CAMB} and \protect\cite{Eisenstein1999} models do not include any magnetic field effects. The enhancements in the matter power spectra due to PMFs can significantly alter the dark matter halo mass function at $z=6$ depending on the spectral slope of the PMF.  These differences occur in the range of masses represented by dwarf galaxies.  Despite the $\sim20\%$ differences between the {\small CAMB} and \protect\cite{Eisenstein1999} at high and low $k$, no significant differences arise in the halo mass function at $z=6$. The turnover in the mass function at low halo masses is due to limited resolution. We show all haloes that are resolved by more than 20 particles.}
\label{ICmag2}
\end{figure}

For our chosen cosmology, the transfer function computed with {\small CAMB} deviates from that of \cite{Eisenstein1999} at both high and low $k$-modes by up to $\sim20\%$ in the case without PMFs (see Figure~\ref{ICmag}). We have confirmed that despite these differences, the impact on the simulation is negligible as there are no noticeable differences in the dark matter halo mass function at $z=6$ (see the dashed and purple lines in the bottom panel of Figure~\ref{ICmag2}). Note that our limited simulation volume inhibits us from testing changes at $k\lesssim0.2$. In contrast, when we include the scalar and tensor perturbations from PMFs, the more the spectral slope of the PMF deviates from $-3$ (i.e. scale-free), the larger the differences in the matter power spectrum compared to standard $\Lambda$CDM. 

\begin{figure}
\centerline{\includegraphics[scale=1]{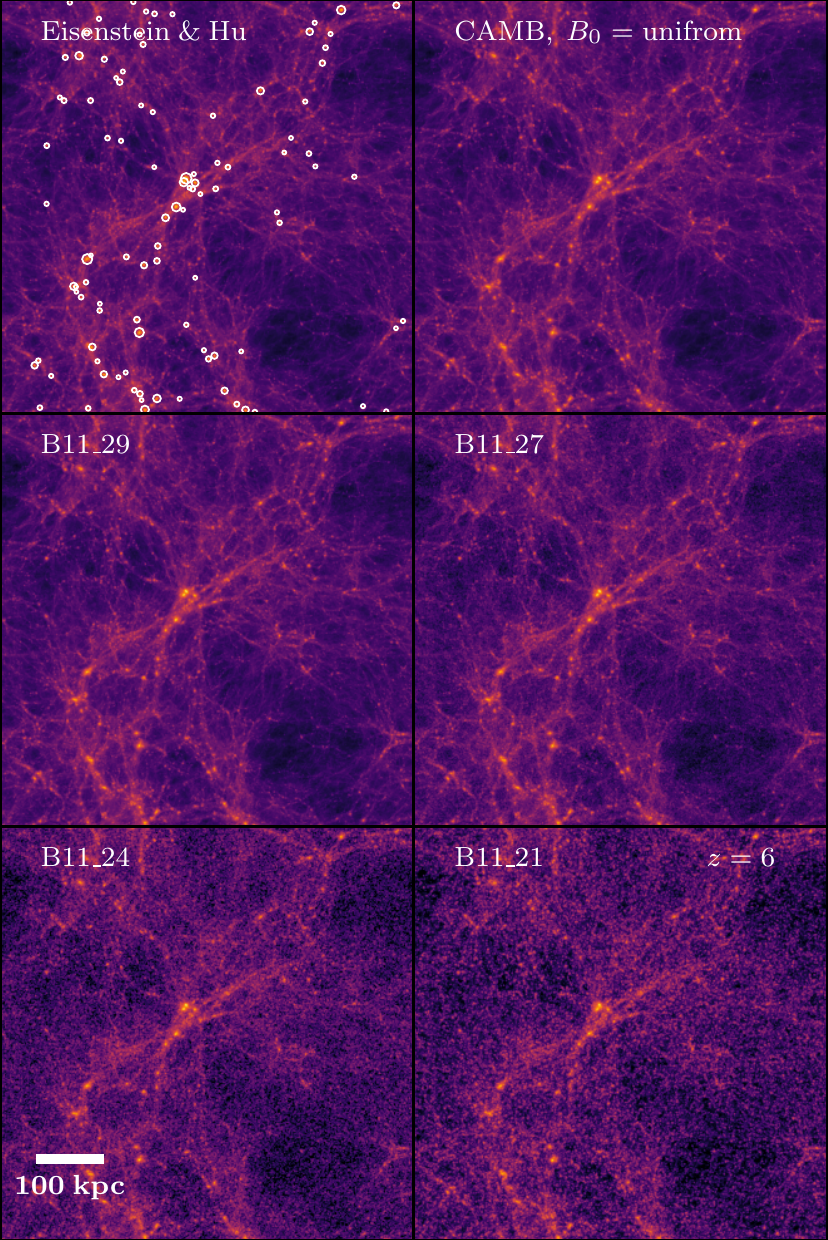}}
\caption{Dark matter column density maps of all of the different initial conditions at $z=6$ in the dark matter-only simulations. The {\small CAMB} and \protect\cite{Eisenstein1999} models do not include any magnetic field effects. As the slope of the PMF power spectrum flattens, more low mass dark matter haloes appear, impacting the filamentary structure of the cosmic web. The locations of the 100 most massive haloes are shown in the top left with each circle representing the virial radius of the halo.}
\label{DMmapsz6}
\end{figure}

In Figure~\ref{ICmag2}, we plot the matter power spectrum for different values of $n_{B}$ for $B_0=0.05\,\text{nG}$. The enhancements in the matter power spectra at high $k$ due to the PMFs have a significant impact on the dwarf halo population at $z=6$ (see the bottom panel of Figure~\ref{ICmag2}). Depending on the spectral slope, an excess of more than an order of magnitude in the number of haloes is seen at certain masses compared with $\Lambda$CDM. These are clearly visible in Figure~\ref{DMmapsz6} where we show the dark matter column density viewed down the $z$-axis of the simulation box at $z=6$ for the set of dark matter-only simulations. For $n_B=-2.1$, much of the filamentary structure is affected. Because the amount of mass in the box is conserved, the increase in the number of dwarf galaxies results in a small reduction at the high-mass end. For PMFs with significantly lower amplitudes, reasonable choices for $n_B$ do not lead to any significant changes in the matter power spectrum; hence, we only adopt these modified initial conditions for different realisations of the B11 simulation.

\subsection{Gravity, Magnetohydrodynamics, \& Radiation}
In order to evolve the simulation with gravity, magnetohydrodynamics, and radiation transfer (RT), we use {\small RAMSES-RT} \citep{Rosdahl2013,Rosdahl2015}, which is a radiation hydrodynamics extension of the {\small RAMSES} code \citep{Teyssier2002}.  The public version of the {\small RAMSES} code includes a constrained transport \citep[CT,][]{Evans1988} implementation of ideal MHD \citep{Teyssier2006,Fromang2006}.  For this work, we have updated the version of the code used for the original {\small SPHINX} simulations \citep{Rosdahl2018} so that it can simultaneously solve the equations for ideal radiation-magnetohydrodynamics (RMHD). 

\subsubsection{Gravity and Hydrodynamics}
The ideal MHD equations are solved using a second-order Godunov scheme based on a MUSCL-Hancock method. In contrast to the original {\small SPHINX} simulation, we employ the HLLD Riemann solver \citep{Miyoshi2005} and the MinMod slope limiter \citep{Roe1986} to construct gas variables at cell interfaces from their cell-centred values. We assume an adiabatic index of $\gamma=5/3$ (i.e. that of an ideal monatomic gas) to close the relation between gas pressure and internal energy.  The motions of collisionless dark matter, star particles and gas are computed by solving the Poisson equation. Dark matter and star particles are projected onto the adaptive grid using a cloud-in-cell interpolation.  A multigrid solver \citep{Guillet2011} is used to solve the Poisson equation up to a refinement level of 12. At more refined levels we adopt a conjugate gradient solver to improve the speed of the simulation.

\subsubsection{Radiative Transfer}
Radiation is advected between cells using a first-order moment method that uses the M1 closure for the Eddington tensor \citep{Levermore1984} and a Global-Lax-Friedrich intercell flux function \citep[e.g.,][]{Toro2009}. In comparison to many other RT solvers, the M1 closure relies only on local quantities and does not scale with the number of radiation sources in the computational volume.  Because of our choice of RT solver, the time steps in our simulations are limited by the RT Courant condition such that $\Delta t<\Delta x/3c_{\rm sim}$, where $\Delta x$ is the size of a cell, and $c_{\rm sim}$ is the speed of light chosen for the simulation. Setting $c_{\rm sim}=c$ would result in a prohibitively small time step that in some cases is $\sim1000\times$ smaller than that of a simulation without RT.  For this reason, we adopt the variable-speed-of-light approximation (VSLA) described in \cite{Katz2017} where we adaptively change $c_{\rm sim}$ depending on the local grid refinement level so that the speed of ionisation fronts is properly captured in both low- and high-density regions.  We adopt a value of $c_{\rm sim}=0.2c$ on the base (coarse) grid of the simulation and divide this quantity by a factor of two on each subsequently refined level. We set a minimum $c_{\rm sim}=0.0125c$ to ensure that the radiation velocity is always greater than that of the gas. Compared to the original implementation of VSLA in \cite{Katz2017}, we employ an updated and much more computationally efficient version of the algorithm \citep{Katz2018,Rosdahl2018} that is integrated with the adaptive time stepping on the AMR grid as well as the RT subcycling scheme present in {\small RAMSES-RT} \citep{Rosdahl2018}. To further reduce the computational demands of the RT, we allow for up to 500 RT subcycles for each hydrodynamic time step by adopting Dirichlet boundary conditions at coarse-fine interfaces \citep{Commercon2014}. In practice, the actual number of RT subcycles is of $\mathcal{O}(<10)$ after the first SNe explode in the simulation.

The radiation in the simulation is tracked in three energy bins: $13.6\,{\rm eV}-24.59\,{\rm eV},\ 24.59\,{\rm eV}-52.54\,{\rm eV},\ \&\ 52.54\,{\rm eV}-\infty$. This allows us to track the ionisation states of hydrogen and helium.  The mean energy of the radiation within each frequency bin is computed every ten coarse time steps as the luminosity-weighted mean energy across all star particles in the simulation. The number of photons in each frequency bin for each cell is updated on every fine time step and we apply the ``smoothing'' technique to reduce the total number of cooling subcycles \citep{Rosdahl2013}.

\subsubsection{Ideal MHD}
\label{ideal_mhd}
\begin{figure}
\centerline{\includegraphics[scale=1,trim={0 0.6cm 0 1.4cm},clip]{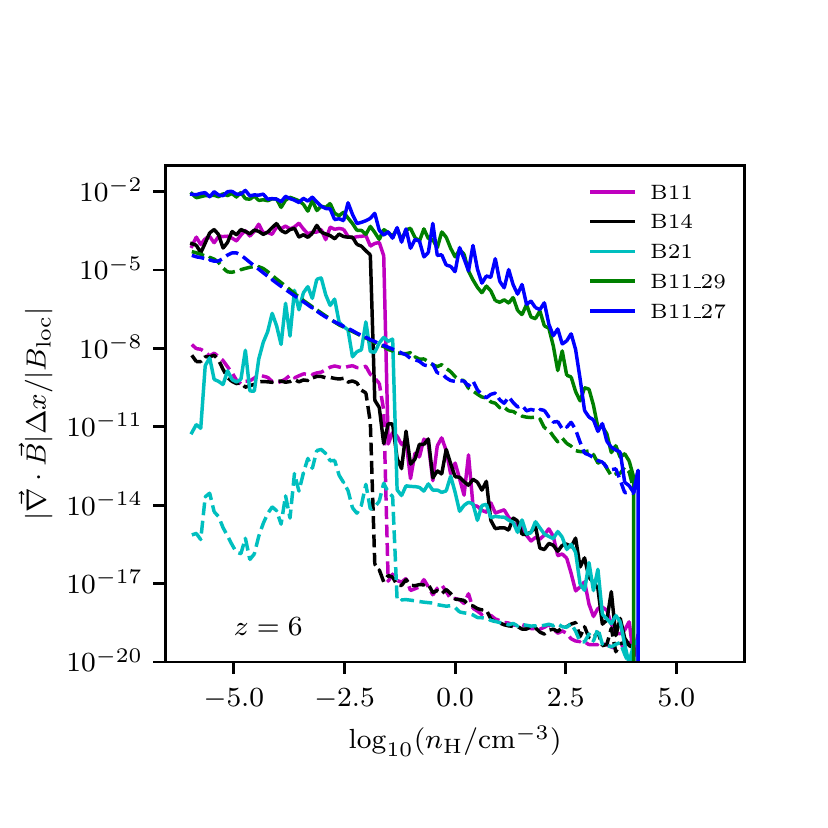}}
\caption{Maximum (solid) and average (dashed) divergence with respect to the maximum value of the local $B$-field ($|\vec{\nabla} \cdot \vec{B}|\Delta x/|B_{\rm loc}|$) as a function of density at $z=6$ for different simulations.  The maximum divergence never becomes greater than $1\%$ of the maximum of the local $B$-field, indicating that it has little impact on the dynamics in our simulation. The sharp feature that occurs in the B11, B14, and B21 simulations at $n_{\rm H}\sim10^{-1.5}{\rm cm^{-3}}$ is due to SN feedback as it only occurs for $T>10^5$K gas. This feature is erased in the B11\_27 and B11\_29 simulations where the divergence is already higher than the values in the other simulation after the jump.}
\label{divergence}
\end{figure}

As stated earlier, the simulation includes an implementation of ideal MHD using a CT method.  Unlike the other hydrodynamic quantities in the simulation whose properties are stored as cell-centred quantities, the induction equation is solved in an integral form that requires the magnetic field properties to be stored on the faces of each cell in the AMR grid \citep{Teyssier2006,Fromang2006}. This consists of storing six $B-$field quantities for each gas cell. The CT method allows us to maintain the solenoidal constraint and conserve $\nabla\cdot B=0$ to machine precision, in contrast to divergence cleaning methods \citep[e.g.,][]{Powell1999,Dedner2002}. These divergence cleaning methods often struggle in certain astrophysical situations which may cause artificial amplification of the magnetic field \citep[e.g.,][]{Hopkins2016} and may no longer conserve physical quantities \citep{Toth2000}. Our code uses a divergence-preserving scheme to interpolate the magnetic field at coarse-fine boundaries of the grid \citep{Balsara2001,Toth2002}. To demonstrate the divergence-less behaviour, in Figure~\ref{divergence} we plot the maximum (solid) and average (dashed) divergence with respect to the maximum value of the local $B$-field ($|\vec{\nabla} \cdot \vec{B}|\Delta x/|B_{\rm loc}|$) as a function of density at $z=6$ for different simulations. For the simulations that are initialised with a uniform magnetic field along the $z$-axis, the maximum divergence never becomes greater than $1\%$ of the maximum $B$-field on the face of any cell, indicating that it has little impact on the dynamics in our simulation. For the simulations initialised with a random magnetic field with a given spectral slope and normalisation, the divergence is significantly greater due to the simulations being initialised using single precision. However, even for these simulations, the divergence errors are not dynamically important. Inside of galaxies, the divergence of the magnetic field is often six to ten orders of magnitude below the strength of the local $B$-field, demonstrating the near-machine precision conservation of the solenoidal constraint provided by the CT algorithm.

In contrast to simulations without magnetic fields, the time step in our simulations has an additional constraint where it can be limited by the Alfv\'en velocity ($v_{A}$), where
\begin{equation}
v_{A}=\frac{B}{\sqrt{\mu_0\rho}},
\end{equation}
where $\rho$ is the gas density and $\mu_0$ is the magnetic permeability of vacuum. In general, the Alfv\'en velocity does not set the strongest limit on the simulation time step, except when the magnetic field strength is very high and the density is low.  This situation can occur after SN events when the density drops considerably in the case of strong PMFs (e.g. the B11 simulation). In these regions, the additional magnetic pressure is extremely efficient in reducing the density of the SN bubbles causing the time step to drop to values of order years. To avoid this issue, we force a density floor of $3\times10^{-9}{\rm H\ cm^{-3}}$ in regions with $B>5\times10^{-8}G$. The value we choose for the density floor is considerably lower than the density reached in any of the simulations that never satisfy this condition; hence the floor does not impact the reionization of the IGM.  Furthermore, the density floor does not impact the escape of ionising photons from galaxies as the conditions for the floor are only met in regions of strong SN feedback where the gas is very far into the optically thin regime.  

\begin{figure}
\centerline{\includegraphics[scale=1,trim={0 0.6cm 0 1.4cm},clip]{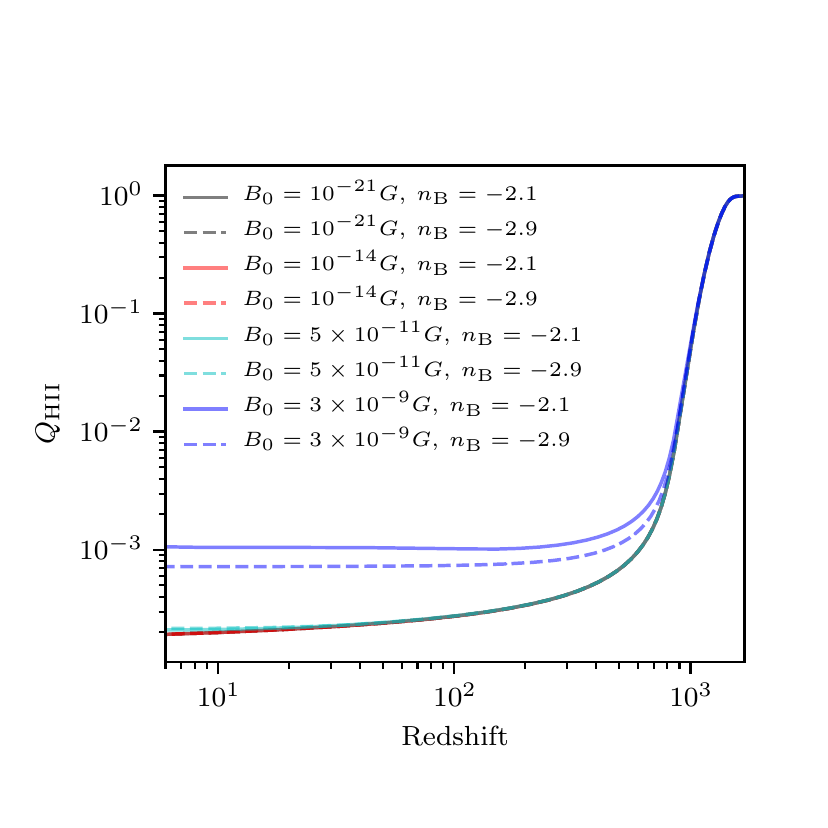}}
\centerline{\includegraphics[scale=1,trim={0 0.6cm 0 1.4cm},clip]{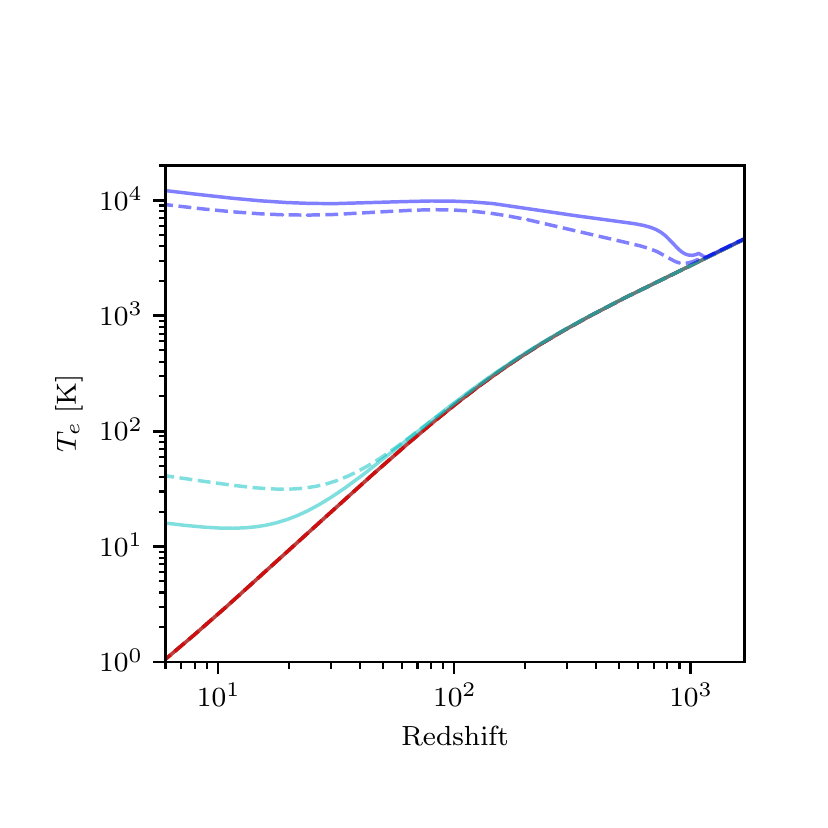}}
\caption{Hydrogen ionisation fraction (top) and electron temperature (bottom) as a function of redshift for a variety of seed field strengths ($B_0$) and spectral indices ($n_{\rm B}$) as a result of ambipolar diffusion and decaying magnetic turbulence.  These calculations were made with {\small recfast++} \protect\citep{Chluba2015}.  The ionisation fraction is only impacted for seed fields that are much stronger than those used in this work.  Similarly the impact of both ambipolar diffusion and decaying magnetic turbulence is very unlikely to have a serious impact on our results as the temperature change is $\ll100$ K.}
\label{recfast}
\end{figure}

Although ideal MHD is the current state-of-the-art for large cosmological simulations, it is well established that non-ideal effects can impact both the dynamics and thermodynamical state of the ISM \citep[e.g.,][]{Machida2008,Duffin2009,Marchand2018} as well as the state of the IGM post recombination \citep[e.g.][]{Sethi2005}. For the simulations presented here, the scales that we resolve are significantly larger than those required for individual star formation where non-ideal effects are important. However, for strong enough magnetic fields, on large scales, non-ideal effects may be important since ambipolar diffusion and decaying magnetic turbulence can both ionise and significantly increase the electron temperature ($T_e$) of the IGM \citep[e.g.,][]{Jedamzik1998,Subramanian1998,Sethi2005}. Using {\small recfast++} \protect\citep{Chluba2015}, in Figure~\ref{recfast}, we plot the ionised fraction $Q_{\rm HII}$ and $T_e$ of the IGM as a function of redshift for various initial values of $B_0$ and $n_B$, in the absence of other ionising sources. For seed fields with $B_0>1\,{\rm n}G$, a significant amount of heating and residual ionisation occurs.  In contrast, for the values of the seed fields considered in the this work (see Table~\ref{sims_tab}), both effects will have limited impact on the simulation; hence we can exclude them.  There is a small temperature enhancement in the IGM for the scenario with $B_0=0.05\,{\rm n}G$ at $z<40$; however, this is much below the ionisation temperature of hydrogen and is unlikely to impact our results.

\subsubsection{Cooling \& Non-Equilibrium Chemistry}
Gas cooling and non-equilibrium chemistry in the SPHINX-MHD simulations follow very closely to the prescriptions used in the original SPHINX simulations.  We employ a six species non-equilibrium chemistry model that tracks $e$, HI, HII, HeII, \& HeII, which are fully coupled to the radiation transfer through photoionization, photoheating, and UV radiation pressure.  For these primordial species, we compute cooling and heating due to photoionization, collisional ionisation, collisional excitation, recombination, bremsstrahlung, Compton cooling/heating off
the cosmic-microwave background, and di-electronic recombination \citep{Rosdahl2013}.  In addition, we compute cooling due to metal lines.  At $T>10^4$ K, cooling from metals is computed by interpolating cooling tables generated with {\small CLOUDY} \citep{Ferland1998} that were calculated assuming photoionization equilibrium with a UV background \citep{Haardt1996}.  At $T<10^4$ K, we use the fine structure cooling rates from \cite{Rosen1995}. A density- and redshift-independent temperature floor of 15 K is used throughout the simulation volume.

\subsection{Refinement}
Taking advantage of the adaptive grid in {\small RAMSES}, we allow the gas cells to refine when certain criteria are fulfilled. A cell is refined into eight equal-volume children cells using a quasi-Lagrangian scheme if its contained dark matter mass summed with its stellar and gas mass multiplied by $\Omega_{\rm m}/\Omega_{\rm b}$ is greater than eight times the dark matter particle mass.  Furthermore, we also allow a cell to refine if the width of the cell is larger than 25\% of the local Jeans length. We allow for up to 16 total levels of refinement which results in a spatial resolution of 7.31$h^{-1}$pc at $z=6$. As in the original {\small SPHINX} simulation, rather than maintain an approximately constant physical resolution by releasing new AMR levels at predefined redshifts, we allow the simulation to refine to level 16 at any redshift.  This implies that the physical cell width is significantly smaller at higher redshifts (e.g,. $\Delta x=2.44h^{-1}$pc at $z=20$).  In order to prevent particle scatterings due to strong two-body interactions, we smooth the dark matter particle density on one refinement level coarser than the maximum.

\begin{figure*}
\centerline{\includegraphics[scale=1,trim={0 0.0cm 0 0.0cm},clip]{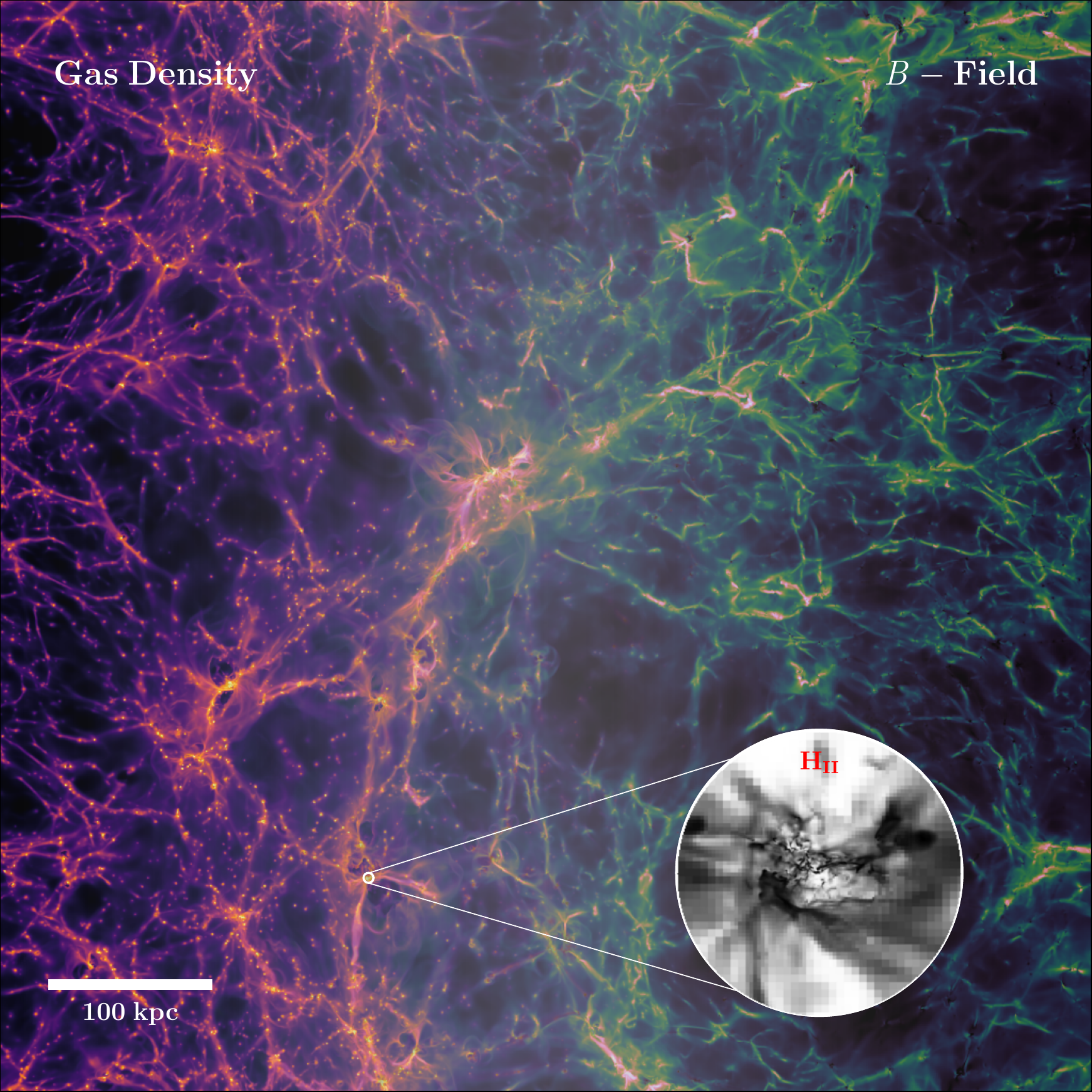}}
\caption{Average gas density (left) and magnetic energy density (right) along the line-of-sight for the B14 simulation at $z=6$.  The structure of the magnetic field follows that of the cosmic web. To illustrate the dynamic range of the simulations, the inset shows a map of the HII fraction for an individual galaxy where black represents neutral regions and white represents ionised regions.}
\label{hero}
\end{figure*}

\begin{figure*}
\centerline{\includegraphics[scale=1,trim={0 0.5cm 0 1.0cm},clip]{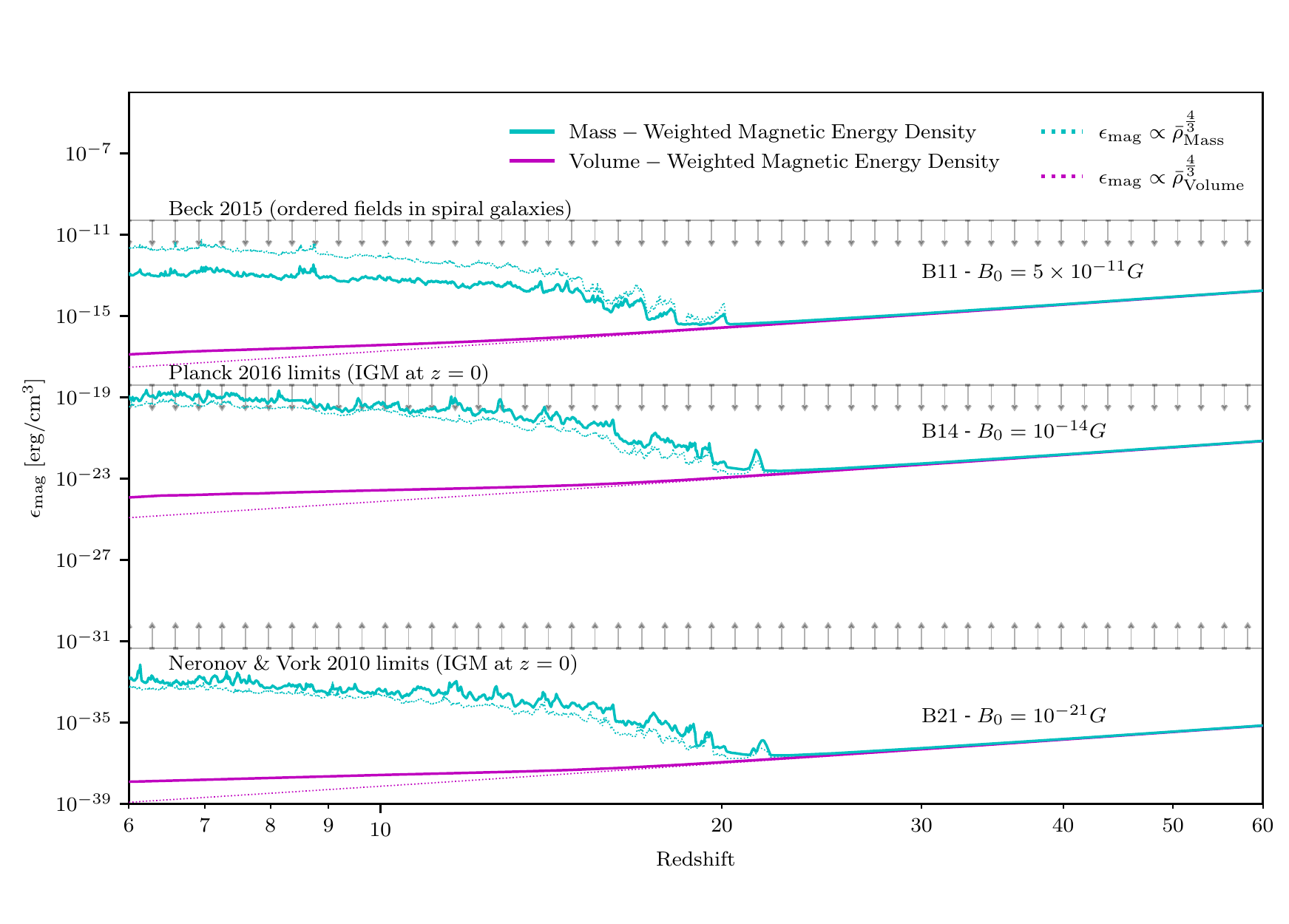}}
\caption{Evolution of the mass-weighted (cyan) and volume-weighted (magenta) energy-density in the magnetic field as a function of redshift for each of the simulations that employ a uniform PMF.  The solid lines represent the values calculated directly from the simulation while the dotted lines represents the expectation from flux conservation (i.e. $B\propto\rho^{2/3}$). The energy density dissipates with time due to cosmological expansion but as soon as the first structures collapse at $z\sim22$, the mass-weighted magnetic energy density deviates from cosmological expansion.  The fact that the solid line is consistently above the dotted line indicates that additional amplification beyond adiabatic compression is occurring in the simulation except for the mass-weighted measurement in B11, indicating that the magnetic energy is saturated in the densest regions of this simulation (i.e. galaxies).}
\label{bf_eve}
\end{figure*}

\subsection{Star Formation}
Collisionless star particles are allowed to form only in gas cells at the maximum level of refinement depending on the local properties of the gas. We employ a magneto-thermo-turbulent (MTT) star formation criteria \citep{Padoan2011,Hennebelle2011,Federrath2012}. We define the MTT Jeans length as:
\begin{equation}
\lambda_\text{J,MTT} = \frac{\pi \sigma_V^2 + \sqrt{36 \pi c_\text{s,eff}^2 G {\Delta x}^2 \rho + \pi \sigma_V^4}}{6 G \rho {\Delta x}},
\end{equation}
where $G$ is the gravitational constant and $\sigma_V$ is the gas turbulent velocity. The effective sound speed, $c_\text{s,eff}$, accounts for small-scale pressure support due to the presence of magnetic fields such that
\begin{equation}
    c_\text{s,eff} = c_\text{s} \sqrt{1 + \beta^{-1}},
\end{equation}
where $\beta = P_\text{thermal} / P_\text{mag}$\footnote{Typical value for $\beta$ in star-forming gas in the B14 and B21 simulations are $10^5$ and $10^{19}$, respectively and hence the magnetic field does not impact the effective sound speed in these simulations. For the B11 simulation, $\beta$ can drop to values $\ll1$ for very cold gas, substantially increasing the effective sound speed.}.
Star particles are only allowed to form when the local gas density is $>10\,{\rm cm^{-3}}$, the cell is a local density maximum compared to its immediate neighbours, the gas velocities are locally convergent, and the MTT Jeans length is unresolved such that $\Delta x > \lambda_\text{J,MTT}$.  When a gas cell satisfies these criteria, stars are formed following a Schmidt law \citep{Schmidt} with a star formation rate of
\begin{equation}
\dot{\rho}_\text{star} = \epsilon_\text{ff} \frac{\rho}{t_\text{ff}}.
\end{equation}
The free-fall time of the gas is defined as
\begin{equation}
t_\text{ff} = \sqrt{\frac{3\pi}{32 G \rho}}.
\end{equation}

The local efficiency of star formation, $\epsilon_\text{ff}$, is computed from the magneto-thermodynamical properties of the host gas cell following the multi-scale PN model \citep{Padoan2011} from \citet{Federrath2012} such that:
\begin{equation}
\epsilon_\text{ff} = \frac{\epsilon_\text{cts}}{2 \phi_t} \exp{\left(\frac{3}{8} \sigma_s^2\right)} \left[1 + \text{erf}\left(\frac{\sigma_s^2 - s_\text{crit}}{\sqrt{2 \sigma_s^2}}\right) \right].
\end{equation}
Here, $\sigma_s = \ln{\left(\rho_{\rm gas} / \left<\rho_{\rm gas}\right>\right)}$, $1/\phi_t = 0.57$ takes into account the uncertainty in free-fall timescales at the mean density of the cloud and that of higher density gas, and $\epsilon_\text{cts}=0.5$ represents the maximum fraction of the gas that can be converted into stars when accounting for proto-stellar feedback. Finally, $s_\text{crit}$ is the critical density beyond which post-shock gas in a magnetised cloud can collapse against magnetic pressure support \citep{Hennebelle2011,Padoan2011} so that
\begin{equation}
    s_\text{crit} = \ln{\left(0.067 \; \theta^{-2} \alpha_\text{vir} \mathcal{M}^2 f\left(\beta\right)\right)}.
\end{equation}
In this equation, $\alpha_\text{vir}=2E_{\rm kin}/|E_{\rm grav}|$ is the virial parameter, $\theta = 0.35$ \citep{Hennebelle2011}, $\mathcal{M}$ is the mach number, and $f \left( \beta \right)$ is a dimensionless quantity defined in \cite{Padoan2011} as 
\begin{equation}
f \left( \beta \right) = \frac{(1+0.925\beta^{-3/2})^{2/3}}{(1+\beta^{-1})^2}.
\end{equation}
Thermo-turbulent star formation prescriptions of this ilk have already been described in \cite{Kimm2017,Trebitsch2017,Rosdahl2018} and the MHD extension has been used by \cite{Katz2019c,Sergio2020,Sergio2020b}. Note that, in contrast to the original SPHINX simulations, we do not include the stellar and dark matter density when calculating the local density to measure the star formation efficiency. Dark matter particles are sparsely sampled in the ISM and the mass of the star particle impacts the local efficiency. Thus for these simulations, we choose to only consider gas density. This leads to reionization occurring slightly later. 

\subsection{Stellar Feedback}
Star particles in the simulation can explode via SN throughout the first 50~Myr of their life. We randomly sample a realistic delay-time distribution over this time period to determine when the SNe occur. We use the mechanical feedback scheme of \cite{Kimm2015} as was also used in {\small SPHINX} to inject momentum into the surrounding cells of the star particle depending on resolution.  The aim is to inject the final snowplow momentum of a SN remnant if the adiabatic phase is unresolved, or to allow the remnant to evolve naturally if the adiabatic phase is resolved. For each individual SN event, the equivalent of $10^{51}$ergs is injected.  We adopt a \cite{Kroupa2001} stellar initial mass function (IMF) and recycle 20\% of the total mass of every star particle back into the simulation as gas.  Some of this material will be metal enriched due to nucleosynthesis in the stars and we assume that 7.5\% of the ejecta is in the form of elements heavier than hydrogen and helium. For a standard \cite{Kroupa2001} IMF, with a maximum stellar mass of $100\,{\rm M_{\odot}}$, we would expect $\sim1$ SN event per $100\,{\rm M_{\odot}}$ in stars (i.e. 10 SN per star particle in our simulation).  However, in order to reproduce a realistic UV luminosity function and stellar mass-halo mass relation in the early Universe \citep{Garel2021}, we follow the approach of \cite{Rosdahl2018} and boost the number of SNe per star particle by a factor of four (i.e. 40 SN explosions per star particle) as was done for {\small SPHINX}.

\begin{figure*}
\centerline{\includegraphics[scale=1,trim={0 0.1cm 0 0.4cm},clip]{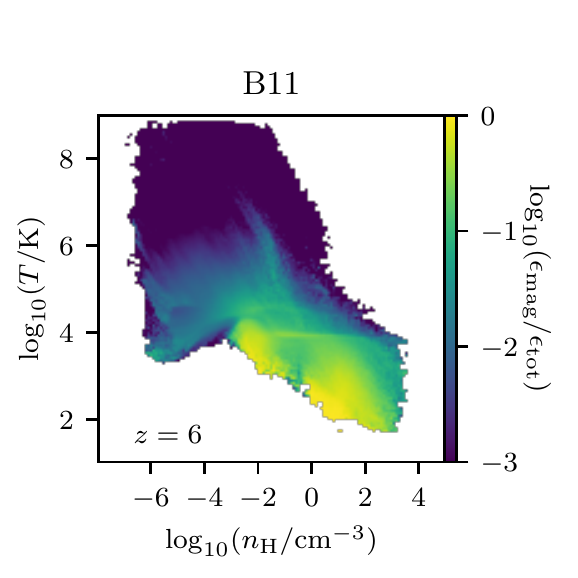}\includegraphics[scale=1,trim={0 0.1cm 0 0.4cm},clip]{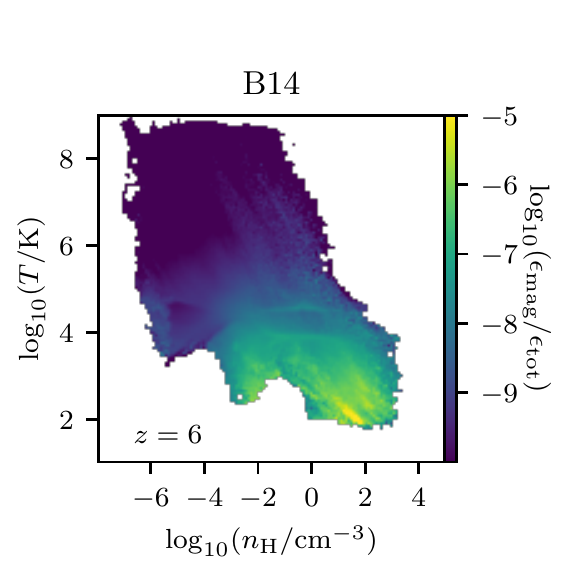}\includegraphics[scale=1,trim={0 0.1cm 0 0.4cm},clip]{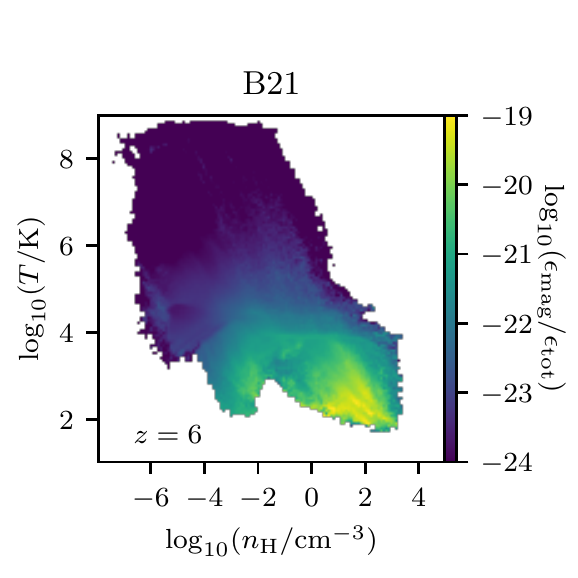}}
\caption{Temperature-density phase space diagrams of the entire simulation volume coloured by the fraction of the total energy in the bin (magnetic, turbulent kinetic, and thermal) that is represented by the magnetic component. For the B14 and B21 simulations, the magnetic energy never reaches equipartition with either the thermal or turbulent kinetic energy across the entire phase-space. In contrast, for the B11 simulation, in certain regions of phase-space, particularly at intermediate densities and temperatures, the magnetic component can become dynamically important and represent $>1\%$ of the total energy. For individual galaxies, this diagram will appear different.}
\label{equip}
\end{figure*}

In addition to momentum, we inject radiation that emanates from every star particle in the simulation.  The amount of radiation injected on each fine time step in the simulation is calculated using spectral energy distributions computed for binary stellar populations \citep[BPASSv2,][]{Eldridge2008,Stanway2016} for a stellar IMF with a maximum mass of $100{\rm M_{\odot}}$. The effects of binary stellar populations are now well studied in the context of reionization \citep{Stanway2016,Ma2016,Rosdahl2018,Ma2020} and our fiducial simulations would not reionize without assuming binary stellar populations \citep{Rosdahl2018} unless we modified the subgrid escape fraction (i.e. a parameter representing the fraction of photons that escape the unresolved molecular cloud\footnote{This value can be $>1$ if the ionised channels through which photons are expected to escape are unresolved by our simulations.}).  For this work, we have set the subgrid escape fraction to 1; hence, we do not modify the total number of photons injected into each cell as computed from the age and metallicity dependent SED. For the simulations with PMFs that reionize early due to the enhancement in the number of dwarf galaxies, adopting an SED with fewer ionising photons or an SED with LyC production more biased towards young stellar ages for the stellar population (as is the case without binary stars) may help bring the simulations in better agreement with observations.  However the primary goal of this work is to study the impact of magnetic fields on reionization rather than to develop a model that is in perfect agreement with all observational constraints.

\subsection{Halo Finding}
To identify haloes in the simulation, we use the {\small ADAPTAHOP} algorithm using the most massive submaxima (MSM) mode \citep{Aubert2004,Tweed2009}. We identify all haloes and subhaloes that are represented by at least 20 dark matter particles; however, for the analysis in the work, we only consider haloes that contain at least 300 dark matter particles (i.e. $M_{\rm halo} > 6.24\times10^6h^{-1}{\rm M_{\odot}}$). The virial mass of each halo is determined by fitting a triaxial ellipsoid to each halo and subhalo with a centre located at the densest region of the halo and iteratively decreasing the volume of the halo until the virial theorem is satisfied.  This method produces haloes that have similar virial masses and radii to adopting a spherical overdensity criteria of 200$\rho_{\rm crit}$  \citep{Rosdahl2018}. We assign star particles to haloes based on their location, i.e. whether they reside within the virial radius of a dark matter halo or subhalo. In case of overlap (i.e. a star particle overlapping with more than one halo) it is assigned only to the closest halo, the distance being measured as $d=r/R_{\rm vir}$, where $r$ is the distance from the halo centre and $R_{\rm vir}$ is the virial radius of the halo. We also ignore all sub-halos that are fully contained within their parent halo, and instead assign their stellar particles to the parent halo.

\subsection{Escape Fractions}
We measure LyC escape fractions from every halo that contains stars in every simulation snapshot using the publicly available {\small RASCAS} code \citep{Leo2020}. We post-process the simulation snapshots by casting 500 rays from each star particle in random directions and measuring the optical depth to neutral hydrogen and helium up to the virial radius of the halo.  The escape fraction for each star particle is measured as the average of the escape fraction along each of the 500 rays and the global escape fraction for each snapshot is taken as the luminosity weighted average among all star particles.  The choice of 500 rays was found to be a robust estimator by \cite{Rosdahl2018,Katz2020b} as minimal differences were found when sampling up to 100,000 rays per star particle. Furthermore, we note that \cite{Trebitsch2017} found that measuring $f_{\rm esc}$ using ray-casting in post-processing resulted in very similar results to directly measuring the LyC flux present in the simulation that crosses the virial boundary of the halo. Because we use the variable-speed-of-light approximation, measuring the delay time between the moment when photons were emitted and when they cross the virial boundary becomes very difficult due to the adaptive nature of the grid. Hence ray-tracing is a more robust approach in our simulations.

\section{Results}
\label{results}
We now present our results on the impact of magnetic fields on reionization.  For each simulation we save snapshots of the volume at $5$~Myr intervals from the point at which the first stars form\footnote{This results in $\sim150$ total snapshots for each simulation, corresponding to $\sim40-100$Tb of data per simulation.} and many of the statistics are computed on-the-fly at every coarse time step in the simulations, providing a time resolution of $\sim1,000-10,000$~years. All simulations have been evolved to $z=6$ except the B11\_24 simulation which was evolved until it reached a volume-weighted ionisation fraction of 50\% at $z\sim55$.

\subsection{Evolution of the magnetic field}
In Figure~\ref{hero} we show a map of the gas column density and density-weighted magnetic field strength at $z=6$ for the B14 simulation. The gas density exhibits a rich filamentary structure that mimics the dark matter density field seen in Figure~\ref{DMmapsz6}, except for the least persistent filaments that are the least self-shielded to reionization \citep{Katz2020}. The large scale magnetic field clearly follows the density field as expected due to flux conservation as the gas collapses onto filaments and is fed into galaxies. Evidence for strong SN feedback is visible around the locations of many galaxies as bipolar outflows and density cavities. For the most massive halo in the simulation that resides in the centre of the image, we can see that the magnetic field is dragged out of the galaxy along with the gas by the SN feedback and into the low-density IGM.

\begin{figure}
\centerline{\includegraphics[scale=1,trim={0 0.5cm 0 1cm},clip]{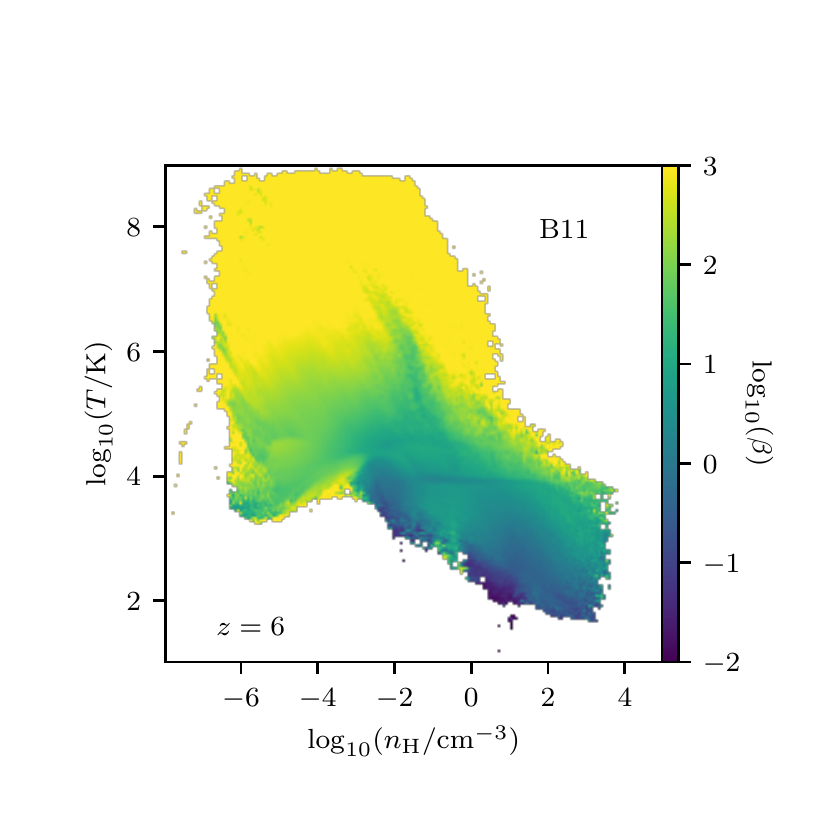}}
\caption{Temperature-density phase space diagram of the B11 simulation at $z=6$ coloured by $\beta$, the ratio of thermal to magnetic pressure. At high densities and low temperatures, $\beta$ drops below one and impacts star formation.}
\label{beta_sf}
\end{figure}

\begin{figure}
\centerline{\includegraphics[scale=1,trim={0 0.5cm 0 1.0cm},clip]{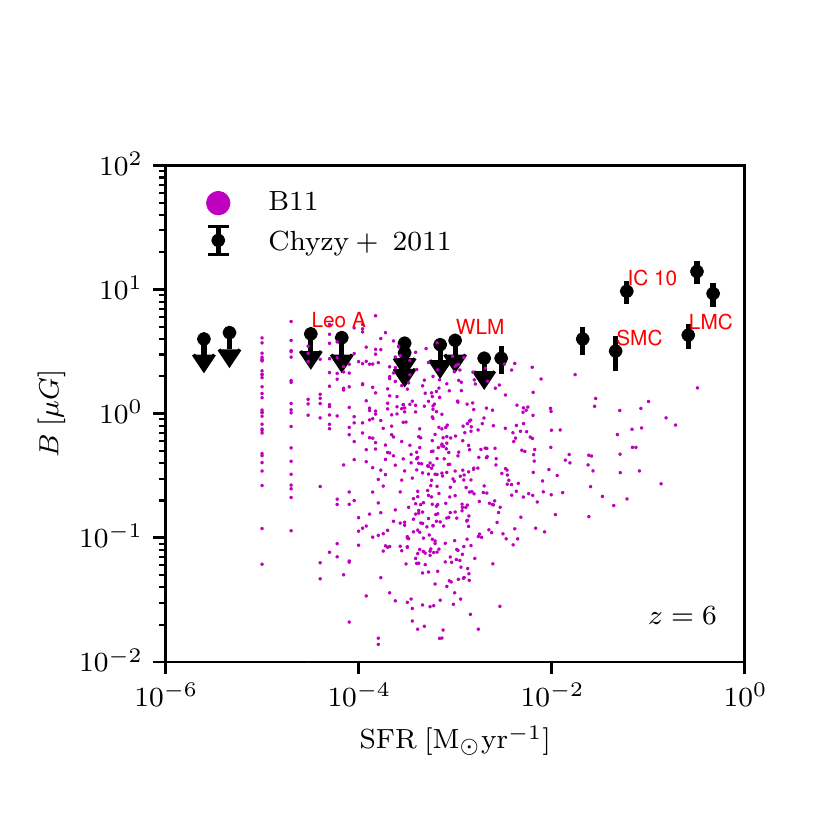}}
\caption{Mass-weighted average magnetic field inside haloes versus 100~Myr-averaged SFR in the B11 simulation at $z=6$ compared to those in local group dwarf galaxies.}
\label{b_sfr}
\end{figure}

\begin{figure*}
\centerline{\includegraphics[scale=1,trim={0 0.0cm 0 0.0cm},clip]{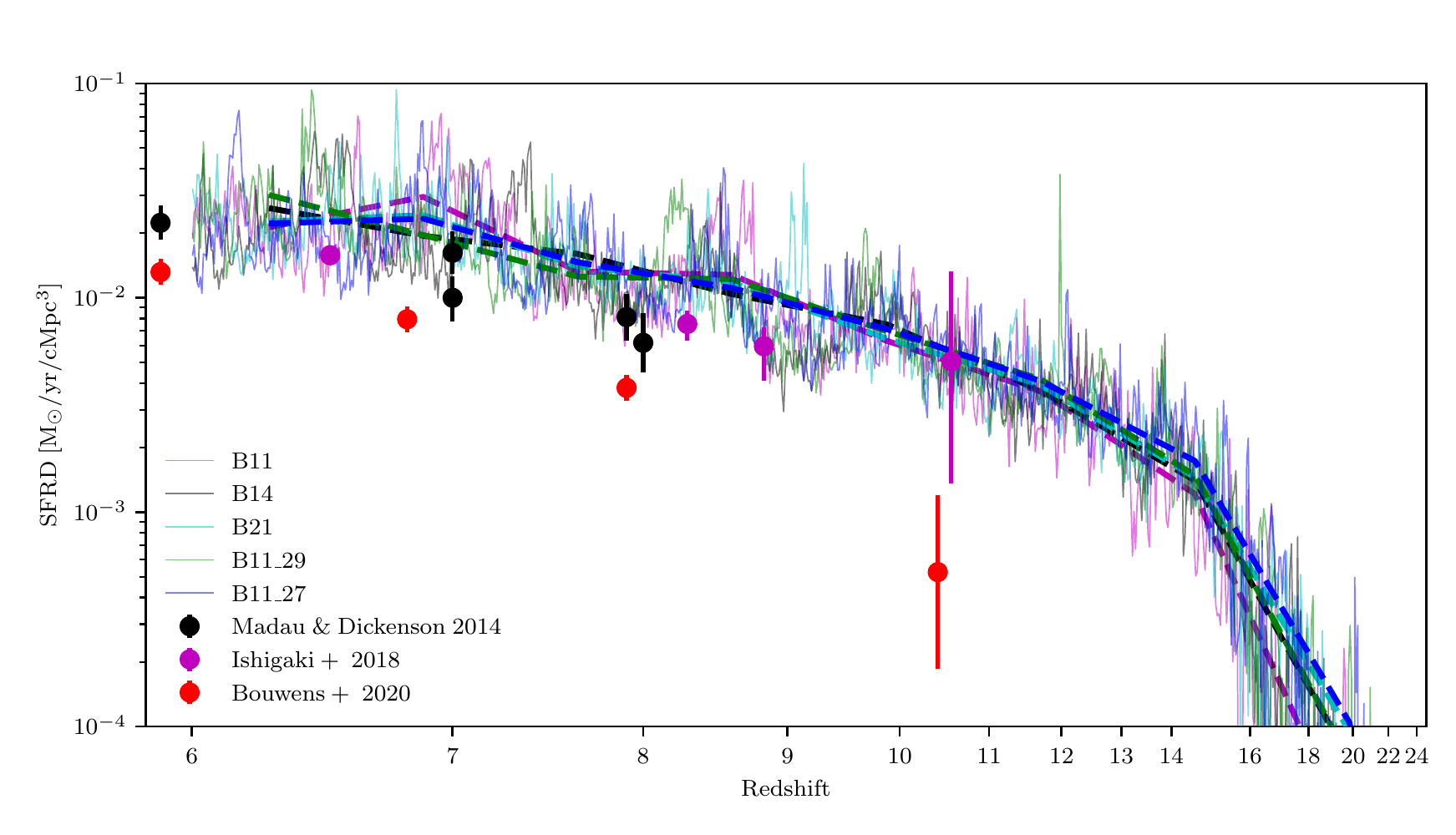}}
\caption{Star formation rate density (SFRD) as a function of redshift for our simulations compared to observational estimates from \protect\cite{Madau2014,Ishigaki2018,Bouwens2020}. Solid lines represent the 1Myr-averaged SFRD while dashed lines represent the 100Myr-averaged SFRD.}
\label{sfrd}
\end{figure*}

In Figure~\ref{bf_eve} we show the evolution of the mass-weighted and volume-weighted magnetic energy density in cyan and magenta, respectively, as a function of redshift for the B21, B11, and B14 simulations. In all simulations, the mean strength of the magnetic field decreases with redshift due to cosmic expansion such that $B\propto (1+z)^{2}$. This holds until the collapse of the first galaxy where in Figure~\ref{bf_eve}, we observe that the mass-weighted magnetic energy density begins to deviate from the volume-weighted magnetic energy density at $z\sim23$ for the B21 and B14 simulations. First collapse is slightly delayed in the B11 simulation where the additional pressure support from the magnetic field temporarily prevents runaway collapse of the gas \citep{Sergio2020}. As the Universe evolves and more galaxies collapse, the mass-weighted magnetic energy density continues to increase in all simulations. The dotted lines in Figure~\ref{bf_eve} represent the expectations from flux conservation (i.e. $\epsilon_{\rm mag}\propto\rho^{-4/3}$, where $\rho$ is either the volume-weighted or mass-weighted average gas density in the simulation). For the B21 and B14 simulations, the mass-weighted magnetic energy density remains consistently higher than the expectation from flux conservation indicating that other mechanisms (e.g., turbulent or rotational motions) act to amplify the magnetic field. We emphasise that numerical viscosity and diffusion as well as limited resolution and an adaptive grid prevent us from reaching the Reynolds numbers needed to sustain significant turbulent amplification. We can estimate the effective Reynolds number as ${\rm Re}=(L/\epsilon \Delta x)^{4/3}$, with $L$ being the typical length scale or turbulent injection scale of the system and $\epsilon\sim7$ \citep{Donnert2018}. Setting $L=0.1r_{\rm vir}$, we find ${\rm Re}\sim4$\footnote{Using an alternative estimate for the Reynolds number, $2L/\Delta x$ \citep{Rieder2017}, we find mean values of Re in our simulation of $\sim40$ (assuming the entire central region of the galaxy is resolved at the highest resolution) which are slightly above the critical value.} for galaxies in our simulation assuming that the entire central region of the galaxy is resolved that the maximum spatial resolution. This mean value is below the critical Reynolds numbers of $30-35$ needed to trigger a turbulent dynamo for a magnetic Prandtl\footnote{The ratio between viscosity and magnetic diffusivity.} number of $\sim1$ \citep[e.g.][]{Haugen2004b,Haugen2004,Brandenburg2005}. Hence we do not expect extreme deviations between the measured magnetic energy and that predicted from flux conservation. The most resolved systems have Reynolds numbers above the estimated critical value which perhaps explains the enhancement above flux conservation that we observe in the B14 and B21 simulations.

Significant deviations between the measured and flux conservation predictions for the mass-weighted magnetic energy density are also observed for the B11 simulation where the PMF is strong enough that the magnetic energy becomes saturated inside haloes.  In all three simulations, after first collapse, the volume-weighted magnetic energy density increases above the predicted value as the galaxies decouple from the expanding background and magnetic energy is ejected from the galaxies and into the IGM. Observational lower limits have been estimated using the TeV emission from blazars and their GeV secondary emission \citep{Neronov2010,Dolag2011}, but their interpretation requires further investigation \citep{Broderick2012,Broderick2018}. For scenarios where the PMF results in an IGM magnetic field weaker than such proposed lower limits (e.g., our B21 simulation), invoking astrophysical phenomena, such as magnetised winds would be required for consistency with such constraints. In contrast, if the filling factor of the IGM magnetic field was shown to be robustly $\ll1$, this may rule out strong magnetic fields where the volume-weighted magnetic field strength is higher than any upper limit set by observations. In all of our simulations, the volume filling factor of magnetic fields is equal to 1 because of how they were initialised. Scenarios accounting for cosmological perturbations of the magnetic field by e.g. modelling an initial magnetic spectrum, such as our B11\_27 and B11\_29 runs, might still be found to be in agreement due to the spatial fluctuations depending on the exact volume filling factors and magnetic field strengths observed in the IGM.

Within the ISM of galaxies, weak magnetic fields are expected to be amplified by a turbulent dynamo until their saturation at a fraction of the turbulent kinetic energy. In our simulations, saturation will not occur at equipartition due to the drastic difference between the Prandtl and viscous and magnetic Reynolds numbers achievable in our simulation compared to the real ISM \citep{Sergio2018}. Numerical simulations show that saturation occurs when the magnetic pressure reaches $\sim 2\%-5\%$ of the turbulent energy for Prandtl numbers of $\mathcal{O}(1)$ \citep[e.g.,][]{Federrath2014,Tricco2016,Rieder2017}, although saturation fractions may exhibit a much wider range depending on Prandtl and Mach numbers \citep[e.g.][]{Schober2015,chirakkara2021efficient}. In Figure~\ref{equip} we show temperature-density phase-space diagrams of all the gas in the box at $z=6$ for the B21, B14, and B11 simulations, coloured by the ratio of magnetic energy to total energy (thermal, turbulent kinetic, and magnetic). We use a simple approximation for the turbulent kinetic energy and define it as $\frac{1}{2}m_{\rm gas}\sigma^2$, where $m_{\rm gas}$ is the gas mass of the cell and $\sigma$ is the velocity dispersion. For the B14 and B21 simulations, the magnetic energy remains far below both the thermal and kinetic energy indicating that it is not dynamically important in any region of the phase space. Interestingly, we can see some striations in the high-temperature and low-density regions of the panels for these simulations. These features represent individual SN events dragging the magnetic field out of galaxies. In contrast, in the B11 simulation, we can identify localised regions of the temperature-density phase space where the magnetic field is dynamically important and represents more than 50\% the total energy. In particular, this occurs often at $T\lesssim10^4$K, inside of galaxies.

Perhaps more interesting in the context of star formation is the value of $\beta$ (i.e. the ratio of thermal to magnetic pressure) in low-temperature, high-density star-forming gas. In Figure~\ref{beta_sf}, we plot a 2D histogram of density and temperature coloured by the values of $\beta$. In star forming regions, $\beta$ drops to values $\ll1$, consistent with observations of local molecular clouds where typical values can range from $\beta\sim10^{-3}-10^{-1}$ \citep{Crutcher2012,Krumholz2019}. Clearly, the magnetic field in the B11 can have a drastic impact on our SF recipe. In contrast, the B14 and B21 simulations exhibit values of $\beta\gg1$ and hence for these simulations, the additional magnetic pressure does not impact star formation.

\begin{figure}
\centerline{\includegraphics[scale=1,trim={0 0.6cm 0 1.4cm},clip]{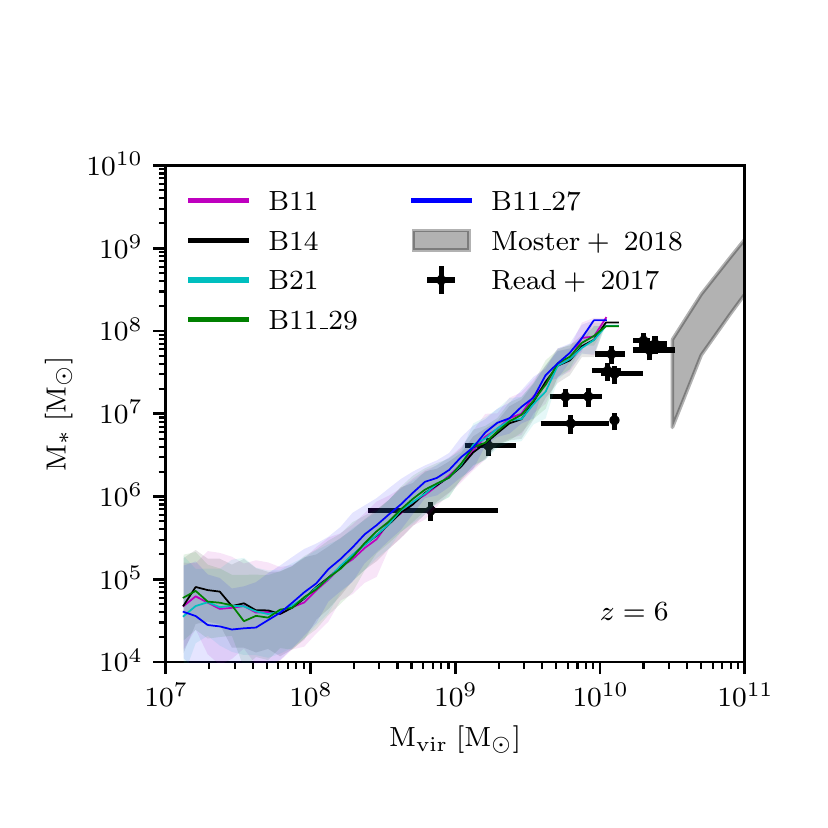}}
\centerline{\includegraphics[scale=1,trim={0 0.6cm 0 1.4cm},clip]{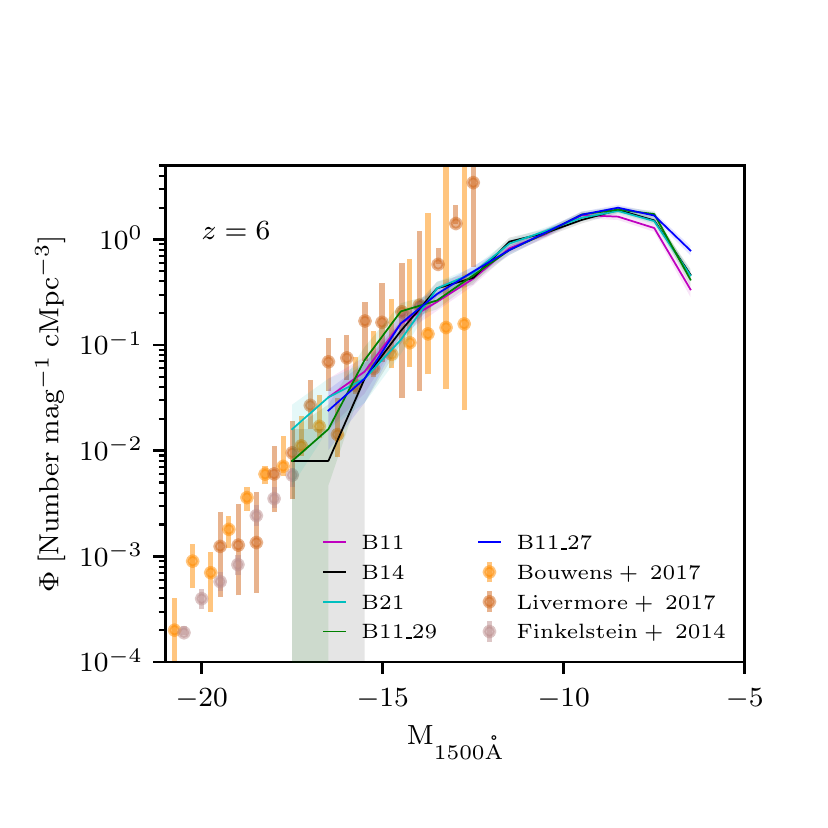}}
\centerline{\includegraphics[scale=1,trim={0 0.6cm 0 1.4cm},clip]{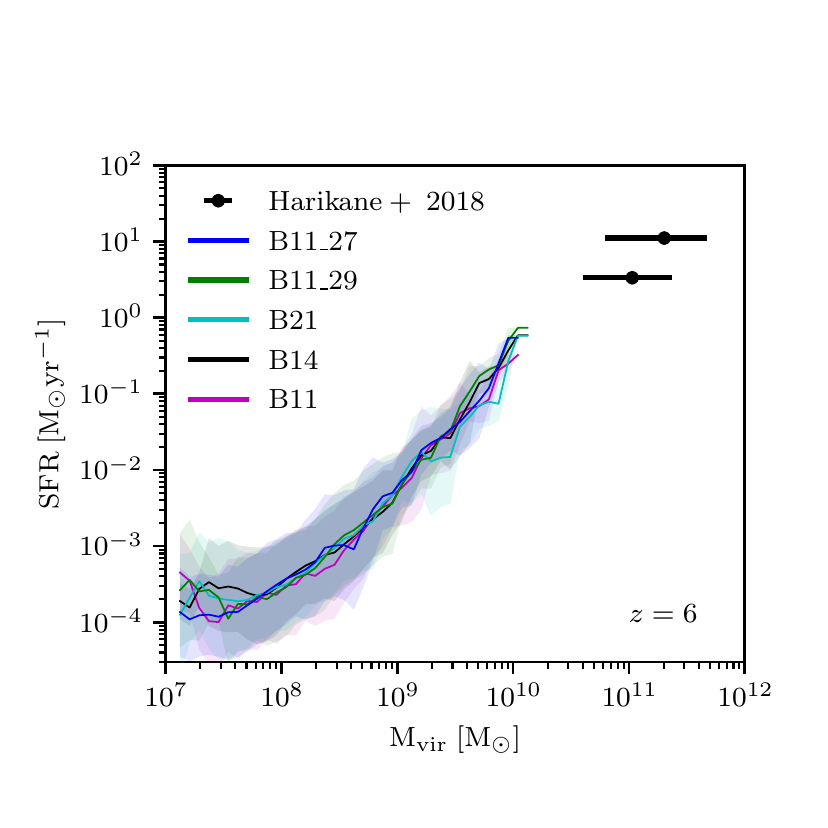}}
\caption{Stellar mass-halo mass relation compared to local group dwarf galaxies \protect\citep{Read2017} and constraints from abundance matching \protect\citep{Moster2018} (top), $1500\angstrom$ UV luminosity function compared to observational constraints from \protect\cite{Finkelstein2015,Livermore2017,Bouwens2017b} (centre), and SFR-halo mass relation compared to observational constraints from \protect\cite{Harikane2018} (bottom), for each of the simulations at $z=6$.  In the top and bottom panels, lines and shaded regions represent the running means and $1\sigma$ scatter around the relation.}
\label{sm}
\end{figure}

In Figure~\ref{b_sfr}, we compare the mass-weighted magnetic field strengths inside of haloes in the B11 simulation at $z=6$ with estimates for the magnetic field strengths of local group dwarf galaxies as a function of SFR \citep{Chyzy2011}.  Interestingly, for the galaxies with low SFR (i.e. SFR$<10^{-2}{\rm M_{\odot}yr^{-1}}$), our measurements appear consistent with the constraints from local dwarfs. In contrast, the galaxies in our simulation with higher SFRs tend to exhibit magnetic field strengths over the whole halo that are lower than local galaxies such as the LMC and SMC. A detailed comparison between observed and simulated magnetic fields would require studying in better detail the properties of the gas inside galaxies and its synchrotron emission, which is beyond the scope of this work. However, the limited resolution of our simulations might be inhibiting both the growth rate and maximal saturation strength of magnetic fields in the simulated galaxies \citep{Rieder2017,Sergio2018}. Furthermore, there is no reason {\it a priori} for dwarf galaxies in the epoch of reionization to exhibit magnetic field strengths similar to those in Local Group dwarf galaxies. ISM conditions are expected to be significantly different as high-redshift galaxies are likely more compact, gas rich, and turbulent, while $z=0$ galaxies have had significantly more time to evolve, amplify, and organise on galactic scales their seed magnetic fields. Even for weak seed fields (i.e. $B\sim10^{-20}G$), saturation at levels of $\gtrsim10^{-6}G$ are predicted within the free-fall time of a primordial halo \citep{Schober2012} which is consistent with the saturation levels achieved in our simulations with the strongest PMFs. 

Observational constraints on the magnetic fields within galaxies at high redshift remain scant; however, for massive galaxies at $z\sim3$ where such observations can be made, it seems that these galaxies exhibit magnetic field strengths consistent with those observed at $z=0$ \citep{Bernet2008}, supporting the possibility of magnetic saturation in galaxies at extremely early times in the evolution of the Universe. 

Thus far our analysis has focused only on the simulations that are initialised with uniform magnetic fields. The magnetic field evolution in the simulations with randomly seeded fields and specific spectral slopes are not fundamentally different from those already discussed.

\subsection{Impact of magnetic fields on star formation}
One of the primary goals of this work is to better understand how the presence of PMFs impacts galaxy formation in the epoch of reionization.  In this Section, we focus on how PMFs impact star formation.  

In Figure~\ref{sfrd}, we show the star formation rate density (SFRD) for each of the simulations compared with the observational estimates from \cite{Madau2014,Ishigaki2018,Bouwens2020}. In all simulations, the onset of star formation occurs at $z\sim20$ and increases rapidly until $z\sim12-14$.  As more haloes form stars, stellar feedback regulates the formation of new stars and the SFRD settles to a value slightly higher than $\sim10^{-2}{\rm M_{\odot}/yr/cMpc^3}$ by $z=6$, which is consistent with observations. Due to the small volume of the simulation, the SFRD continues to fluctuate by factors up to an order of magnitude, even at $z<7$ as intense star formation events in the massive galaxies can dominate the star formation rate of the simulation volume. It should be noted that the observational estimates of the SFRD from \cite{Madau2014,Bouwens2020} are calculated up to a limiting magnitude which is barely probed by our simulation as a consequence of the small volume. Hence the observational data point at $z=10.5$ falls far below our simulated data, where lower luminosity galaxies are expected to contribute significantly to the star formation rate \citep[e.g.][]{Behroozi2020}. In contrast, the data from \cite{Ishigaki2018} are integrated to a limiting magnitude of -11 and is much more consistent with our simulations. It is perhaps a coincidence that the brighter galaxies at redshifts closer to 6 probed by observations have a similar SFRD compared to the low mass galaxies in our computational volume.

For the strengths of the PMF sampled in this work, we do not see a significant impact on the total amount of star formation in the simulations. The evolution of the SFRD is consistent across simulations, regardless of the strength of the PMF. Similarly, the total stellar masses formed in the simulations are all within a factor of two by $z=6$. This behaviour is consistent with our earlier work \cite{Katz2019c,Sergio2020} as well as others that have demonstrated that the stellar content of galaxies is not impacted for PMFs with $\lesssim10^{-9}G$ \citep[e.g.][]{Marinacci2016} nor does seeding mechanism play a substantial role \citep{Garaldi2020}.  

\begin{figure}
\centerline{\includegraphics[scale=1,trim={0 0.6cm 0 1.4cm},clip]{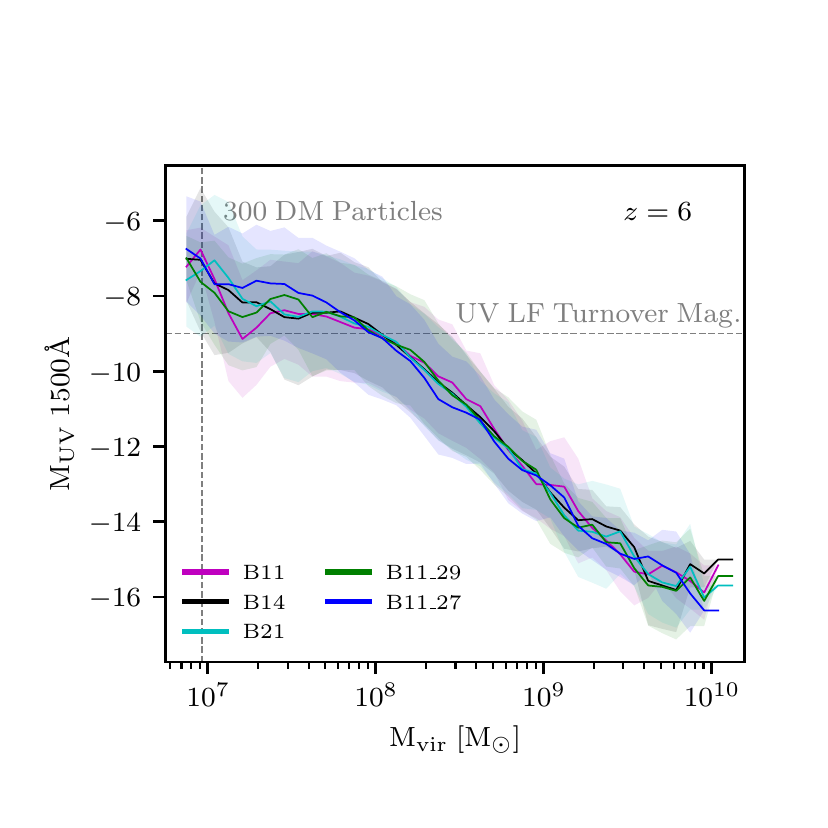}}
\caption{Virial mass versus $1500\angstrom$ UV luminosity for each of the simulations at $z=6$. The lines and shaded regions represent the running means and $1\sigma$ scatter around the relation.}
\label{hm_mag}
\end{figure}

In the top panel of Figure~\ref{sm} we show the stellar mass-halo mass relation at $z=6$ for each simulation compared to $z=6$ estimates from abundance matching \citep{Moster2018} as well as individual local group dwarf galaxies \citep{Read2017}. The lines and shaded regions represent the running mean of stellar mass and the $1\sigma$ standard deviation for haloes that host at least one star particle. Note that the fraction of haloes that host a stellar population at a halo mass of $10^8{\rm M_{\odot}}$ is only 50\%.  In general, the simulations are in very good agreement with each other and exhibit a slightly steeper relation compared with local dwarf galaxies.  Below halo masses of $10^8{\rm M_{\odot}}$ there is a small deviation between the B11\_27 simulation and the other runs such that, on average, the stellar mass for a given halo mass is lower in this run. 

This is more easily observed in the second panel of Figure~\ref{sm} where we show the $1500\angstrom$ luminosity function for each of the simulations at $z=6$ compared to observations from \cite{Finkelstein2015,Livermore2017,Bouwens2017b}.  At UV magnitudes brighter than $-13$, where the simulations overlap with observations, we find very good agreement between our predictions and observations.  There is some disagreement in the observational data at magnitudes fainter than $-15$ where observational estimates are derived from lensed galaxies where magnification uncertainties can be substantial and observational volumes are small.  Nevertheless, our predictions seem to fall in between the different observational constraints.  At the faintest magnitudes (i.e. $>-9$), the luminosity function turns over. We caution that this is partially due to limited resolution in the simulation, both because the haloes are resolved by fewer than $10,000$ DM particles, and the minimum stellar mass in the simulations is $1000\,{\rm M_{\odot}}$. We demonstrate this further in Figure~\ref{hm_mag} where we plot the relation between virial mass an $1500\angstrom$~UV magnitude. The turnover in the LF occurs at a UV magnitude of $-9$ which corresponds to a virial mass of $\sim10^8{\rm M_{\odot}}$. Such haloes are resolved by $\sim3,000$ DM particles which is sufficient to resolve some of their DM and gas properties but only marginally resolve the stellar content \citep[e.g.][]{Brooks2007}. The feedback from the process of reionization has the capacity to induce this turn-over; however, the simulations may not have not reached the point where we expect this to occur since the galaxies can self-shield long after reionization \citep{Katz2020b}.   

At faint magnitudes, the B11\_27 simulation exhibits a slightly higher number density of galaxies, consistent with the enhancement in the number of low mass dark matter haloes. The differences at the faint-end are larger than the $1\sigma$ uncertainty. Furthermore, comparing the B11\_27 and B14 simulations UV magnitude distributions with a two-sided KS tests results in a $p$-value of 0.0038 indicating that they were likely drawn from a different underlying distribution. The enhancement in faint galaxies combined with the tendency of strong magnetic fields to delay star formation in haloes translates to a small decrease in the average stellar mass at halo masses $<10^8{\rm M_{\odot}}$ in the top panel of Figure~\ref{sm}.  Similar behaviour was observed in the zoom-in simulations of \cite{Sanati2020} at $z=0$ where data from the Local Group was used to constrain both the strength and slope of the PMF.

Note that the UV luminosity functions presented here represent the intrinsic UV magnitudes of the galaxies and do not include the impact from dust. While this is likely a good approximation at such faint magnitudes \citep{Ma2016}, we cannot rule out the possibility that the intrinsically brighter galaxies, which do not exist in our small volume, would be reddened to the brightest magnitude bins represented in our simulations, and hence increase the simulated luminosity function upwards and out of the range of observational constraints. This effect is expected to be mild \citep{Garel2021}.

In the bottom panel of Figure~\ref{sm} we show the $z=6$ relation between halo mass and the 100~Myr-averaged SFR compared with observations from \cite{Harikane2018}.  Once again, we observe that all simulations exhibit similar behaviour, regardless of the strength of slope of the PMF.  This holds true even for the lowest mass haloes with ${\rm M_{vir}<10^8M_{\odot}}$. Unfortunately, the halo masses where observational constraints exist for the ${\rm M_{vir}-SFR}$ relation are an order of magnitude larger than those probed by our simulation.  Nevertheless, consistent with our earlier work \citep{Rosdahl2018}, extrapolating the trends seen in the simulations presented here would slightly over-predict the few constraints that we have from observations. However, including dust obscuration may bring the simulations into agreement with observations (Garel~et~al.~Submitted). 

\begin{figure}
\centerline{\includegraphics[scale=1,trim={0 0.6cm 0 1.4cm},clip]{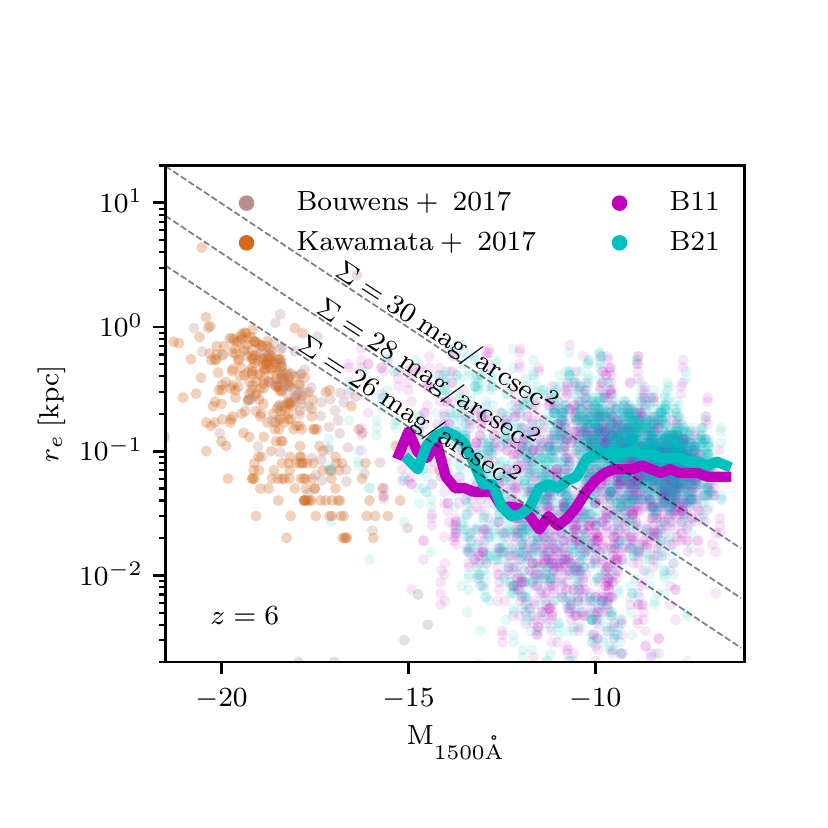}}
\centerline{\includegraphics[scale=1,trim={0 0.6cm 0 1.4cm},clip]{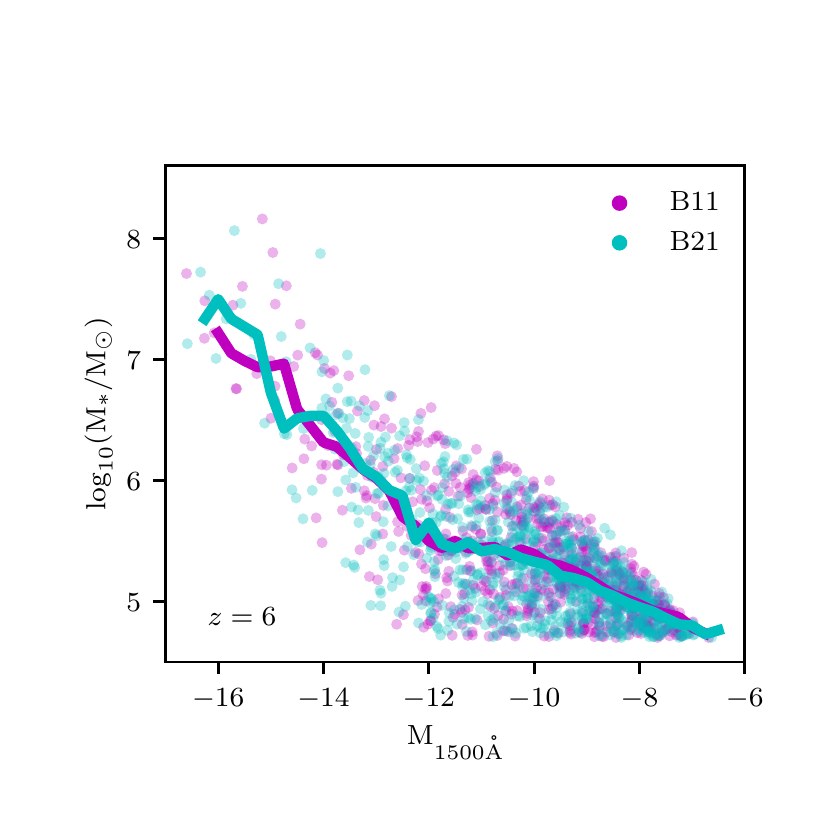}}
\centerline{\includegraphics[scale=1,trim={0 0.6cm 0 1.4cm},clip]{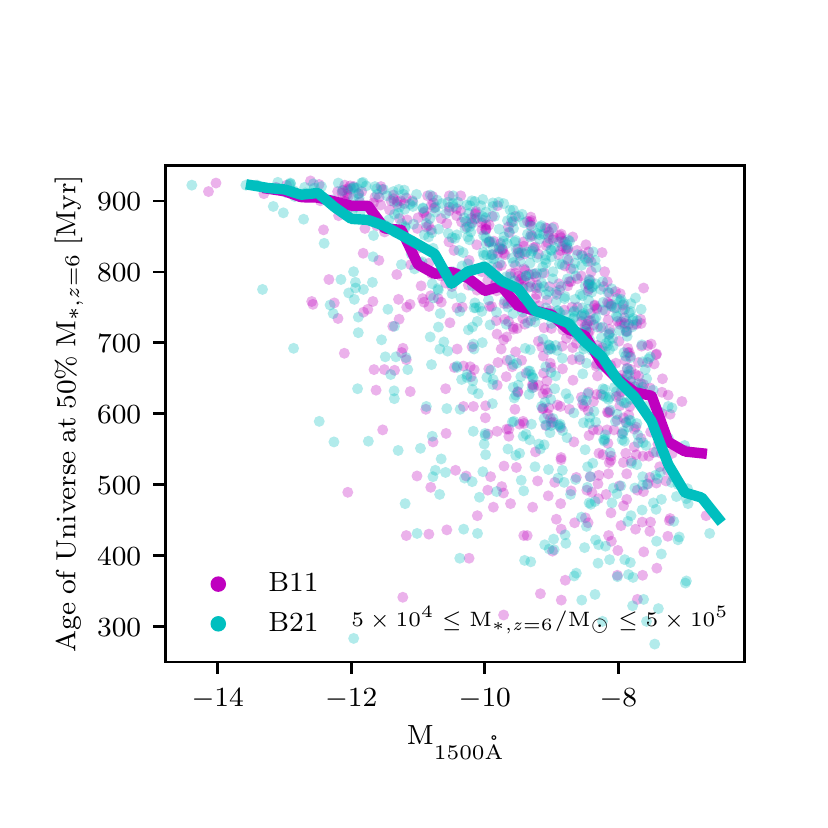}}
\caption{(top) Size-luminosity relation ($1500\angstrom$ UV magnitude versus effective radius) at $z=6$ for the B11 (strongest magnetic field) and B21 (weakest magnetic field) simulations compared with the observations from  \protect\cite{Kawamata2018,Bouwens2017}. We calculate the effective radius $r_e$ as the 2D radius that encapsulates half of the light for three different viewing angles of each galaxy that contains at least 50 star particles. The solid lines represent the running medians for each simulation and a bin is only included if it contains at least 30 galaxies. The effective radii in the B11 simulation are generally smaller than those in the B21 simulation indicating that strong galactic magnetic fields can shrink galaxies. Dashed black lines represent constant surface brightness as indicated. (centre) $1500\angstrom$ UV magnitude versus $z=6$ stellar mass. (bottom) $1500\angstrom$ UV magnitude versus the age of the Universe at which the galaxy formed 50\% of its stars for galaxies with $5\times10^4\leq{\rm M_{*,{\it z}=6}/M_{\odot}}\leq5\times10^5$. }
\label{sl}
\end{figure}

\begin{figure}
\centerline{\includegraphics[scale=1,trim={0 0.6cm 0 1.4cm},clip]{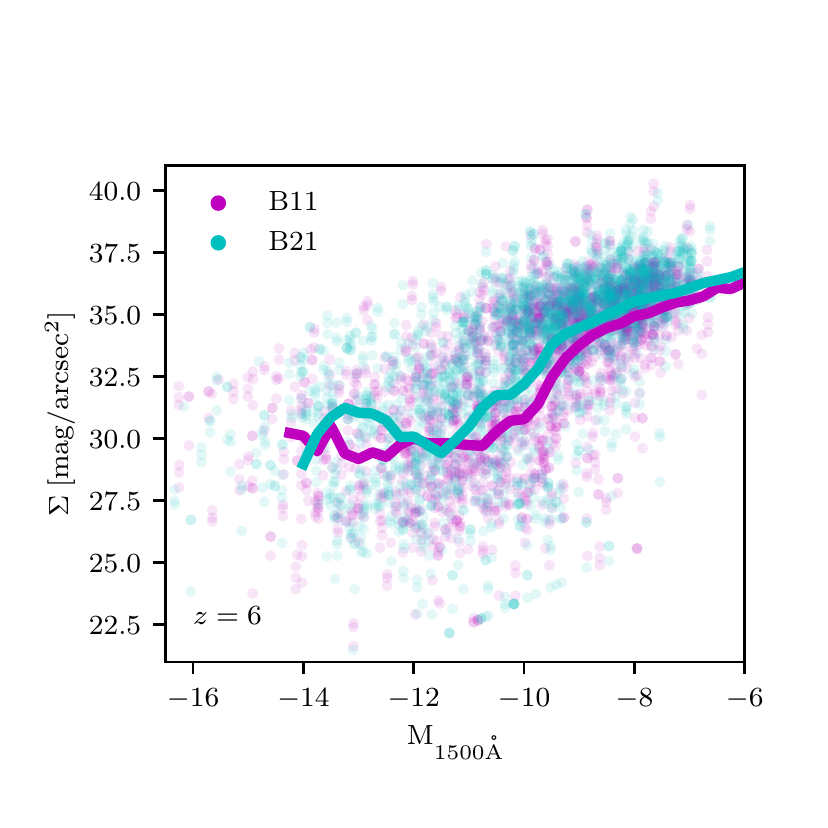}}
\centerline{\includegraphics[scale=1,trim={0 0.6cm 0 1.4cm},clip]{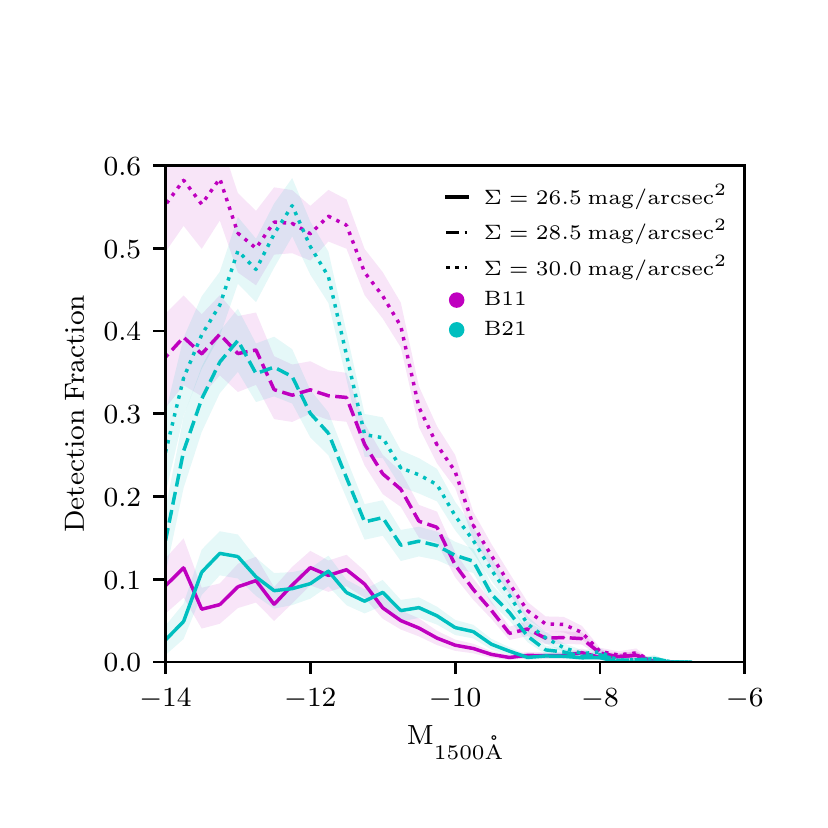}}
\caption{(top) $1500\angstrom$ UV magnitude versus apparent surface brightness at $z=6$ for galaxies in the B11 and B21 simulations. (bottom) $1500\angstrom$ UV magnitude versus the fraction of galaxies that could be detected in a surface-brightness limited survey with limits of 26.5 (solid), 28.5 (dashed), and 30.0 (dotted) mag/arcsec$^2$. This does not account for the aperture and magnitude limit of a telescope and only considers surface brightness. The shaded regions represent the $1\sigma$ standard deviation around the mean for 1,000 bootstrap re-samples from the data-set. Because stronger magnetic fields tend to shrink galaxies, they have higher average surface-brightness and are more easily detectable.}
\label{s2}
\end{figure}

\subsection{Magnetic Fields and Galaxy Structure}
\label{magnetic-fields-galaxy-structure}
The galaxy size-luminosity relation is one of the primary uncertainties in determining the completeness of galaxy surveys at high redshift. This in turn impacts our estimates of the UV luminosity function, especially at the faint-end where current surveys are expected to be incomplete \citep[e.g.][]{Bouwens2017}. Hence, constraining this relation is important, both for current and upcoming surveys.

\cite{Sergio2020} demonstrated that strong PMFs can shrink galaxies by nearly a factor of two as well as reduce the spin parameter of the galaxy. Numerous physical mechanisms can drain angular momentum from the gas and deposit it further away from the centre of the galaxy such as Maxwell stresses \citep{Sparke1982} or deceleration of inflows \citep{Birnboim2009}. This drives gas closer to the centre of the galaxy; hence reducing its size. In the top panel of Figure~\ref{sl} we plot the $1500\angstrom$ UV magnitude versus the effective radius\footnote{Effective radius, $r_e$, is measured by choosing a projection angle for the galaxy centred on the centre of light and finding the radius that encloses half the total light of the galaxy. For each galaxy, we choose three different projection angles along the principle axes of the simulation. Note that we do not include the impact of dust on $r_e$ which could impact our results, especially if dust obscuration varies with radius.} for galaxies in the B11 (strongest magnetic field) and B21 (weakest magnetic field) simulations at $z=6$ compared with observations from \cite{Bouwens2017} and \cite{Kawamata2018}. The observations are surface brightness-limited, as shown by the dashed black lines, and thus the observed relation appears to be steep. There is good agreement between our simulations and observations in the regime where the data overlap; however, most of the simulated galaxies exhibit UV magnitudes that are far dimmer than what is currently observed.

In general, the median galaxy in the simulation with weak PMFs (B21) has an effective radius that is 44\% larger than galaxies of similar absolute magnitude in the simulation with a strong PMF (B11). The scatter is indeed significant due to differences in star formation history, feedback, tidal effects, and halo accretion history and the mean difference between the effective radii, although systematic, are well within the scatter. The large scatter is consistent with expectations from similar reionization-era simulations \citep{Ma2018}. Furthermore, the trend between magnetic field strength and galaxy size expected from the zoom simulations of \cite{Sergio2020} is consistent with what we find in this work. Since surface brightness is proportional to the square of the galaxy radius, the effective decrease in surface brightness for galaxies in the B21 simulation is about a factor of two compared to those in the B11 simulation. This implies that the true completeness fraction of galaxy surveys at faint magnitudes is smaller for a Universe with weak PMFs compared to strong PMFs.

Such an effect is more easily visualised in Figure~\ref{s2} where in the top panel we show the mean surface brightness of the galaxies within the effective radius $r_e$ in the B11 and B21 simulations at $z=6$ for three different viewing angles. Once again we can see the trend that the median galaxy in the B11 simulation has a higher average surface brightness by $\sim0.7$ magnitudes/arcsec$^2$ since they tend to be more compact. As indicated earlier, these differences are still well within the scatter of the relation. In the bottom panel of Figure~\ref{s2} we show the fraction of galaxies that would be detected for a given surface-brightness threshold. For surface brightness limits of 26.5 mag/arcsec$^2$ (approximately equivalent to that of the Hubble Frontier Fields assuming a $5\sigma$ detection threshold and a 0.4'' diameter aperture \citealt{Lotz2017}), fewer than 10\% of galaxies could be detected. Such few galaxies are visible at this surface brightness limit that no difference is seen between the B11 and B21 simulations.  However, for a surface-brightness limit of 30 mag/arcsec$^2$, considerably more galaxies could be detected at magnitudes fainter than -12 in a Universe with strong PMFs. The deviations seen at higher magnitudes should be interpreted with caution due to the limited number of galaxies in these magnitude bins. Hence stochastic effects from star formation play a role in the detection fraction. Note that real surveys are also magnitude limited so realising this effect will be difficult, even with JWST. 

Although galaxies tend to be larger in simulations with weaker PMFs, the general trends between effective radius and UV magnitude hold between the simulations.  Effective radius tends to decrease towards fainter magnitudes until ${\rm M_{UV}}\sim-12$, where the effective radius slightly increases and then remains flat.  The negative slope observed at the brighter magnitudes is due to the fact that the galaxies are more massive at brighter magnitudes and thus angular momentum conservation makes them larger \citep[e.g.][]{Mo1998}. In the central panel of Figure~\ref{sl}, we plot ${\rm M_{UV}}$ versus galaxy stellar mass at $z=6$ and it is clear that for magnitudes brighter than -12, stellar mass increases towards brighter UV luminosities.  In contrast, at magnitudes fainter than -12, the slope becomes much more shallow.

\begin{figure}
\centerline{\includegraphics[scale=1,trim={0 0.6cm 0 1.4cm},clip]{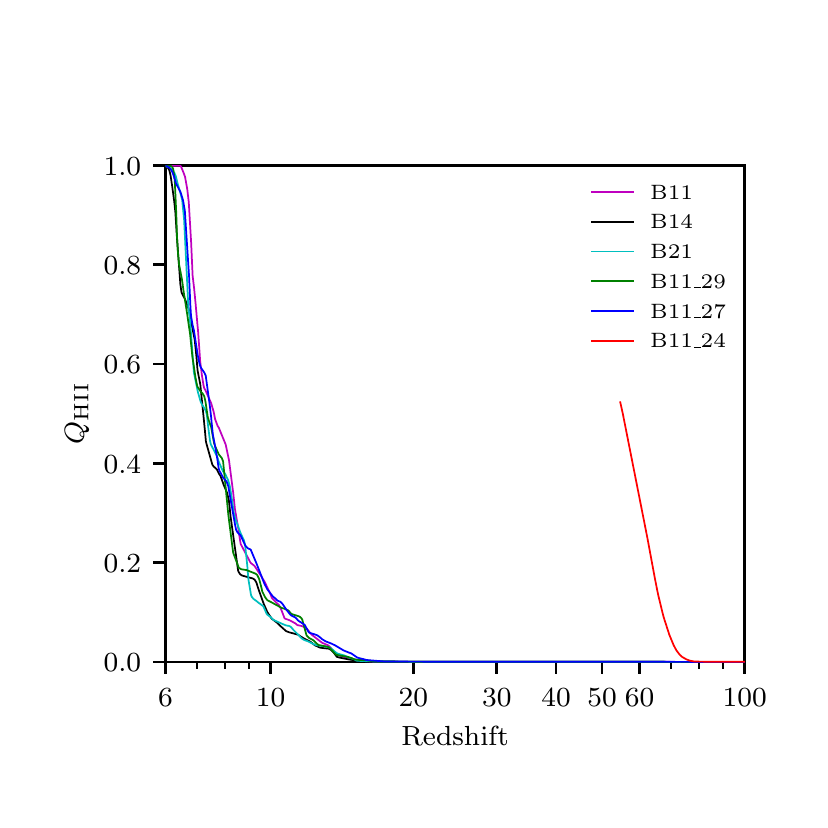}}
\centerline{\includegraphics[scale=1,trim={0 0.6cm 0 1.4cm},clip]{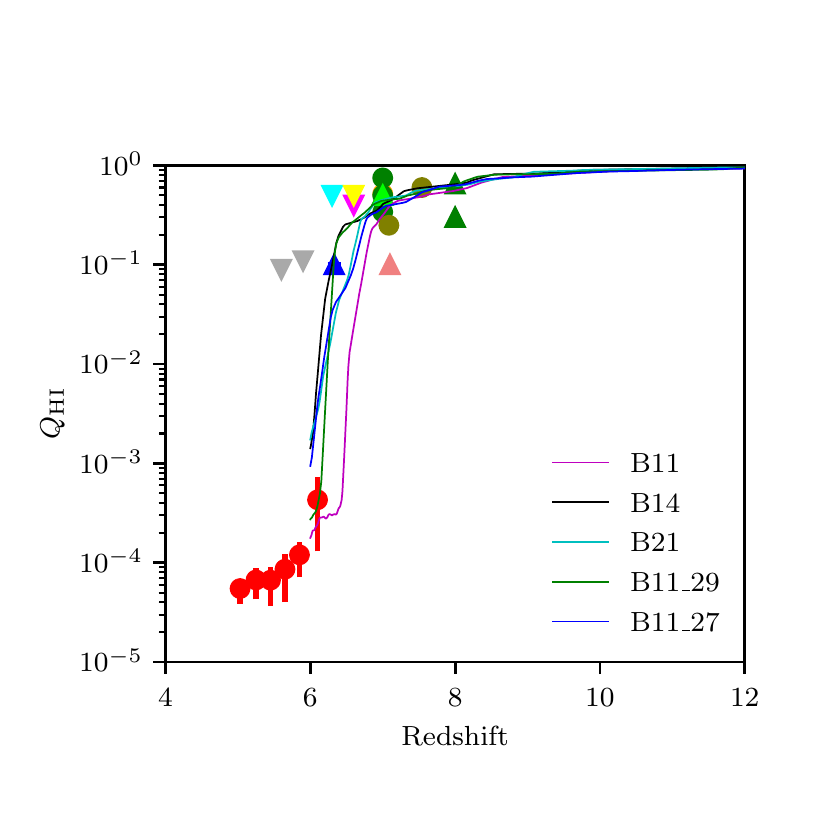}}
\centerline{\includegraphics[scale=1,trim={0 0.6cm 0 1.4cm},clip]{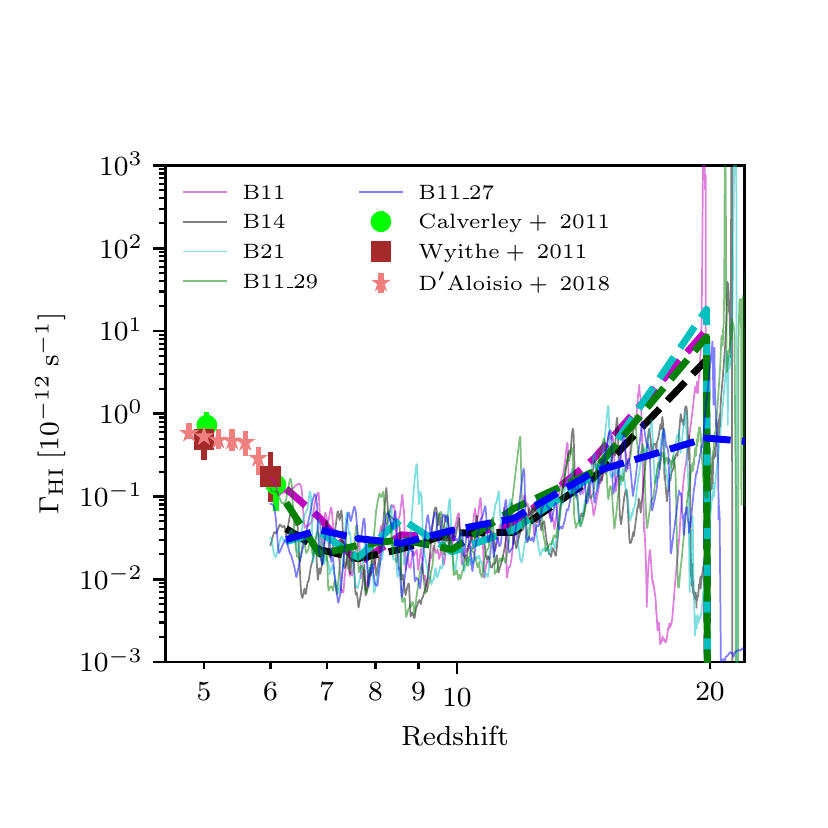}}
\caption{Evolution of the volume-weighted ionised fraction (top), volume-weighted neutral fraction (centre), and the volume-weighted HI photoionization rate (bottom) as a function of redshift for each of the simulations. The dashed lines in the bottom panel represent the 100Myr-averaged values of $\Gamma_{\rm HI}$. Neutral fraction data points are from the observations and modelling of \protect\cite{Fan2006,McGreer2015,Schroeder2013,Schenker2014,Caruna2014,Ono2012,Pentericci2014,Robertson2013,Tilvi2014,Totani2006,McQuinn2008,McQuinn2007,Ouchi2010,Ota2008,Sobacchi2015,Mortlock2011,Bolton2011,Durovcikova2020}. Photoionization rate data points and models have been compiled from \protect\cite{Calverly2011,Wyithe2011,Daloisio2018}.}
\label{reionization}
\end{figure}

\begin{figure*}
\centerline{\includegraphics[scale=1,trim={0 0.0cm 0 0.0cm},clip]{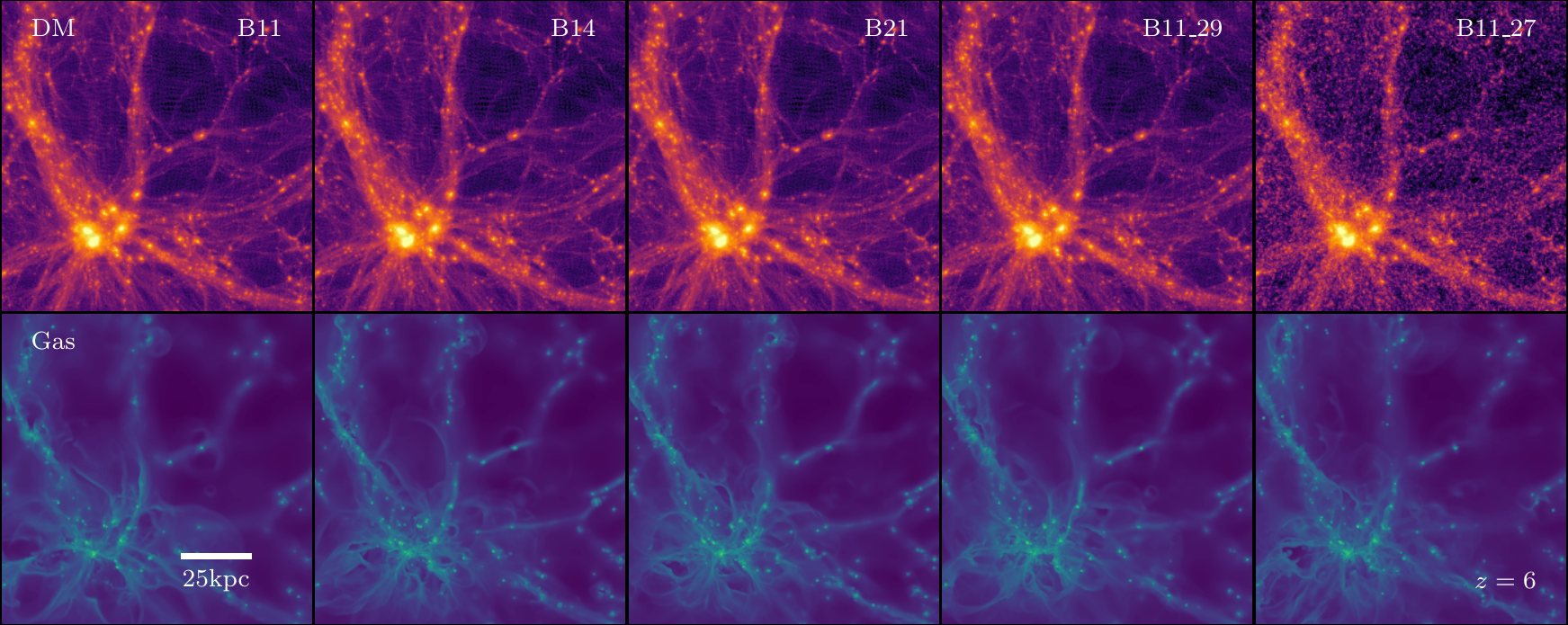}}
\caption{Surface density maps of dark matter (top) and gas (bottom) for the central 25\% of the computation volume at $z=6$ for each of the simulations. The enhancement in number density of low mass dark matter haloes can be seen in the B11\_29 and B11\_27 simulations; however, such differences are not observed in the baryon distribution.}
\label{lymaps}
\end{figure*}

We argue that the increase in effective radius towards fainter magnitudes above ${\rm M_{UV}}\sim-12$ is due to galaxies of approximately similar stellar masses ageing towards fainter magnitudes, and expanding due to feedback reducing the gravitational potential.  In the bottom panel of Figure~\ref{sl} we plot the UV magnitude against the age of the Universe at which the galaxy formed 50\% of its $z=6$ stellar population. We only show galaxies that have a stellar mass within the range $5\times10^4{\rm M_{\odot}}-5\times10^5{\rm M_{\odot}}$, which approximately selects for the galaxies with magnitudes fainter than -12. The upper envelope on this plot results from the magnitude of a fixed stellar population ageing over time while the scatter is due to variations in star formation history. It is clear that the galaxies that formed their stars earlier are much fainter. These galaxies that formed earlier will also be initially more compact; however, stellar feedback can quickly overwhelm the gravitational potential of these galaxies and cause expansion \citep[e.g.][]{Dekel1986,Pontzen2012}, consistent with the increase in effective radius seen in the top panel of Figure~\ref{sl}. It is clear from Figure~\ref{sl} that our simulations do not predict a monotonic trend between size and luminosity. Thus the relations in the literature \citep[e.g.][]{Shibuya2015} derived from brighter galaxies are not predicted to hold at faint luminosities.

\subsection{Evolution of the intergalactic medium}
Because PMFs can impact both the ISM of galaxies as well as the dark matter halo mass function, we expect that the history of reionization and the properties of the intergalactic medium can change depending on the characteristics of the PMF. \cite{Pandey2015} and \cite{Sanati2020} have already demonstrated that the history of reionization can be used as a constraint on the properties of the PMFs using analytical models and zoom-in simulations, respectively. Furthermore, for strong enough PMFs, we expect to be able to observe their signatures in the properties of the Ly$\alpha$ forest \citep{Meiksin2014}. Our simulations are among the first to study these effects in cosmological radiation magnetohydrodynamics simulations and in this section, we explore the impact of various PMFs on the history of reionization and the properties of the IGM. 

In the top panel of Figure~\ref{reionization}, we show the ionised fraction ($Q_{\rm HII}$) as a function of redshift from $z=100$ to $z=6$ for all of the simulations. Most notably, the B11\_24 simulation is already substantially reionized at $z>50$, inconsistent with nearly every observational probe of reionization. The strong enhancement of the power spectrum at high $k-$modes in the B11\_24 simulation leads to the early collapse of a substantial number of dwarf galaxies and thus a very early reionization history. Due to computational limitations, we have only evolved this simulation until $Q_{\rm HII}=50\%$. Our results demonstrate that we can rule out PMFs with $B_0=5\times10^{-11}G$ and spectral indices $n_B>-2.4$. Such findings are qualitatively consistent with \cite{Sanati2020} who used standard hydrodynamics simulations with a modified initial density field and also found that initial conditions produced with magnetic fields with $B_0=5\times10^{-11}G$ and $n_B>-2.4$ resulted in reionization histories that are inconsistent with observations. However, our simulations seem to reionize considerably earlier than their semi-analytic approach based on the SFHs from their cosmological hydrodynamic simulations. We provide more detailed constraints on the properties of the PMF in Section~\ref{pmf_constraint}.

In contrast to the B11\_24 simulation, all other models exhibit a reasonable reionization histories that complete at $z\sim6$. In the centre panel of Figure~\ref{reionization} we plot the neutral fraction for each of the simulations as a function of redshift where we see that nearly all simulations reach $Q_{\rm HI}\lesssim10^{-3}$ by $z=6$. This is consistent with the majority of models and observations in the literature, although this is perhaps slightly early compared to more recent models \citep{Kulkarni2019,Keating2020} that seem to point towards a later reionization, completing at $z\lesssim5.5$.

As was discussed earlier, there are few differences between the simulations in terms of their stellar content. Hence it is not surprising that the reionization histories are consistent between the simulations. The B11 simulation, which has the highest amount of energy stored in the magnetic field, reionizes very slightly early compared to the other models, possibly due to differences in ISM structure of the galaxies as shown in Section~\ref{magnetic-fields-galaxy-structure}. We note that, due to the small volume of the simulation, re-running the simulation with a different random seed for the star formation subroutine would also result in small changes in the reionization history due to differences in the star formation history. More interestingly, despite the significant increase in the number density of dwarf galaxies with ${\rm M_{vir}<10^7M_{\odot}}$ in the B11\_27 simulation, the reionization history does not deviate significantly from the other simulations, consistent with expectations from \cite{Sanati2020}. Our stellar and dark matter particle mass resolution can suppress star formation in these low-mass haloes if they are under-resolved. However, previous work using similar star formation and feedback models and a standard $\Lambda$CDM primordial matter power spectrum shows that reionization is primarily driven by haloes with ${\rm M_{vir}>10^{8}M_{\odot}}$ \citep{Kimm2017} and we therefore expect these results to hold even for higher resolution simulations.

In the bottom panel of Figure~\ref{reionization} we show the volume-weighted HI photoionisation rate ($\Gamma_{\rm HI}$) in regions that are more than 50\% ionised as a function of redshift. Once again, we find good agreement between the simulations, as expected given their consistent reionization histories. Many of the small fluctuations in $\Gamma_{\rm HI}$ correlate between the simulations. These are indicative of individual haloes collapsing or galaxy mergers resulting in large bursts of star formation that temporarily dominate $\Gamma_{\rm HI}$. For larger volume simulations, the evolution of $\Gamma_{\rm HI}$ would be much more smooth. Consistent with other work \citep[e.g.][]{Katz2018,Rosdahl2018}, after the formation of the first stars, $\Gamma_{\rm HI}$ is high as the ionised regions only consist of the volume very close to galaxies.  As the HII regions expand into the IGM, $\Gamma_{\rm HI}$ decreases. Once the HII bubbles begin to merge, percolation occurs and $\Gamma_{\rm HI}$ begins to increase again.

By $z=6$, the three simulations with the lowest neutral fractions (B11, B11\_29, and B11\_27) have photoionisation rates consistent with observations, while the B14 and B21 simulations exhibit a $\Gamma_{\rm HI}$ that is approximately five times weaker. $\Gamma_{\rm HI}$ is expected to rapidly increase right at the end of reionization, when there is significant overlap of ionised bubbles (see Figure~10 of \citealt{Rosdahl2018}), and thus such a difference is expected between simulations with mildly different reionization histories.

\begin{figure}
\centerline{\includegraphics[scale=1,trim={0 0.0cm 0 0.0cm},clip]{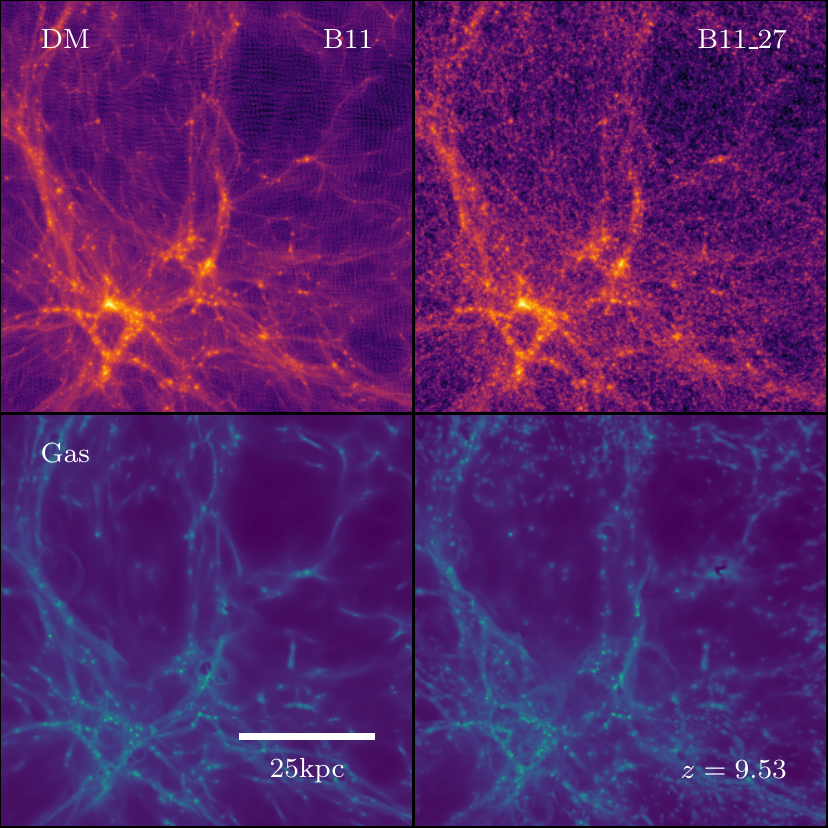}}
\caption{Surface density maps of dark matter (top) and gas (bottom) for the central 25\% of the computation volume at $z=9.53$ for the B11 (left) and B11\_27 (right) simulations. The enhancement in number density of low mass dark matter haloes can be seen in the B11\_27 simulations and the baryon distribution is also clearly modified at this redshift.}
\label{lymaps_z9}
\end{figure}

\begin{figure}
\centerline{\includegraphics[scale=1,trim={0 0.2cm 0 1.1cm},clip]{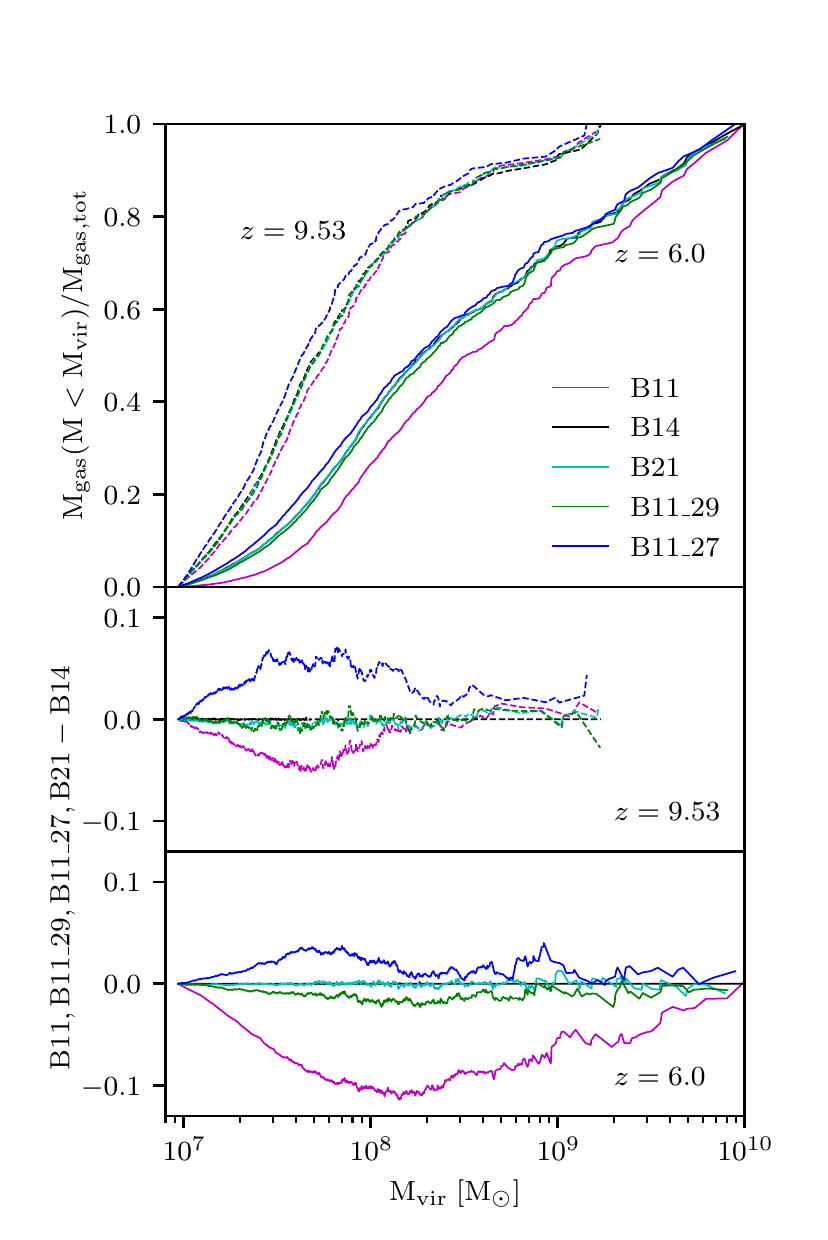}}
\caption{Fraction of the total gas mass in haloes that is contained in haloes with ${\rm M<M_{vir}}$ at $z=6$ (solid) and $z=9.53$ (dashed) for each of the simulations (top). We show the difference between each of the simulations and the B14 simulation for this quantity at $z=9.53$ (centre) and at $z=6$ (bottom). There is an excess in gas contained in low mass haloes in the B11\_27 simulation compared to the B14 simulation while there is a reduction in the gas content of low mass galaxies in the B11 simulation.}
\label{vwdf}
\end{figure}

Due to the differences in initial matter power spectrum and the resulting modifications to the dark matter halo mass function, we might expect that the presence of PMFs may have a significant impact on the Ly$\alpha$ forest \citep[e.g.][]{Pandey2013,Meiksin2014}. In Figure~\ref{lymaps} we show projected surface-density maps of the dark matter and gas distribution for the central 25\% of the computational volume at $z=6$ for each of the simulations. Consistent with what was observed in the dark matter-only simulations, we can clearly see a strong enhancement in the number of dark matter clumps in the B11\_27 simulation compared to any of the fiducial models that do not include any modification to the matter power spectrum. Most noticeable is that the filamentary structure in the B11\_27 simulation is significantly less smooth.

\begin{figure*}
\centerline{\includegraphics[scale=1,trim={0 0cm 0 0.0cm},clip]{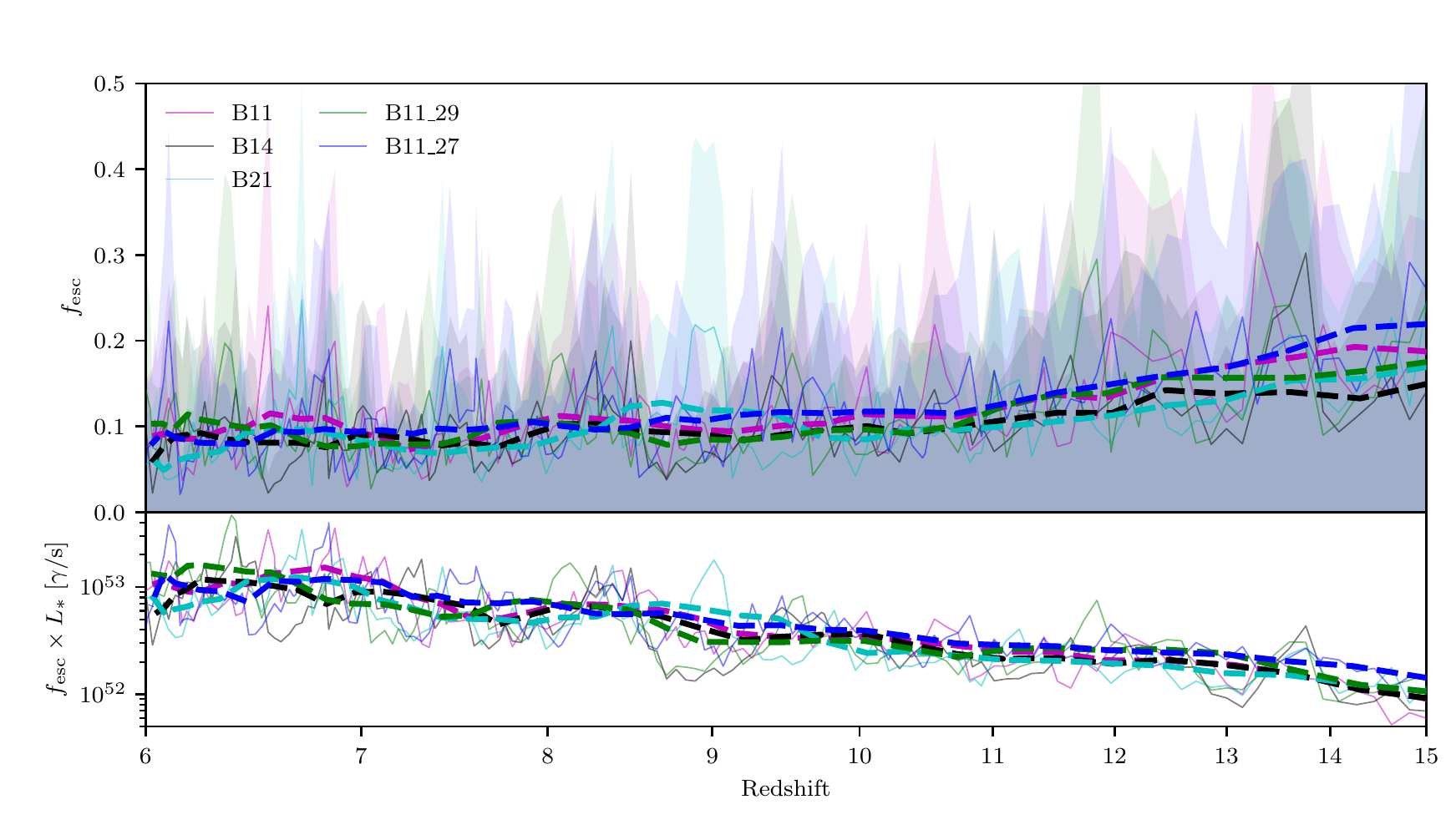}}
\caption{Luminosity-weighted mean escape fraction (top) and number of escaping ionising photons (bottom) as a function of redshift for each of the simulations.  The solid lines and shaded region represent the mean and $1\sigma$ standard deviation of $f_{\rm esc}$ as measured at each individual snapshot.  The dashed lines represent the luminosity-weighted average $f_{\rm esc}$ over the previous 100~Myr.}
\label{fesc}
\end{figure*}

Despite the differences in the $z=6$ dark matter distribution, the gas distribution appears visually very similar in the bottom row of Figure~\ref{lymaps}, albeit small differences near galaxies where stellar feedback has clearly had an impact. These results are consistent with \cite{Pandey2013} who require either stronger magnetic fields or larger spectral indices in order to obtain observable differences in the effective optical depth, some of which are inconsistent with the history of reionization \citep{Sanati2020}.

More quantitatively, we can compare the surface density distributions of gas and dark matter between the different simulations. For simulations without modified initial conditions, one would expect stronger differences in the gas compared to the dark matter due to feedback and other hydrodynamic processes. When compared to the B14 simulation the mean difference in the region shown in Figure~\ref{lymaps} is 25\% and 7.5\% for the dark matter for the B11 and B21 simulations, respectively. It is not surprising that there is a larger deviation for the B11 simulation compared to the B21 simulation as the additional magnetic pressure from the strong PMF can impact the dark matter indirectly by modifying the baryon distribution. In contrast, we find mean deviations of 59\% and 54\% in the gas surface density for the B11 and B21 simulations, respectively. The situation is reversed for the B11\_29 and B11\_27 simulations. For the dark matter, we find a difference of 880\% and 1924\% for the B11\_29 and B11\_27 simulations, respectively compared to the B14 simulation while for the gas, the difference is only 53\% and 66\%, which is much more consistent with what we found for the B11 and B21 simulations.

For the B11\_29 and B11\_27 simulations, the modifications to the power spectrum only have a strong impact on haloes with ${\rm M_{vir}\lesssim3\times10^{7}M_{\odot}}$ (see Figure~\ref{ICmag2}). It is well established that photoionization and photoheating from the process of reionization can starve, photoevaporate, and prolong cooling times around low-mass dwarf galaxies \citep{Rees1986,Efstathiou1992,Okamoto2008,Gnedin2014,Dawoodbhoy2018,Katz2020}, particularly those with virial temperatures below the atomic cooling threshold. Furthermore, reionization can reduce the gas masses of filaments by more than 80\% \citep{Katz2020}. Hence the structures that are most modified by the presence of the primordial magnetic fields sampled in this work are also those that are most sensitive to radiation feedback. If reionization is indeed the process that is smoothing the gas density field and erasing the small-scale structure, at higher redshifts, prior to reionization, we should observe differences in the gas density field. In Figure~\ref{lymaps_z9}, we show the dark matter and gas density field in the same central region of the volume as Figure~\ref{lymaps}; however, here, we show the results at $z=9.53$ when the simulation volume is less than 20\% ionised. Here we can see that there are significantly more small clumps of gas in the B11\_27 simulation.

We illustrate this more quantitatively in Figure~\ref{vwdf} where we show the fraction of the total gas mass in haloes that is contained in haloes with ${\rm M<M_{vir}}$ at $z=6$ and $z=9.53$. At $z=9.53$, the B11\_27 simulations exhibits a $\sim7\%$ excess in gas mass in haloes with ${\rm M_{vir}<3\times10^7M_{\odot}}$ compared to the B14, B21, and B11\_29 simulations. This excess decreases by more than a factor of two by $z=6$ where reionization has smoothed the density field. In contrast, the B11 simulation exhibits a $\sim5\%$ and $\sim10\%$ reduction in the gas content of haloes with ${\rm M_{vir}<10^8M_{\odot}}$ at $z=9.53$ and $z=6$, respectively. The reduction at high redshift is likely due to the enhanced pressure support that may slow gas accretion. By $z=6$ the B11 simulation is more ionised than the other volumes and hence the difference grows. This effect can also be seen in the B11\_29 simulation. At $z=9.53$, the cumulative distribution function of the B11\_29 simulation is nearly identical to that of the B14 and B21 simulations. However, this simulation is more ionised at $z=6$ and hence there is a reduction in the gas content of low mass haloes compared to these other simulations.

\subsection{Impact of primordial magnetic fields on the LyC escape fraction}
Constraining the LyC escape fraction as a function of both redshift and galaxy properties has been the subject of numerous observational and theoretical studies as constraining this parameter will allow for the determination of both the reionization history and the primary sources responsible \citep[e.g.][]{Nakajima2020,Kimm2017,Barrow2020}. While the global, luminosity-weighted escape fractions cannot be substantially different between the simulations with different PMFs because of the observed consistency in star formation and reionization histories, we may expect slight differences due to the variations in ISM properties.

\begin{figure}
\centerline{\includegraphics[scale=1,trim={0 0cm 0 0cm},clip]{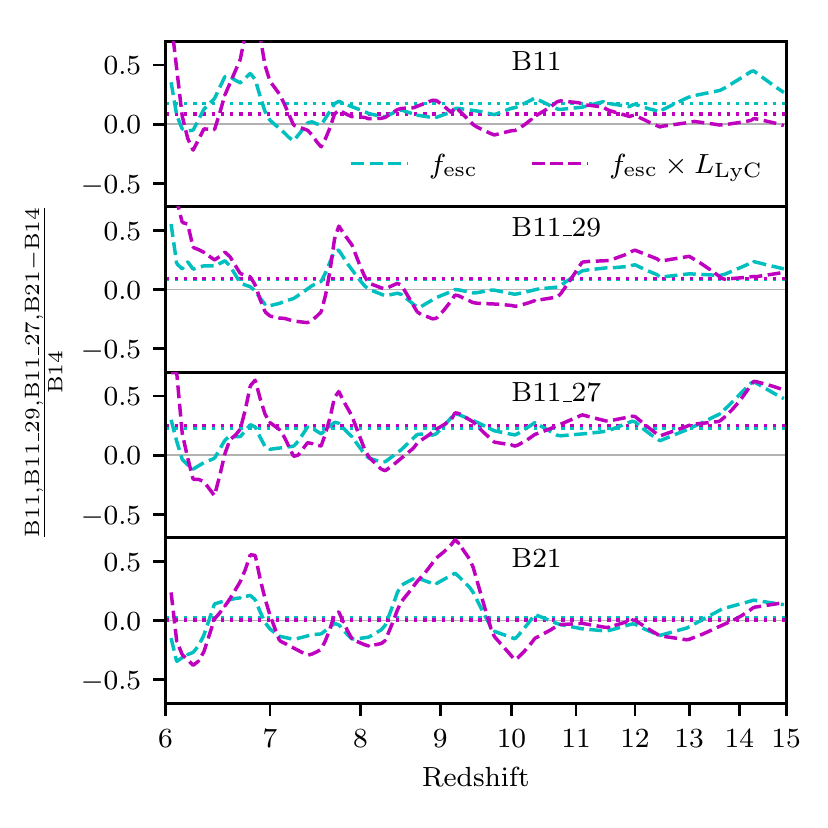}}
\caption{Fractional difference between the 100Myr averaged $f_{\rm esc}$ (cyan) and the average of $f_{\rm esc}\times L_{\rm LyC}$ (magenta) in the B11 (top), B11\_29 (second panel), B11\_27 (third panel), and the B21 (bottom) simulations and the B14 simulation. The horizontal magenta and cyan lines represent the mean difference between $z=6$ and $z=15$. In general, the simulations with stronger PMFs exhibit higher LyC escape fractions as well as a greater number of leaking LyC photons but the effect is only $\sim25\%$, whereas the average escape fractions are nearly identical in the simulations with weaker PMFs.}
\label{fesc_res}
\end{figure}

In Figure~\ref{fesc} we show the luminosity-weighted $f_{\rm esc}$ as a function of redshift for each of the simulations as well as the $1\sigma$ scatter about the relation. The escape fractions are slightly higher at high redshifts where low-mass dwarf galaxies dominate the photon budget. We note that $f_{\rm esc}$ is also a function of metallicity \citep{Yoo2020} which evolves with redshift. $f_{\rm esc}$ decreases as a function of redshift from $>15\%$ at $z=15$ to $f_{\rm esc}\sim5\%-10\%$ at $z=6$. The $z=6$ values for $f_{\rm esc}$ in our simulations are in complete agreement with the larger volume non-MHD simulations of \cite{Rosdahl2018} that also employ a stellar SED that includes binary stars \citep{Eldridge2008,Stanway2016}.

\begin{figure}
\centerline{\includegraphics[scale=1,trim={0 0.6cm 0 1.4cm},clip]{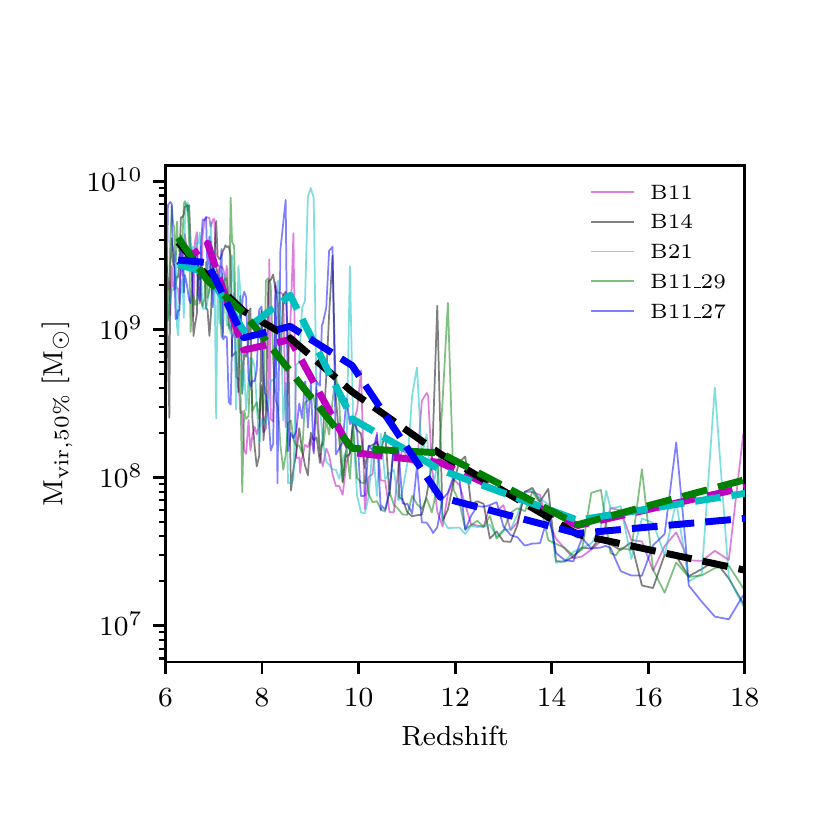}}
\centerline{\includegraphics[scale=1,trim={0 0.6cm 0 1.4cm},clip]{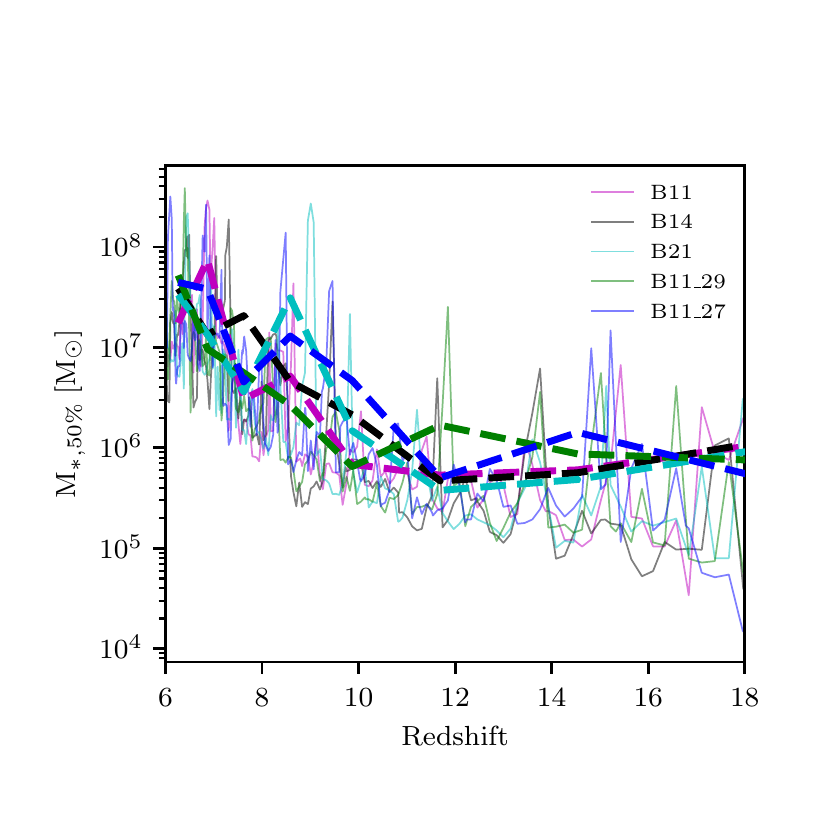}}
\centerline{\includegraphics[scale=1,trim={0 0.6cm 0 1.4cm},clip]{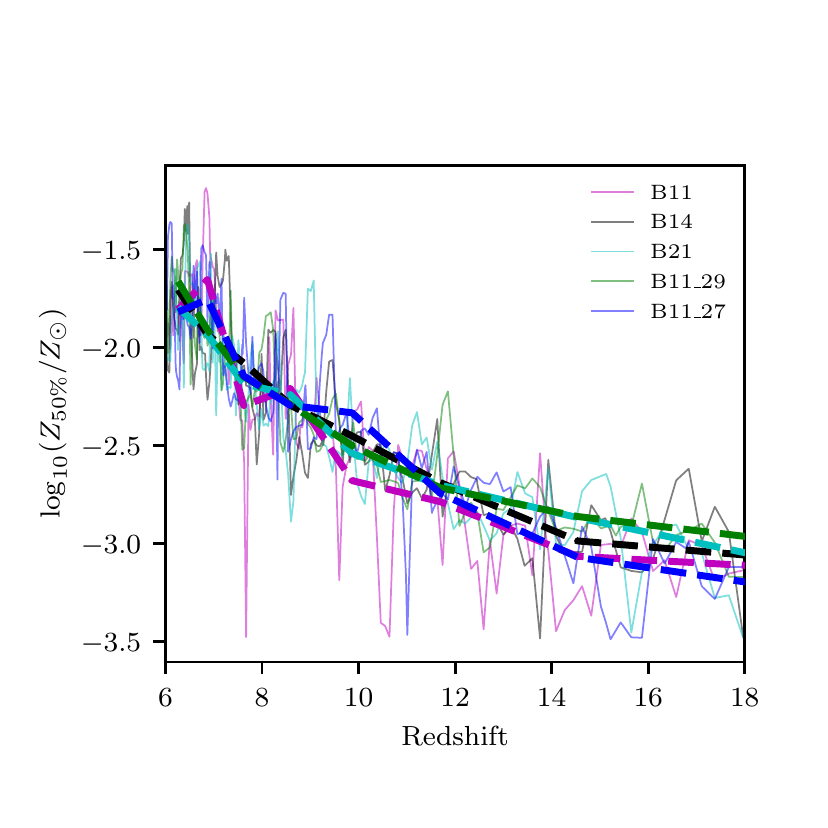}}
\caption{Halo mass (top), galaxy stellar mass (centre), and stellar metallicity (bottom) below which 50\% of the leaking ionising photons originate as a function of redshift. For example, at $z=10$, 50\% of the ionising photons that leak into the IGM are emitted by haloes with ${\rm M_{vir}<10^8M_{\odot}}$. Dashed lines represent the 100Myr-averaged values of the individual quantities. As the universe evolves, most of the photons that reionize the Universe originate in more massive galaxies and more metal enriched stars.}
\label{reionization_cdfs}
\end{figure}

\begin{figure}
\centerline{\includegraphics[scale=1,trim={0 0.6cm 0 1.4cm},clip]{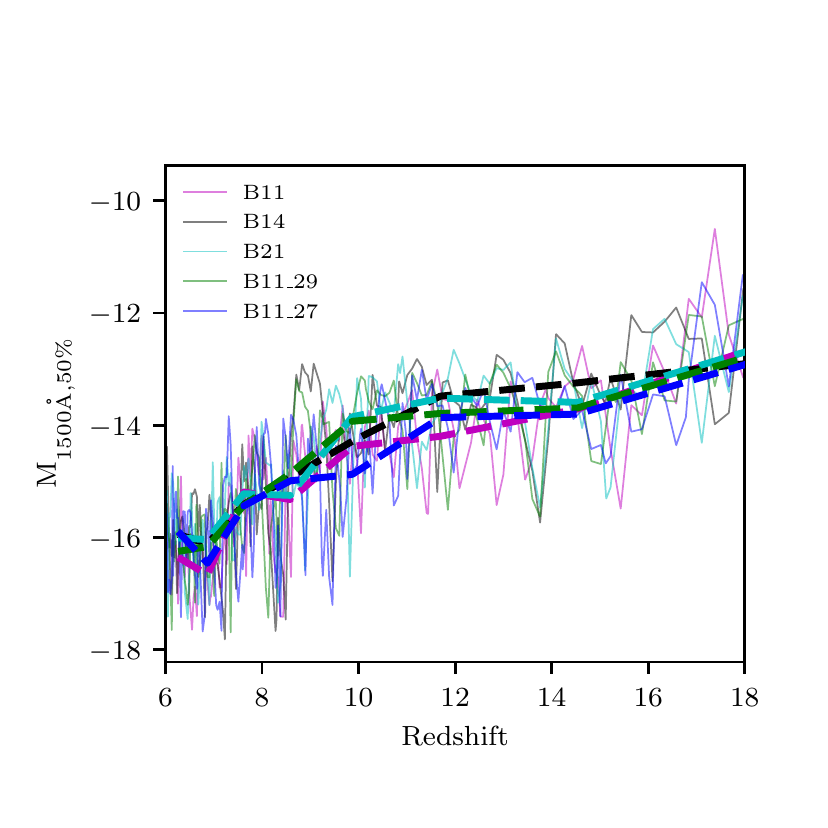}}
\caption{$1500\angstrom$ absolute magnitude below which 50\% of the leaking ionising photons originate as a function of redshift. For example, at $z=6$, 50\% of the ionising photons that leak into the IGM are emitted by galaxies with UV magnitudes fainter than $-16$. Dashed lines represent the 100Myr-averaged values of ${\rm M_{1500\angstrom,50\%}}$.}
\label{reionization_cdfs2}
\end{figure}

As expected from the reionization histories, the simulations all exhibit luminosity-weighted escape fractions that are well within each others 1$\sigma$ standard deviation. Because the galaxies in the simulations with strong magnetic fields are more compact, we might expect small differences in $f_{\rm esc}$ due to the changes to the ISM. In Figure~\ref{fesc_res} we show the fractional difference between the B11, B11\_29, B11\_27, and B21 simulations and the B14 simulation as the dashed cyan lines. The average values of $f_{\rm esc}$ in the simulations with the stronger PMFs are $10\%-25\%$ higher in the B11 series simulations compared to B14 while the average value of $f_{\rm esc}$ is nearly identical between B14 and B21. The enhancement is not particularly large (e.g. compared to the difference between single star and binary star SEDs which is upwards of $300\%$ \citealt{Rosdahl2018}) and it also fluctuates. Because the density field is different in the B11\_27 and B11\_29 simulations, the cleanest comparison is between the B11 and B14/B21 simulations. There is a small difference in reionization history; however, the could also be due to small differences in the star formation history. LyC photon production is the other important variable in determining reionization history and thus in Figure~\ref{fesc_res}, we also show a comparison of the 100Myr average $f_{\rm esc}\times L_{\rm LyC}$ where $L_{\rm LyC}$ is the LyC photon luminosity.  We can see that the increase in escape fraction is not compensated by a decrease in LyC luminosity for the simulations with strong PMFs. This explains why the HI photoionization rates are higher in the B11, B11\_29, and B11\_27 simulations at z=6 compared to the B14 and B21 simulations. We emphasise that this effect is small and therefore caution over-interpreting this result.

In Figure~\ref{reionization_cdfs} we explore which sources are responsible for providing the photons that drive reionization as a function of redshift. For each simulation snapshot, we calculate the cumulative distribution function of the escaping ionising luminosity as a function of halo mass (top), galaxy stellar mass (centre), and stellar metallicity (bottom) and plot the 50th percentile as a function of redshift. In the top panel of Figure~\ref{reionization_cdfs} we can see that at $z=18$, 50\% of all of the escaping ionising photons are emitted by haloes with ${\rm M_{vir}\sim3\times10^7M_{\odot}}$. This value increases to well above ${\rm 10^9M_{\odot}}$ by $z=6$. Similarly, the stellar masses of the galaxies driving reionization increases from ${\rm \sim10^5M_{\odot}}$ at $z=18$ to ${\rm >10^7M_{\odot}}$ at $z=6$. As the universe evolves and galaxies become more massive, the sources driving reionization also increase in mass. 

The metallicity of the stars driving reionization also strongly increases with redshift. Even at $z=18$, the stars that are contributing most of the escaping photons to reionization have been enriched above our primordial metallicity floor of $10^{-3.5}Z_{\odot}$. This is consistent with \cite{Xu2016} who found that the contribution from Population~III stars is expected to dominate at $z>16$.

We find no significant differences between the sources of reionization in any of the simulations with different PMFs at redshifts up to $z=18$. Note that out limited mass and spatial resolution prohibits us from self-consistently modelling the formation of Population~III stars. Given the strong enhancement in the number of low mass galaxies in the B11\_27 simulation, a higher resolution simulation may find some differences at $z>16$ where these haloes are expected to dominate.

A particularly interesting question is whether we can observe the sources that are driving reionization. In Figure~\ref{reionization_cdfs2}, we plot the intrinsic $1500\angstrom$ absolute UV magnitude below which 50\% of the escaping ionising photons originate as a function of redshift. We find that this magnitude decreases from $-12$ at $z=18$ to $\sim-16$ at $z=6$, consistent with \cite{Wise2014}. Using strong lensing from galaxy clusters, the Hubble Frontier Fields can already probe UV magnitudes much fainter than $-16$ at $z=6$ \citep[e.g.][]{Livermore2017} and thus, current observations have already detected the galaxies that are primarily responsible for keeping the Universe reionized at $z=6$. Because the photoionization and photoheating from reionization are incredibly efficient at suppressing star formation in low mass dwarf galaxies at halo masses that we resolve \citep{Okamoto2008,Xu2016,Katz2020}, this result is expected to hold even for higher resolution simulations. Furthermore, larger boxes will only probe brighter galaxies which could only bias our measurements in Figure~\ref{reionization_cdfs2} towards brighter magnitudes (see \citealt{Lewis2020}), hence strengthening our argument.

\subsection{Constraints on the properties of primordial magnetic fields}
\label{pmf_constraint}
The properties for the PMFs chosen in this work bracket the range of having no impact on the history of reionization to having an ionised fraction of 50\% at $z>50$. Due to computational resource limitations, we cannot conduct radiation magnetohydrodynamics simulations for an entire grid of values of $B_0$ and $n_B$. In this section, we construct analytic models based on the results of our simulations to improve our constraints on the properties of the PMF.

\subsubsection{Ionisation history and electron optical depth constraints}
The evolution of the ionised hydrogen fraction ($Q_{\rm HII}$) can be modelled using the following ordinary differential equation:
\begin{equation}
\label{qode}
\frac{dQ_{\rm HII}}{dt} = \frac{\dot{n}_{\rm ion}}{\langle n_{\rm H} \rangle} - \frac{Q_{\rm HII}}{\bar{t}_{\rm rec}(C_{\rm HII})},
\end{equation}
\citep[e.g.][]{Madau1999}, where $\dot{n}_{\rm ion}$ is the ionising photon production rate (${\rm \gamma/s/cMpc^3}$), $\langle n_{\rm H} \rangle$ is the mean comoving hydrogen number density, $C_{\rm HII}$ is the clumping factor of ionised hydrogen, and $\bar{t}_{\rm rec}$ is the volume-averaged recombination timescale of HII. $\bar{t}_{\rm rec}$ is a function of temperature and the clumping factor such that
\begin{equation}
\bar{t}_{\rm rec} = \frac{1}{C_{\rm HII}\alpha_{\rm B}(T)\left(1+\frac{Y}{4X}\right)\langle n_{\rm H} \rangle(1+z)^3}
\end{equation}
\citep[e.g.][]{Kuhlen2012}, where $\alpha_{\rm B}(T)$ is the temperature-dependent case-B HII recombination rate \citep{Hui1997}, $Y$ is the primordial helium mass fraction, and $X$ is the primordial hydrogen mass fraction. We choose a fixed value of $2\times10^4$K for the IGM temperature\footnote{Note that this temperature is consistent with the local reionization temperature of the {\small SPHINX} simulations. See Figures 9 and 10 of \cite{Katz2020}.} \citep[e.g.][]{Trac2008} and to be consistent with our simulations, we set $X=0.76$ and $Y=0.24$. For the clumping factor, we follow \cite{Kimm2017} and adopt a redshift-dependent value of $C_{\rm HII}=1+e^{-0.28z+3.59}$ at $z\geq10$ and $C_{\rm HII}=3.2$ at $z<10$ \citep{Pawlik2009}. We note that the values of $Q_{\rm HII}$ at high redshifts are especially sensitive to this parameter.

$\dot{n}_{\rm ion}$ as a function of redshift can be modelled as
\begin{equation}
\dot{n}_{\rm ion} = \xi_{\rm ion}(z)f_{\rm esc}(z)\rho_{\rm SFR}(z),
\end{equation}
where $\rho_{\rm SFR}(z)$ is the star formation rate density as a function of redshift, $f_{\rm esc}(z)$ is the escape fraction of ionising photons as a function of redshift, and $ \xi_{\rm ion}(z)$ is the ionising photon production efficiency per stellar mass as a function of redshift. To obtain the star formation rate density, we first generate the matter power spectrum for each PMF for different values of $n_B$ and $B_0$ using the modified version of {\small CAMB} \citep{Shaw2012}. We then use {\small COLOSSUS} \citep{Diemer2018} to calculate the redshift-dependent halo mass function \citep{Press1974}. Each halo is populated with stars using the stellar mass-halo mass relation from our B11 simulation as a function of redshift. Note that this function evolves only mildly with redshift up to $z\sim20$. We extrapolate the function to both lower and higher mass haloes than what are resolved or sampled by our simulation and additional systematic uncertainty may result from this extrapolation. Not all low mass haloes host stellar populations (i.e. the occupation fraction, $f_{\rm occ}$, is less than 1). To account for this, we use the fitting formula for the occupation fraction based on the {\small SPHINX} simulations \citep{Katz2020},
\begin{equation}
f_{\rm occ}=\frac{1}{1+e^{-k[\log_{10}({\rm M_{vir}})-{\rm M_c}(z)]}},
\end{equation}
where $k=8.20$ and ${\rm M_c}(z)=-0.05c+8.32$. The total star formation rate per unit volume is measured as the growth in total stellar mass as a function of redshift. $f_{\rm esc}(z)$ is calculated using a linear fit to the redshift evolution of the luminosity-weighted $f_{\rm esc}$ in the B11 simulation such that
\begin{equation}
f_{\rm esc}(z) = {\rm Min}(0.0096z+0.0246,50\%).
\end{equation}
We have capped $f_{\rm esc}$ to prevent it from reaching unrealistic values at very high redshifts such that the value is consistent with the escape fractions measured in simulated mini-haloes \citep{Wise2014,Xu2016,Kimm2017}. Note that we apply the escape fraction to the total $\rho_{\rm SFR}(z)$ rather than that of each halo mass as the luminosity-weighted value of $f_{\rm esc}$ should capture any mass dependence. Furthermore, the {\small SPHINX} simulations do not show a strong evolution of $f_{\rm esc}$ with halo mass (Rosdahl~et~al.~{\it in prep}). Finally, the value of $ \xi_{\rm ion}$ is fixed to $10^{53.36}\ {\rm \gamma/s/(M_{\odot}/yr)}$, consistent with the lowest metallicity bin of the {\small BPASSv2.0} SED that was used in the simulations. From the bottom panel of Figure~\ref{reionization_cdfs}, we can see that most of the stars that provide ionising photons have metallicities lower than the lowest metallicity bin of the {\small BPASSv2.0} SED. Therefore, we do not apply a redshift dependence to this value. 

\begin{figure}
\centerline{\includegraphics[scale=1,trim={0 0.6cm 0 1.4cm},clip]{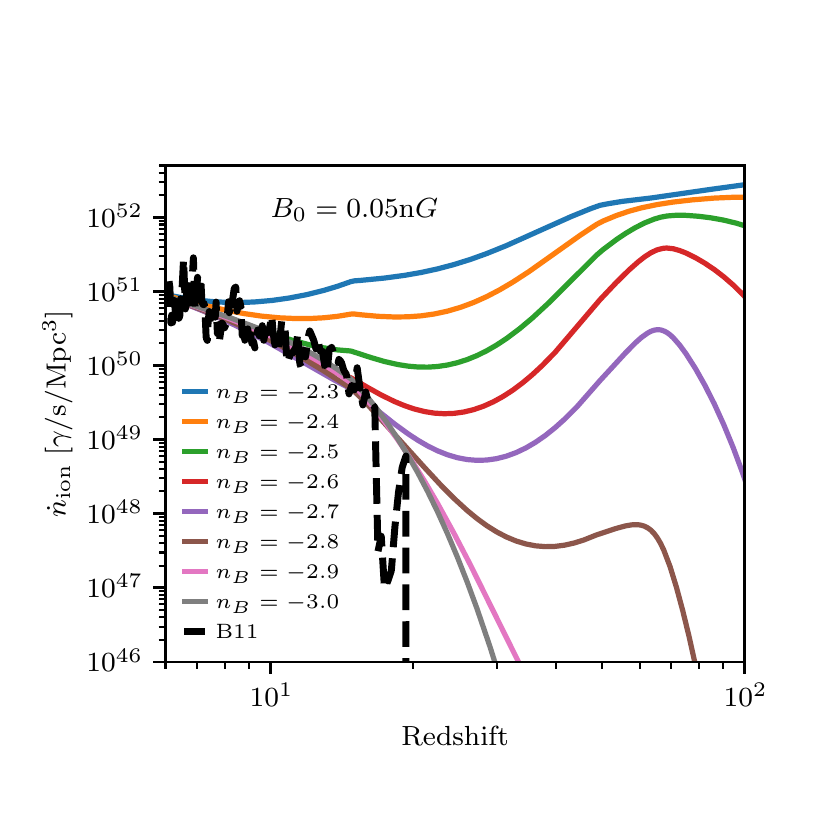}}
\centerline{\includegraphics[scale=1,trim={0 0.6cm 0 1.4cm},clip]{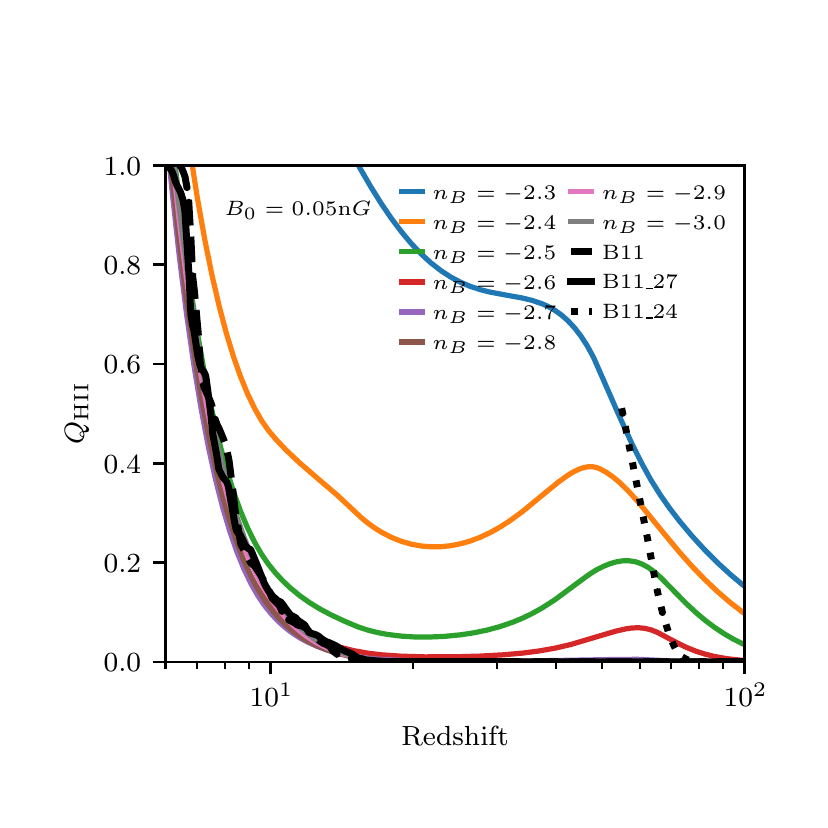}}
\centerline{\includegraphics[scale=1,trim={0 0.6cm 0 1.4cm},clip]{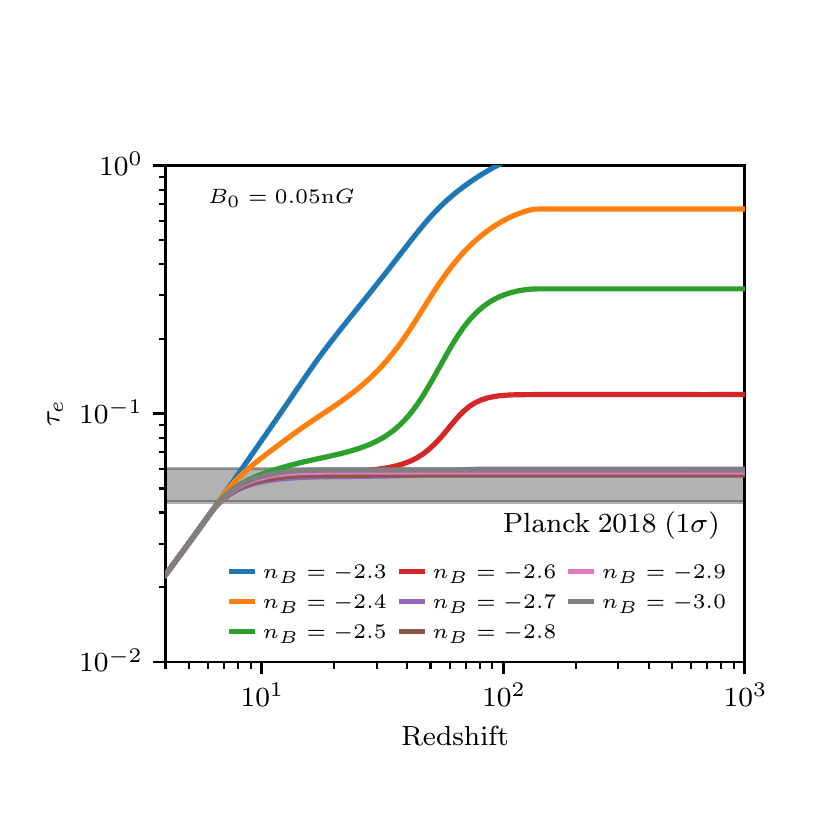}}
\caption{(top) Ionising photon production rate, $\dot{n}_{\rm ion}$, as a function of redshift for PMF models with $B_0=0.05{\rm n}G$ and $n_B$ in the range $[-3.0,-2.3]$.  The dashed black line shows the results from the B11 simulation. (centre) Ionised fraction $Q_{\rm HII}$ as a function of redshift for each PMF model compared to the results from the B11 (dashed), B11\_27 (solid), and B11\_24 (dotted) simulations represented by black lines. (bottom) Electron optical depth for each PMF compared to the constraints from \protect\cite{Planck2018}.}
\label{nion}
\end{figure}

\begin{figure}
\centerline{\includegraphics[scale=1,trim={0 0.0cm 0 1.4cm},clip]{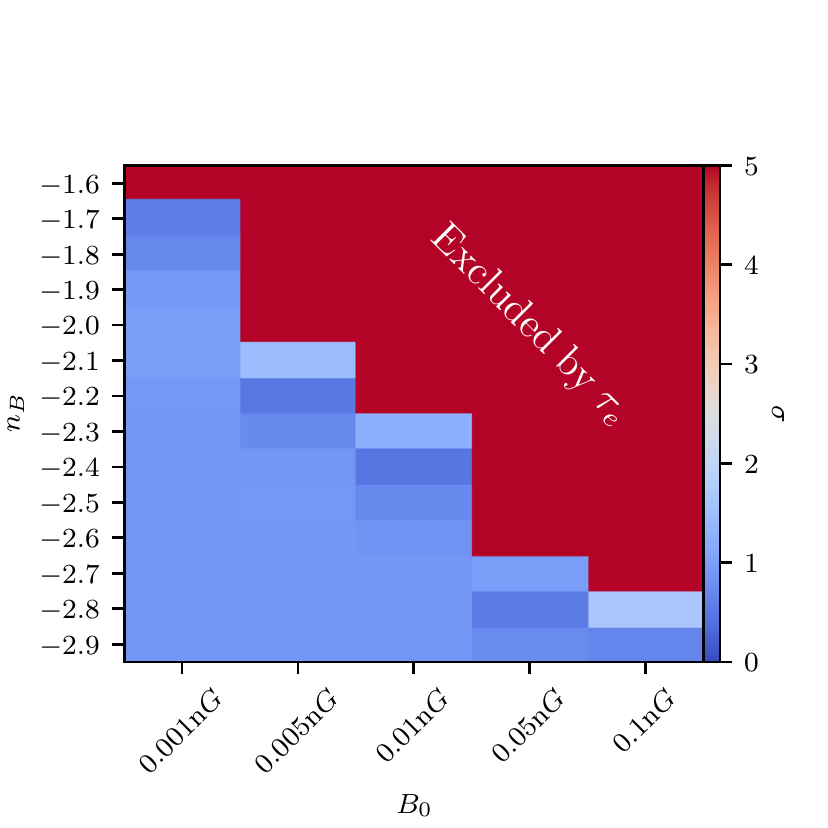}}
\caption{Constraints on $n_B$ as a function of $B_0$ based on the electron optical depth $\tau_e$. The colour at each grid point indicates by how many $\sigma = |\tau_{e,{\rm model}}-\tau_{e, {\rm Planck}}|/\sigma_{\tau_{e,{\rm Planck}}}$ the measured value of $\tau_e$ deviates from the \protect\cite{Planck2018} value. }
\label{bgrid}
\end{figure}

In the top panel of Figure~\ref{nion}, we plot $\dot{n}_{\rm ion}$ as a function of redshift for PMFs with $B_0=0.05{\rm n}G$ and $n_B$ in the range $[-3.0,-2.3]$. The grey line which represents $n_B=-3.0$ reproduces the results from the B11 simulation confirming that our analytic model is well calibrated with the simulation.

As $n_B$ increases, $\dot{n}_{\rm ion}$ deviates more from the fiducial case of $n_B=-3.0$. Modifications to the matter power spectrum drive earlier collapse of more massive haloes. Extreme models (e.g. $n_B=-2.4$) have a decreasing $\dot{n}_{\rm ion}$ as a function of redshift. The shapes of these lines are dictated by the competition between the collapse of new haloes and the decreasing occupation fraction of low mass haloes as a function of redshift. Flatter values of $n_B$ result in enhancements in the number densities of higher mass haloes as well (see Figure~\ref{ICmag2}) which are only impacted by the evolving occupation fraction at lower redshifts.

In the centre panel of Figure~\ref{nion}, we show $Q_{\rm HII}$ as a function of redshift for $B_0=0.05{\rm n}G$ and $n_B$ in the range $[-3.0,-2.3]$. For comparison, we also show $Q_{\rm HII}$ as a function of redshift for the B11 (dashed), B11\_27 (solid), and B11\_24 simulations (dotted). As expected, the model with $n_B=-3.0$ reionizes at $z\sim6$, consistent with the B11 simulation. Similarly, the model with $n_B=-2.7$ closely follows the model with $n_B=-3.0$ and exhibits a reionization history consistent with the B11\_27 simulation. The model with $n_B=-2.4$ exhibits an ionisation fraction of $40\%$ at $z=50$, which is consistent with the early evolution of $Q_{\rm HII}$ in the B11\_24 simulation; however, the analytic model exhibits a more gradual evolution in $Q_{\rm HII}$ compared to the simulation. This may be due to the fact that the simulation does not have high enough resolution to resolve some of the low mass haloes that may be contributing ionising photons at these redshifts. Furthermore, the recombination timescales at high redshift are very short and we note again that $Q_{\rm HII}$ at these high redshifts is very sensitive to the chosen clumping factor. The final reionization redshift for the model with $n_B=-2.4$ is only slightly higher than that of the other models because the short recombination timescale and the evolving occupation fraction result in a decreasing $\dot{n}_{\rm ion}$. Unfortunately, computational limitations prevent us from evolving the B11\_24 simulation to the redshifts where the analytic models predict a turnover in $Q_{\rm HII}$. For $n_B=-2.3$, the reionization redshift is completely incompatible with observations as it occurs at $z\sim15$. Our results are once again qualitatively consistent with the analytic models of \citep{Sanati2020} who demonstrated that strong differences in the reionization history only appear for models with $n_B\gtrsim-2.4$. 

For models with $B_0=0.05{\rm n}G$ and $n_B\geq-2.6$, $Q_{\rm HII}$ deviates noticeably from zero at very high redshift due to early onset star formation. Although $Q_{\rm HII}$ decreases again due to the short recombination timescale and the evolution of $f_{\rm occ}$, the free electrons at high redshift can impact the electron optical depth, $\tau_e$. In the bottom panel of Figure~\ref{nion}, we plot $\tau_e$ for each of the PMF models.  While the models with $n_B<-2.6$ have a $\tau_e$ that is consistent with observations of the CMB \citep{Planck2018}, flatter spectral indices result in electron optical depths that are inconsistent with observational constraints. Hence $\tau_e$ places a stronger constraint on the properties of PMFs compared to observations of the reionization history.

Using our calibrated model, we run a grid of calculations varying both $B_0$ and $n_B$ and use $\tau_e$ to put constraints on the allowable values of $n_B$ for a given $B_0$. In Figure~\ref{bgrid} we plot $\sigma = |\tau_{e,{\rm model}}-\tau_{e, {\rm Planck}}|/\sigma_{\tau_{e,{\rm Planck}}}$ for each of the models in the grid. As the value of $B_0$ decreases, the allowable values of $n_B$ can deviate more from the scale-free case. At $B_0<10^{-12}G$, we have exhausted our grid and all  $n_B<-1.6$ are in the allowed region. The boundary for allowable values of $n_B$ as a function of $B_0$ is nicely fit by the relation:
\begin{equation}
n_{B,{\rm allowed}} \leq -0.562\log_{10}\left(\frac{B_0}{1{\rm n}G}\right) - 3.35.
\end{equation}

\begin{figure}
\centerline{\includegraphics[scale=1,trim={0 0.6cm 0 1.4cm},clip]{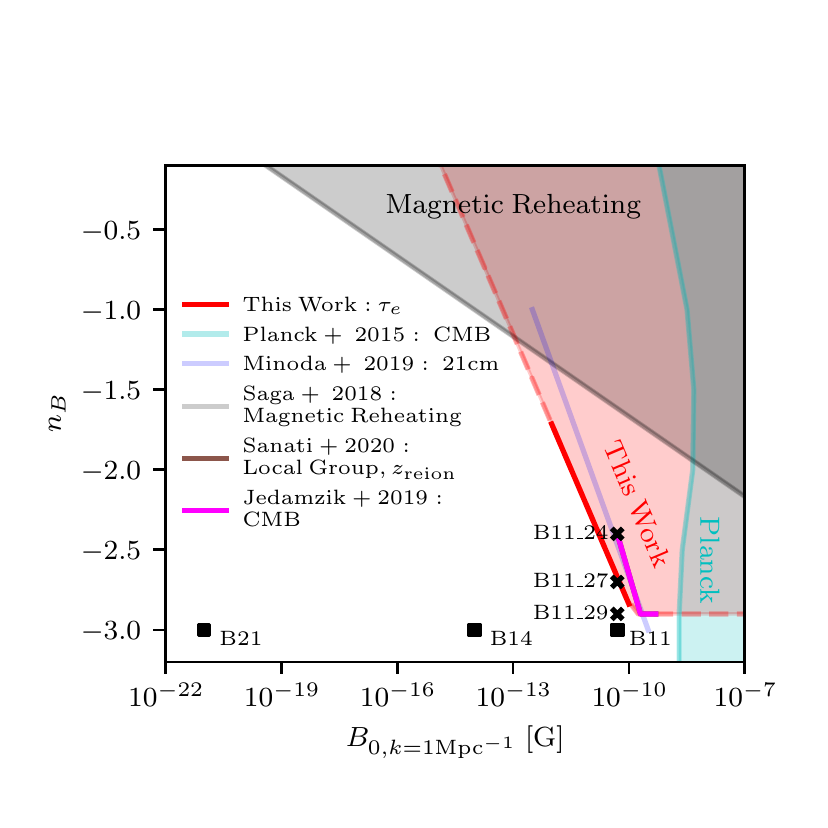}}
\caption{Constraints on $n_B$ as a function of $B_0$ from $\tau_e$ derived in this work (red) compared to other constraints from the literature. The shaded regions represents $B_0$, $n_B$ combinations that are excluded. The dashed red line shows where our constraints have been extrapolated. For comparison, we show constraints from \protect\cite{PlanckPMF2016} from observations of the CMB (cyan), from \protect\cite{Saga2018} based on changes to the baryon-photon number ratio resulting from heating caused by the dissipation of PMFs on small scales (magnetic reheating, grey), from \protect\cite{Jedamzik2019} from CMB anisotropies, and from \protect\cite{Sanati2020} due to the impact of PMFs on the reionization history and the number of luminous local group dwarf galaxies (brown). We also show constraints from \protect\cite{Minoda2019} based on the impact of heating from ambipolar diffusion and decaying magnetic turbulence (blue) on the 21cm signal. Note that these constraints only apply if the signal observed by EDGES \protect\cite{Bowman2018} is confirmed. Data points represent the parameters of our numerical simulations. ``X's" show the locations of the B11\_29, B11\_24, and B11\_24 and squares represent the B21, B14, and B11 simulations. We have placed the latter three simulations at $n_B=-3.0$ for visualisation purposes. because there is no impact to the matter power spectrum, but we stress that these simulations have been initialised with a uniform magnetic field.}
\label{pmf_constraints}
\end{figure}

In Figure~\ref{pmf_constraints} we compare our constraints on the properties of the PMFs based on $\tau_e$ to others in the literature. In general, our constraints are tighter than those based on spectral distortions and other effects that PMFs have on the CMB \citep{PlanckPMF2016}. Similarly, our results constrain a different part of $B_0-n_B$ space compared with \cite{Saga2018} who studied the impact of PMFs on the baryon-photon ratio. As stated previously, our constraints are consistent with those of \cite{Sanati2020} due to the impact of PMFs on the reionization history and the number of luminous local group dwarf galaxies. Finally, our constraint runs nearly parallel to but slightly stronger than that of \cite{Minoda2019} who used the EDGES signal \citep{Bowman2018} to constrain the properties of PMFs based on the heating from ambipolar diffusion and decaying magnetic turbulence (see Section~\ref{sec:21cm}).

We emphasise that these analytic models are subject to limitations. We have had to extrapolate certain relations (e.g. the occupation fraction and the stellar mass-halo mass relation) to redshifts that are not probed by our suite of simulations. Furthermore, the ionisation fraction at high redshift is sensitive to the assumptions that go into calculating the recombination timescale. This may differ in simulations with modified density fields. For strong enough PMFs, gas accretion onto low mass dwarf galaxies may be affected, which will impact our estimate of the star formation rate density. Similarly, the star formation histories of galaxies in our simulation tend to be bursty whereas our model implicitly assumes that they are smooth. Given that the analytic model is able to reproduce both $\dot{n}_{\rm ion}$ and $Q_{\rm HII}$ for the B11, B11\_29, and B11\_27 simulation, we are confident that the model produces reasonable results.

\subsubsection{Constraints from the high-redshift global 21cm signal}
\label{sec:21cm}
Observations of the 21cm line at high redshift are one of the most promising probes of the thermal history of the early Universe \citep[e.g.][]{Pritchard2012}. The brightness temperature of the signal relative to the CMB is
\begin{equation}
\Delta T_b = 27Q_{\rm HI}(1+\delta)\left(\frac{\Omega_bh^2}{0.023}\right)\left(\frac{0.016}{\Omega_mh^2}\right)\left(\frac{1+z}{10}\right)^{0.5}\left(1-\frac{T_{\rm \gamma}}{T_s}\right){\rm mK},
\end{equation}
where $Q_{\rm HI}$ is the neutral fraction, $\delta$ is the fractional overdensity, $T_s$ is the spin temperature of the gas, and $T_{\gamma}$ is the CMB radiation temperature \citep{Furlanetto2006}. The standard picture for the signal usually follows that at $200\lesssim z\lesssim1100$, there is no signal as the gas kinetic temperature and spin temperature are coupled to the radiation temperature due to Compton scattering. As the Universe expands, the gas cools adiabatically and $T_s$ drops below $T_{\gamma}$ due to collisional coupling, hence the signal is seen in absorption. When the density drops below a certain value, collisional coupling becomes inefficient and radiative coupling sets $T_s\sim T_{\gamma}$, diminishing the signal. After the formation of the first stars (and possibly black holes), Ly$\alpha$ and X-ray photons once again couple $T_s$ to the gas temperature which remains below $T_{\gamma}$, resulting in a second absorption signal. Eventually heating from reionization raises the gas above $T_{\gamma}$ and the 21cm signal can be seen in emission until reionization completes \citep{Pritchard2012}. 

The presence of primordial magnetic fields can disrupt this standard model in numerous ways. As discussed in Section~\ref{ideal_mhd}, ambipolar diffusion an decaying magnetic turbulence can increase the ionisation fraction and temperature of the IGM early on in the evolution of the Universe. These two effects are well studied in the literature \citep[e.g.][]{Tashiro2006,Sethi2009,Schleichler2009}. Because of these effects, for strong enough PMFs, $T_s$ never drops below $T_{\gamma}$ and the 21cm signal can only be seen in emission \citep{Sethi2009}. This is in possible conflict with the recent observations from EDGES \citep{Bowman2018} that exhibit a strong absorption profile in the range $14\lesssim z \lesssim 22$. Such results indeed place strong constraints on the properties of PMFs \citep[e.g.][]{Minoda2019,Bera2020,Natwariya2020}. 

While the heating from ambipolar diffusion and decaying magnetic turbulence is well studied, the impact on the global 21cm signal from the modification to structure formation due to PMFs has been less explored. In this section, we employ an analytic model for the global 21cm signal to study how the modification to the matter power spectrum from PMFs impact the global 21cm signal.

To compute the global 21cm signal, we use the {\small ARES} code\footnote{https://github.com/mirochaj/ares} \citep{Mirocha2014,Ares2020} and calibrate the parameters of the models to the results from our simulations. The key components of the model are determining the evolution of the radiation temperature ($T_{\gamma}$), spin temperature ($T_s$), and gas kinetic temperature ($T_K$). $T_{\gamma}$ is set to follow the CMB (i.e. $T_{\gamma}=2.725(1+z)$). $T_K$ is calculated via
\begin{equation}
\frac{3}{2}\frac{d}{dt}\left(\frac{k_BT_kn}{\mu}\right)=\epsilon_X+\epsilon_B+\epsilon_{\rm compton}+\Lambda_C,
\end{equation}
where $n$ is the gas density, $\epsilon_X$ is the photoionisation heating rate, $\epsilon_{\rm B}$ is the heating rate due to ambipolar diffusion and decaying magnetic turbulence, $\epsilon_{\rm compton}$ is the compton heating rate, and $\Lambda_C$ is the cooling rate that encapsulates adiabatic cooling, collisional ionisation cooling, recombination cooling, and collisional excitation cooling \citep{Fukugita1994}. In order to calculate $T_K$ (as well as $T_s$ and $\Delta T_b$), we need to know the emissivity and heating rates from astrophysical sources and derive the ionisation fraction as a function of redshift. For simplicity, we only consider the impact of Ly$\alpha$, HI-ionising, and X-ray photons in our calculation. We have modified {\small ARES} to include the heating rates for ambipolar diffusion and decaying magnetic turbulence following the equations presented in the Appendix of \cite{Chluba2015}.

\begin{figure}
\centerline{\includegraphics[scale=0.99,trim={0.25cm 0.6cm 0 1.4cm},clip]{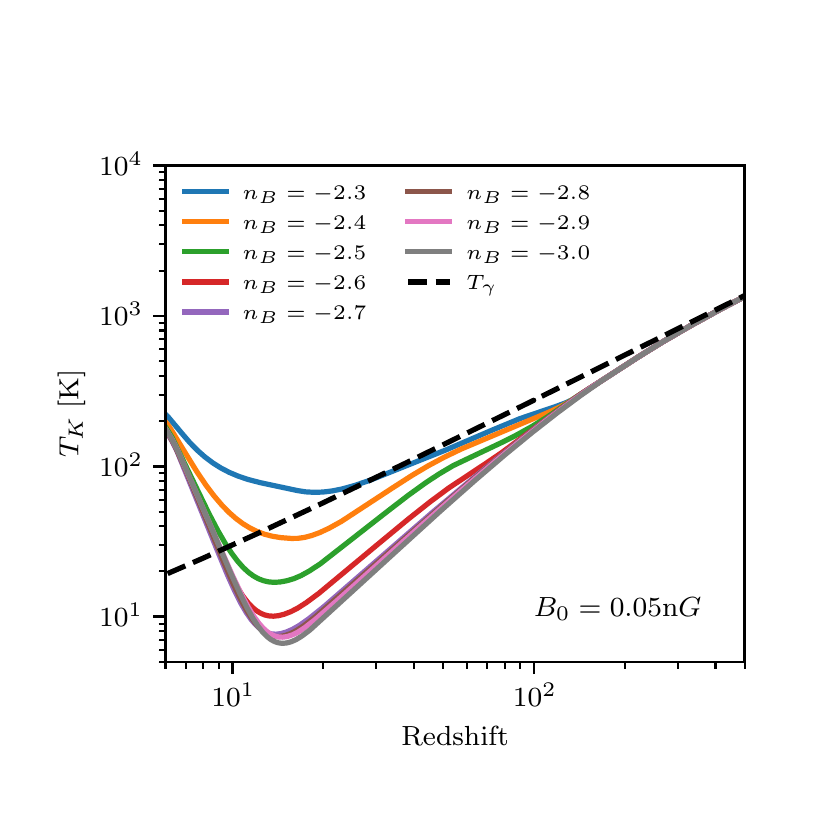}}
\centerline{\includegraphics[scale=0.99,trim={0 0.6cm 0 1.4cm},clip]{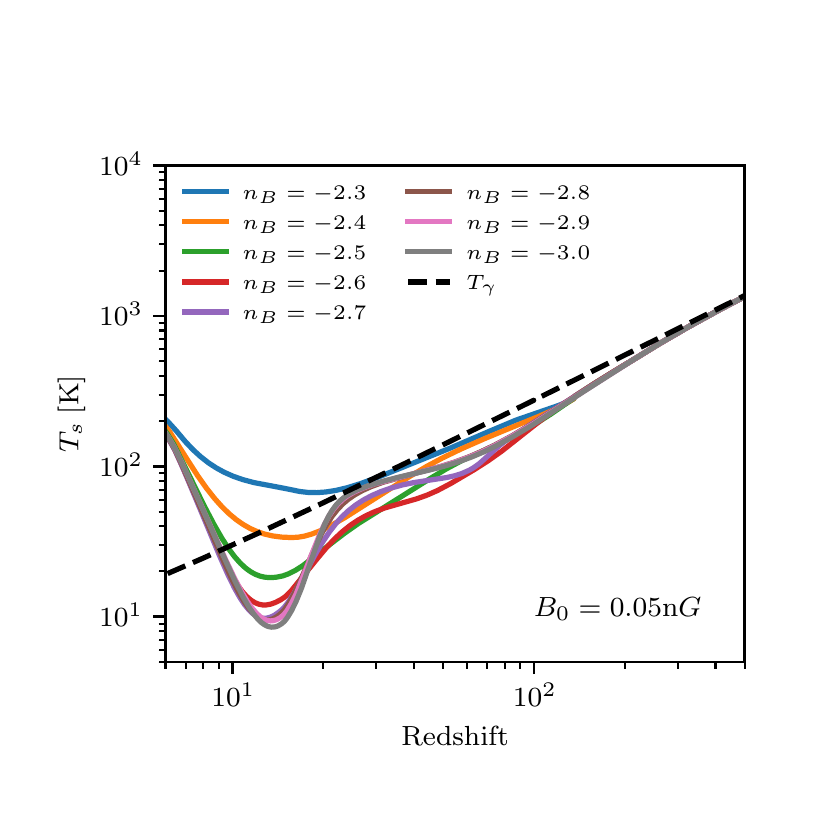}}
\centerline{\includegraphics[scale=0.99,trim={0 0.6cm 0 1.4cm},clip]{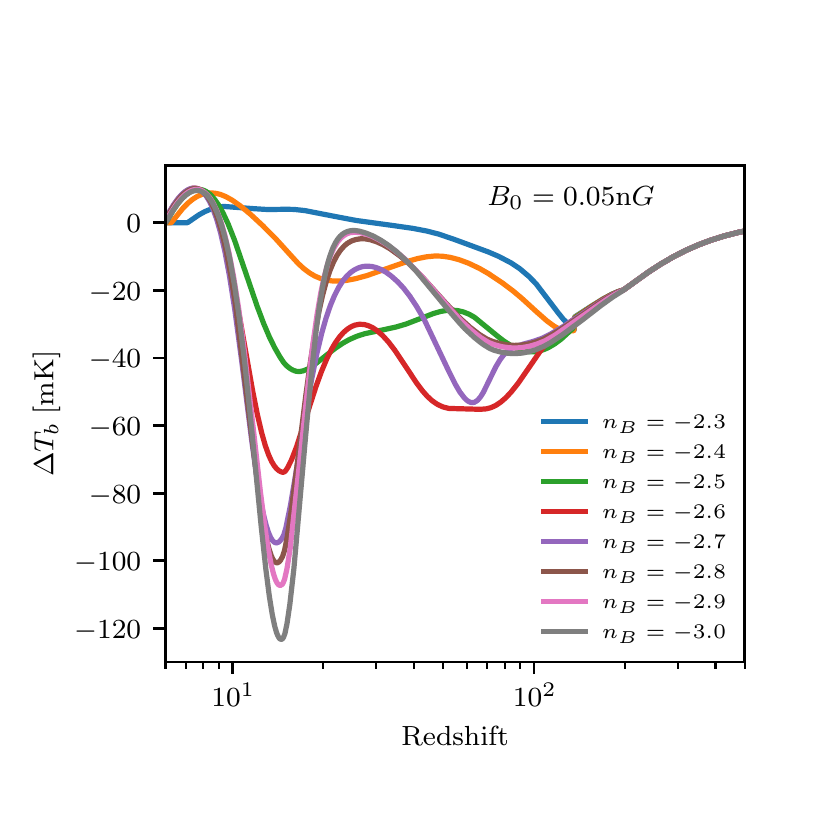}}
\caption{Gas kinetic temperature (top), spin temperature (centre) and global 21cm signal (bottom) as a function of redshift for $B_0=0.05{\rm n}G$ and various values of $n_B$. The dashed black line in the top and centre panels represent the evolution of the radiation temperature, $T_{\gamma}$. }
\label{21cm}
\end{figure}

For each value of $B_0$ and $n_B$, we provide {\small ARES} the values of $\dot{n}_{\rm ion}(z)$ computed in the previous section. Similarly, we also provide {\small ARES} with the number of photons between the Lyman limit and Ly$\alpha$ wavelengths as a function of redshift which are used to compute the Ly$\alpha$ flux. These values were found by integrating the lowest metallicity bin of the BPASSv2.0 SED and assuming an escape fraction of 1.0 at these wavelengths. The X-ray luminosity density is parametrised by the star formation rate density $\dot{\rho}_*(z)$ such that $L_X=f_xc_X\dot{\rho}_*(z)$, where $c_X=2.6\times10^{39}{\rm erg\ s^{-1}}$ \citep{Mineo2012} and $f_X=1$ is a scale factor for the relation\footnote{Note that the depth of the second absorption trough in the global 21cm signal is very sensitive to this parameter.}. We assume that 20\% of the X-ray luminosity is deposited as heat into the IGM \citep{Shull1985,Mirocha2013} and that X-rays also contribute to the ionisation fraction of the gas via secondary ionisations \citep{Shull1985}. For consistency with the previous section, we have applied the same clumping factor evolution, assumed that each ionisation heats the gas to $2\times10^4$K, ignored the contribution from helium, and have edited the recombination rates in {\small ARES} to be consistent with the values in {\small RAMSES-RT}. With these assumptions, we have confirmed that the reionization histories calculated by {\small ARES} are consistent with those in our simulations and analytic models.

The spin temperature of neutral hydrogen is set such that
\begin{equation}
T^{-1}_s = \frac{T^{-1}_{\gamma}+x_{\alpha}T^{-1}_{\alpha}+x_cT^{-1}_{K}}{1+x_a+x_c},
\end{equation}
where $T_{\alpha}\approx T_K$ is the colour temperature of the gas and $x_c$ and $x_{\alpha}$ represent the collisional and radiative coupling coefficients, respectively. $x_c$ is computed by interpolating tabulated values from \cite{Zygelman2005} and coupling due to the Wouthuysen-Field effect \citep{Wouthuysen1952,Field1958} is computed following \cite{Hirata2006} as 
\begin{equation}
x_{\alpha} = \frac{8\pi\lambda^2_{\rm Ly\alpha}\gamma T_*}{9A_{10}T_{\gamma}}S_{\alpha}J_{\alpha},
\end{equation}
where $\lambda_{\rm Ly\alpha}$ is the Ly$\alpha$ wavelength, $T_*=0.068$K is the temperature difference between the HI hyperfine states, $A_{10}=2.85\times10^{-15}$ is the 21cm spontaneous emission coefficient, $\gamma=50$MHz is the half-width at half maximum of the Ly$\alpha$ resonance, $J_{\alpha}$ is the flux of Ly$\alpha$ photons as computed earlier, and $S_{\alpha}$ is a suppression factor that accounts for radiative transfer effects near line centre of the Ly$\alpha$ line \citep[e.g.][]{Furlanetto2006b}. We use the approximation for $S_{\alpha}$ given in \cite{Hirata2006}.

In Figure~\ref{21cm}, we plot $T_K$ (top), $T_s$ (centre), and the global 21cm signal (bottom) as a function of redshift for $B_0=0.05{\rm n}G$ and various values of $n_B$. At $200\lesssim z \lesssim 300$, no stars have formed in any of the models and thus $T_s$ and $T_K$ drop below $T_{\gamma}$, regardless of $n_B$ due to adiabatic expansion and collisional coupling.  By $z\sim150$, the first sources have formed in our most extreme models. This affects the signal in two ways. For models where the ionisation fraction becomes substantial (e.g. $n_B\geq-2.4$), the gas becomes hotter and quickly approaches $T_{\gamma}$. This depletes the strength of the first absorption feature expected at high redshifts. In contrast, for $n_B=-2.6\ {\rm or}\ -2.7$, the heating is not so efficient that the absorption feature is erased, rather, radiative coupling sets in earlier and the spin temperature is coupled to the kinetic temperature for a longer redshift interval with leads to a larger and deeper absorption feature at high redshift (compare the red line with the grey line in the bottom panel of Figure~\ref{21cm}). In both of these models, the spin temperature never fully couples back to the radiation temperature and the global 21cm signal can always be seen in absorption until reionization fully sets in. However, excess heating in the models with $n_B\geq-2.6$ weakens the depth of the second absorption trough to the point where, when $n_B=-2.3$, the second absorption feature is absent and the signal can only be seen in emission. 

We reiterate that none of the models with $n_B\geq-2.6$ are consistent with $\tau_e$ and thus the effects from these models on the global 21cm are unlikely to hold. In contrast, the B11\_27 simulation exhibits a reionization history and $\tau_e$ that are consistent with observational constraints while the 21cm signal, in particular the first absorption feature, is strongly affected. For values of $-3.0<n_B\leq-2.7$, depth of the second absorption feature is only mildly reduced as $n_B$ increases. However, this behaviour is opposite to what is needed to be compatible with the EDGES signal.

The deep absorption feature seen by EDGES is already in conflict with predictions from standard astrophysical scenarios \citep{Bowman2018}. Numerous solutions have been proposed in order to explain the depth of the signal such as more exotic physics, for example, baryon-dark matter interactions or deviations from CDM \citep[e.g.][]{Barkana2018,Munoz2018}, an excess radio background \citep{Fialkov2019}, or an enhanced Ly$\alpha$ background \citep[e.g.][]{Meiksin2020,Mittal2020}. It is also possible that the EDGES signal is entirely spurious \citep{Hills2018}. Unless the PMF is strong enough to prevent accretion onto low mass dwarf galaxies and inhibit star formation in the first collapsing haloes, the presence of strong PMFs with flatter spectral indices will reduce the depth of the global 21cm signal and perhaps result in emission across the redshift range of the EDGES result. Should the EDGES result be validated, this would be indicative of either PMFs that have weak amplitudes or close to scale-free spectral indices, or other physics must counterbalance the impact of the early structure formation.

We note that the global 21cm signal is sensitive to the chosen cosmological parameters in addition to the astrophysical modelling at high redshift. Our models have neglected effects such as baryon streaming velocities \citep{Tseliakhovich2010} which can reduce both the number density and gas content of early mini-haloes \citep[e.g.][]{Naoz2012,Naoz2013}. For large enough streaming velocities, the reduction in gas content and halo number density of low mass haloes may oppose the impact of the modification to the power spectrum, allowing for flatter a $n_B$ at a given $B_0$.

\section{Discussion \& Conclusions}
\label{dc}
We have presented the first results from the {\small SPHINX-MHD} simulations, a suite of cosmological radiation-magnetohydrodynamics simulations designed to study the impact of primordial magnetic fields on reionization and the formation of the first galaxies. We employ two different models for the properties of the PMFs: the first assumes a uniform magnetic field of a given strength as is common technique in cosmological MHD simulations \citep[e.g.][]{IllustrisTNG} and the second assumes Gaussian random magnetic fields with a given strength and spectral index. The former is useful for understanding the impact of magnetic fields on the ISM while the latter has been used to study how the modifications to the matter power spectrum induced by the PMFs \citep{Shaw2012} impacts the history of reionization. 

All of our simulations apart from the B11\_24 simulation result in a $\tau_e$ that is consistent with \cite{Planck2018}. This parameter provides the tightest constraint on the properties of PMFs in the space of $B_0-n_B$ that we have sampled. For PMFs that satisfy the $\tau_e$ constraint, differences between the simulations, while often systematic, become small and comparable to those due to differences in feedback/astrophysics/numerical limitations as described below.

Like all numerical simulations, certain limitations must be considered when interpreting the results. Limited mass and spatial resolution prohibits us from resolving the full spectrum of galaxy formation down to the lowest mass haloes that form Population~III stars and thus their effects on the early Universe. Similarly, finite resolution and numerical diffusivity and viscosity suppress the amplification of magnetic fields in our simulations and reduces their strength in high-redshift galaxies where they are expected to efficiently amplify \citep{Schober2012}. Our simulation volume is far smaller than the cosmological homogeneity scale and therefore we rely on extrapolation to predict the properties of higher mass galaxies where physics might behave differently. We attempted to mitigate the effect of a small volume on the reionization history by choosing an ``average" region (see \citealt{Rosdahl2018}); however, we must keep in mind that there may be effects that are driven by physics on scales larger than the computational volume.

Furthermore, we assumed a set of feedback models aimed at reproducing the effects of energetic processes that are not explicitly resolved by our simulations. Different SN feedback models are well known to impact galaxy properties \citep[e.g.][]{Rosdahl2017} as is the SED that we choose for star particles \citep{Rosdahl2018}, and radiative cooling routines applied to the gas \citep[e.g.][]{Gnedin2012}. How we model this physics can impact star formation and the escape of ionising radiation and we therefore note that certain conclusions are subject to systematic errors. Nevertheless, we have highlighted in many cases where our work is consistent with other numerical simulations presented in the literature that employ different subgrid models providing additional confidence in our results.

With these caveats in mind our main results are as follows:
\begin{itemize}
    \item PMFs with $n_B > -0.562\log_{10}\left(\frac{B_0}{1{\rm n}G}\right) - 3.35$ as well as physical models that generate such fields pre-recombination can be ruled out due to the fact that such models result in an electron optical depth that is inconsistent with constraints from the CMB. 
    \item Early structure formation due to PMFs can strengthen the depth of the first absorption feature in the global 21cm signal at $z\sim60-100$ due to the excess number of X-ray and Ly$\alpha$ photons. This allows the spin temperature to be coupled to the kinetic temperature for a longer period of time. Hence, this regime is particularly sensitive to PMF models that cannot be ruled out by the electron optical depth.
    \item The studied PMF scenarios do not significantly impact the stellar mass-halo mass relation nor the intrinsic luminosity function of galaxies in the epoch of reionization, despite the modifications to the matter power spectrum and the additional pressure support from magnetic fields in the ISM.
    \item Galaxies in the simulations with weak PMFs (B21) have effective radii that are on average 44\% larger than galaxies of similar magnitude in the simulations with strong PMFs (B11). Although this value is well within the scatter in the relation, the result is systematic. This implies that completeness fractions of galaxy surveys at faint magnitudes are lower for a Universe with a weak PMF, which will impact the determination of the high-redshift UV luminosity function.
    \item At UV magnitudes fainter than -12, we do not expect extrapolations of monotonic relations between galaxy size and luminosity to hold due to galaxy ageing and expansion from SN feedback. These effects should be considered in future surveys that can probe such faint magnitudes when estimating the UV luminosity function.
    \item The primordial magnetic fields sampled in this work do not have a strong impact on the gas density field in the post-reionization era due to photoevaporation and photoheating smoothing the density field. Thus it will prove challenging to find any impact on the post-reionization Ly$\alpha$ forest. In contrast, the dark matter density field maintains memory of the modifications to the matter power spectrum and such features could impact other observables such as weak lensing.
    \item Strong PMFs with shallow spectral indices may impact Population~III star formation and the subsequent metal enrichment that is not explicitly resolved by our simulation. However, this will only impact the reionization history at $z\gtrsim18$ where the majority of ionising photons originate from primordial metallicity haloes.
    \item LyC escape fractions are $10\%-25\%$ higher in the simulations with strong PMFs likely due to differences in ISM properties.
    \item By $z=6$, the sources that are leaking half of the ionising photons have $1500\angstrom$ UV magnitudes brighter than $-16$ and are thus well within reach of our deepest surveys. Future surveys that can reliably detect sources with ${\rm M_{UV} < -13}$ will be able to probe the sources of reionization up to $z=12$.
\end{itemize}

In this work we have only considered the impact of primordial magnetic fields on reionization and the formation of the first galaxies. However, there are numerous other mechanisms for seeding cosmic magnetic fields based on astrophysical origins, such as the Biermann battery or in supernova remnants as discussed earlier. In future work (e.g. \citealt{Attia2021}), we will look to model these alternative scenarios to determine their impact on the history of reionization and the first galaxies. Our simulations are among the first to simulate the epoch of reionization with fully coupled radiation-magnetohydrodynamics and thus represent an exciting milestone for cosmological numerical simulations.

\section*{Acknowledgements}
We thank the referee for their detailed revision of the manuscript. HK thanks Richard Shaw for providing us his customised version of {\small CAMB} which allowed us to create the matter initial conditions. This work was supported by the Programme National Cosmology et Galaxies (PNCG) of CNRS/INSU with INP and IN2P3, co-funded by CEA and CNES. MGH  acknowledges support from the UKRI  Science and Technology Facilities Council (grant numbers ST/N000927/1 and ST/S000623/1). Support by ERC Advanced Grant 320596 “The Emergence of Structure during the Epoch of reionization” is gratefully acknowledged. SMA acknowledges support by ERC Starting Grant 638707 “Black holes and their host galaxies: co-evolution across cosmic time”. TK was supported by the National Research Foundation of Korea (NRF-2019K2A9A1A0609137711 and NRF-2020R1C1C100707911). TG acknowledges support from the European Research Council under grant agreement ERC-stg757258 (TRIPLE). The research of AS and JD is supported by Adrian Beecroft and STFC. 
The authors gratefully acknowledge the Gauss Centre for Supercomputing e.V. (www.gauss-centre.eu) for funding this project by providing computing time on the GCS Supercomputer JUWELS \citep{JUWELS} at Jülich Supercomputing Centre (JSC). Computing time for this work was provided by the Partnership for Advanced Computing in Europe (PRACE) as part of the ``First luminous objects and reionization with SPHINX (cont.)'' (2019215124) project. We additionally acknowledges support and computational resources from the Common Computing Facility (CCF) of the LABEX Lyon Institute of Origins (ANR-10-LABX-66).

\section*{Data Availability}
The data underlying this article will be shared on reasonable request
to the corresponding author.



\bibliographystyle{mnras}
\bibliography{example} 

\begin{thebibliography}{}
\makeatletter
\relax
\def\mn@urlcharsother{\let\do\@makeother \do\$\do\&\do\#\do\^\do\_\do\%\do\~}
\def\mn@doi{\begingroup\mn@urlcharsother \@ifnextchar [ {\mn@doi@}
  {\mn@doi@[]}}
\def\mn@doi@[#1]#2{\def\@tempa{#1}\ifx\@tempa\@empty \href
  {http://dx.doi.org/#2} {doi:#2}\else \href {http://dx.doi.org/#2} {#1}\fi
  \endgroup}
\def\mn@eprint#1#2{\mn@eprint@#1:#2::\@nil}
\def\mn@eprint@arXiv#1{\href {http://arxiv.org/abs/#1} {{\tt arXiv:#1}}}
\def\mn@eprint@dblp#1{\href {http://dblp.uni-trier.de/rec/bibtex/#1.xml}
  {dblp:#1}}
\def\mn@eprint@#1:#2:#3:#4\@nil{\def\@tempa {#1}\def\@tempb {#2}\def\@tempc
  {#3}\ifx \@tempc \@empty \let \@tempc \@tempb \let \@tempb \@tempa \fi \ifx
  \@tempb \@empty \def\@tempb {arXiv}\fi \@ifundefined
  {mn@eprint@\@tempb}{\@tempb:\@tempc}{\expandafter \expandafter \csname
  mn@eprint@\@tempb\endcsname \expandafter{\@tempc}}}

\bibitem[\protect\citeauthoryear{{Attia}, {Teyssier}, {Katz}, {Kimm},
  {Martin-Alvarez}, {Ocvirk}  \& {Rosdahl}}{{Attia} et~al.}{2021}]{Attia2021}
{Attia} O.,  {Teyssier} R.,  {Katz} H.,  {Kimm} T.,  {Martin-Alvarez} S.,
  {Ocvirk} P.,   {Rosdahl} J.,  2021, \mn@doi [\mnras]
  {10.1093/mnras/stab1030}, \href
  {https://ui.adsabs.harvard.edu/abs/2021MNRAS.504.2346A} {504, 2346}

\bibitem[\protect\citeauthoryear{{Aubert}, {Pichon}  \& {Colombi}}{{Aubert}
  et~al.}{2004}]{Aubert2004}
{Aubert} D.,  {Pichon} C.,   {Colombi} S.,  2004, \mn@doi [\mnras]
  {10.1111/j.1365-2966.2004.07883.x}, \href
  {https://ui.adsabs.harvard.edu/abs/2004MNRAS.352..376A} {352, 376}

\bibitem[\protect\citeauthoryear{{Balsara}}{{Balsara}}{2001}]{Balsara2001}
{Balsara} D.~S.,  2001, \mn@doi [Journal of Computational Physics]
  {10.1006/jcph.2001.6917}, \href
  {https://ui.adsabs.harvard.edu/abs/2001JCoPh.174..614B} {174, 614}

\bibitem[\protect\citeauthoryear{{Barkana}}{{Barkana}}{2018}]{Barkana2018}
{Barkana} R.,  2018, \mn@doi [\nat] {10.1038/nature25791}, \href
  {https://ui.adsabs.harvard.edu/abs/2018Natur.555...71B} {555, 71}

\bibitem[\protect\citeauthoryear{{Barrow}, {Robertson}, {Ellis}, {Nakajima},
  {Saxena}, {Stark}  \& {Tang}}{{Barrow} et~al.}{2020}]{Barrow2020}
{Barrow} K. S.~S.,  {Robertson} B.~E.,  {Ellis} R.~S.,  {Nakajima} K.,
  {Saxena} A.,  {Stark} D.~P.,   {Tang} M.,  2020, \mn@doi [\apjl]
  {10.3847/2041-8213/abbd8e}, \href
  {https://ui.adsabs.harvard.edu/abs/2020ApJ...902L..39B} {902, L39}

\bibitem[\protect\citeauthoryear{{Basu} \& {Roy}}{{Basu} \&
  {Roy}}{2013}]{Basu2013}
{Basu} A.,  {Roy} S.,  2013, \mn@doi [\mnras] {10.1093/mnras/stt845}, \href
  {https://ui.adsabs.harvard.edu/abs/2013MNRAS.433.1675B} {433, 1675}

\bibitem[\protect\citeauthoryear{{Beck}}{{Beck}}{2007}]{Beck2007}
{Beck} R.,  2007, \mn@doi [\aap] {10.1051/0004-6361:20066988}, \href
  {https://ui.adsabs.harvard.edu/abs/2007A&A...470..539B} {470, 539}

\bibitem[\protect\citeauthoryear{{Beck}}{{Beck}}{2015}]{Beck2015}
{Beck} R.,  2015, \mn@doi [\aap] {10.1051/0004-6361/201425572}, \href
  {https://ui.adsabs.harvard.edu/abs/2015A&A...578A..93B} {578, A93}

\bibitem[\protect\citeauthoryear{{Beck}, {Dolag}, {Lesch}  \&
  {Kronberg}}{{Beck} et~al.}{2013}]{Beck2013}
{Beck} A.~M.,  {Dolag} K.,  {Lesch} H.,   {Kronberg} P.~P.,  2013, \mn@doi
  [\mnras] {10.1093/mnras/stt1549}, \href
  {https://ui.adsabs.harvard.edu/abs/2013MNRAS.435.3575B} {435, 3575}

\bibitem[\protect\citeauthoryear{{Behroozi} et~al.,}{{Behroozi}
  et~al.}{2020}]{Behroozi2020}
{Behroozi} P.,  et~al., 2020, \mn@doi [\mnras] {10.1093/mnras/staa3164}, \href
  {https://ui.adsabs.harvard.edu/abs/2020MNRAS.499.5702B} {499, 5702}

\bibitem[\protect\citeauthoryear{{Bera}, {Datta}  \& {Samui}}{{Bera}
  et~al.}{2020}]{Bera2020}
{Bera} A.,  {Datta} K.~K.,   {Samui} S.,  2020, \mn@doi [\mnras]
  {10.1093/mnras/staa1529}, \href
  {https://ui.adsabs.harvard.edu/abs/2020MNRAS.498..918B} {498, 918}

\bibitem[\protect\citeauthoryear{{Bernet}, {Miniati}, {Lilly}, {Kronberg}  \&
  {Dessauges-Zavadsky}}{{Bernet} et~al.}{2008}]{Bernet2008}
{Bernet} M.~L.,  {Miniati} F.,  {Lilly} S.~J.,  {Kronberg} P.~P.,
  {Dessauges-Zavadsky} M.,  2008, \mn@doi [\nat] {10.1038/nature07105}, \href
  {https://ui.adsabs.harvard.edu/abs/2008Natur.454..302B} {454, 302}

\bibitem[\protect\citeauthoryear{{Biermann}}{{Biermann}}{1950}]{Biermann1950}
{Biermann} L.,  1950, Zeitschrift Naturforschung Teil A, \href
  {https://ui.adsabs.harvard.edu/abs/1950ZNatA...5...65B} {5, 65}

\bibitem[\protect\citeauthoryear{{Birnboim}}{{Birnboim}}{2009}]{Birnboim2009}
{Birnboim} Y.,  2009, \mn@doi [\apjl] {10.1088/0004-637X/702/2/L101}, \href
  {https://ui.adsabs.harvard.edu/abs/2009ApJ...702L.101B} {702, L101}

\bibitem[\protect\citeauthoryear{{Blasi}, {Burles}  \& {Olinto}}{{Blasi}
  et~al.}{1999}]{Blasi1999}
{Blasi} P.,  {Burles} S.,   {Olinto} A.~V.,  1999, \mn@doi [\apjl]
  {10.1086/311958}, \href
  {https://ui.adsabs.harvard.edu/abs/1999ApJ...514L..79B} {514, L79}

\bibitem[\protect\citeauthoryear{{Bolton}, {Haehnelt}, {Warren}, {Hewett},
  {Mortlock}, {Venemans}, {McMahon}  \& {Simpson}}{{Bolton}
  et~al.}{2011}]{Bolton2011}
{Bolton} J.~S.,  {Haehnelt} M.~G.,  {Warren} S.~J.,  {Hewett} P.~C.,
  {Mortlock} D.~J.,  {Venemans} B.~P.,  {McMahon} R.~G.,   {Simpson} C.,  2011,
  \mn@doi [\mnras] {10.1111/j.1745-3933.2011.01100.x}, \href
  {https://ui.adsabs.harvard.edu/abs/2011MNRAS.416L..70B} {416, L70}

\bibitem[\protect\citeauthoryear{{Bonvin}, {Caprini}  \& {Durrer}}{{Bonvin}
  et~al.}{2013}]{Bonvin2013}
{Bonvin} C.,  {Caprini} C.,   {Durrer} R.,  2013, \mn@doi [\prd]
  {10.1103/PhysRevD.88.083515}, \href
  {https://ui.adsabs.harvard.edu/abs/2013PhRvD..88h3515B} {88, 083515}

\bibitem[\protect\citeauthoryear{{Bouwens}, {Illingworth}, {Oesch}, {Maseda},
  {Ribeiro}, {Stefanon}  \& {Lam}}{{Bouwens} et~al.}{2017a}]{Bouwens2017}
{Bouwens} R.~J.,  {Illingworth} G.~D.,  {Oesch} P.~A.,  {Maseda} M.,  {Ribeiro}
  B.,  {Stefanon} M.,   {Lam} D.,  2017a, arXiv e-prints, \href
  {https://ui.adsabs.harvard.edu/abs/2017arXiv171102090B} {p. arXiv:1711.02090}

\bibitem[\protect\citeauthoryear{{Bouwens}, {Oesch}, {Illingworth}, {Ellis}  \&
  {Stefanon}}{{Bouwens} et~al.}{2017b}]{Bouwens2017b}
{Bouwens} R.~J.,  {Oesch} P.~A.,  {Illingworth} G.~D.,  {Ellis} R.~S.,
  {Stefanon} M.,  2017b, \mn@doi [\apj] {10.3847/1538-4357/aa70a4}, \href
  {https://ui.adsabs.harvard.edu/abs/2017ApJ...843..129B} {843, 129}

\bibitem[\protect\citeauthoryear{{Bouwens} et~al.,}{{Bouwens}
  et~al.}{2020}]{Bouwens2020}
{Bouwens} R.,  et~al., 2020, arXiv e-prints, \href
  {https://ui.adsabs.harvard.edu/abs/2020arXiv200910727B} {p. arXiv:2009.10727}

\bibitem[\protect\citeauthoryear{{Bowman}, {Rogers}, {Monsalve}, {Mozdzen}  \&
  {Mahesh}}{{Bowman} et~al.}{2018}]{Bowman2018}
{Bowman} J.~D.,  {Rogers} A. E.~E.,  {Monsalve} R.~A.,  {Mozdzen} T.~J.,
  {Mahesh} N.,  2018, \mn@doi [\nat] {10.1038/nature25792}, \href
  {https://ui.adsabs.harvard.edu/abs/2018Natur.555...67B} {555, 67}

\bibitem[\protect\citeauthoryear{{Brandenburg} \& {Subramanian}}{{Brandenburg}
  \& {Subramanian}}{2005}]{Brandenburg2005}
{Brandenburg} A.,  {Subramanian} K.,  2005, \mn@doi [\physrep]
  {10.1016/j.physrep.2005.06.005}, \href
  {https://ui.adsabs.harvard.edu/abs/2005PhR...417....1B} {417, 1}

\bibitem[\protect\citeauthoryear{{Broderick}, {Chang}  \&
  {Pfrommer}}{{Broderick} et~al.}{2012}]{Broderick2012}
{Broderick} A.~E.,  {Chang} P.,   {Pfrommer} C.,  2012, \mn@doi [\apj]
  {10.1088/0004-637X/752/1/22}, \href
  {https://ui.adsabs.harvard.edu/abs/2012ApJ...752...22B} {752, 22}

\bibitem[\protect\citeauthoryear{{Broderick}, {Tiede}, {Chang}, {Lamberts},
  {Pfrommer}, {Puchwein}, {Shalaby}  \& {Werhahn}}{{Broderick}
  et~al.}{2018}]{Broderick2018}
{Broderick} A.~E.,  {Tiede} P.,  {Chang} P.,  {Lamberts} A.,  {Pfrommer} C.,
  {Puchwein} E.,  {Shalaby} M.,   {Werhahn} M.,  2018, \mn@doi [\apj]
  {10.3847/1538-4357/aae5f2}, \href
  {https://ui.adsabs.harvard.edu/abs/2018ApJ...868...87B} {868, 87}

\bibitem[\protect\citeauthoryear{{Brooks}, {Governato}, {Booth}, {Willman},
  {Gardner}, {Wadsley}, {Stinson}  \& {Quinn}}{{Brooks}
  et~al.}{2007}]{Brooks2007}
{Brooks} A.~M.,  {Governato} F.,  {Booth} C.~M.,  {Willman} B.,  {Gardner}
  J.~P.,  {Wadsley} J.,  {Stinson} G.,   {Quinn} T.,  2007, \mn@doi [\apjl]
  {10.1086/511765}, \href
  {https://ui.adsabs.harvard.edu/abs/2007ApJ...655L..17B} {655, L17}

\bibitem[\protect\citeauthoryear{{Butsky}, {Zrake}, {Kim}, {Yang}  \&
  {Abel}}{{Butsky} et~al.}{2017}]{Butsky2017}
{Butsky} I.,  {Zrake} J.,  {Kim} J.-h.,  {Yang} H.-I.,   {Abel} T.,  2017,
  \mn@doi [\apj] {10.3847/1538-4357/aa799f}, \href
  {https://ui.adsabs.harvard.edu/abs/2017ApJ...843..113B} {843, 113}

\bibitem[\protect\citeauthoryear{{Calverley}, {Becker}, {Haehnelt}  \&
  {Bolton}}{{Calverley} et~al.}{2011}]{Calverly2011}
{Calverley} A.~P.,  {Becker} G.~D.,  {Haehnelt} M.~G.,   {Bolton} J.~S.,  2011,
  \mn@doi [\mnras] {10.1111/j.1365-2966.2010.18072.x}, \href
  {https://ui.adsabs.harvard.edu/abs/2011MNRAS.412.2543C} {412, 2543}

\bibitem[\protect\citeauthoryear{{Caprini} \& {Durrer}}{{Caprini} \&
  {Durrer}}{2002}]{Caprini2002}
{Caprini} C.,  {Durrer} R.,  2002, \mn@doi [\prd] {10.1103/PhysRevD.65.023517},
  \href {https://ui.adsabs.harvard.edu/abs/2002PhRvD..65b3517C} {65, 023517}

\bibitem[\protect\citeauthoryear{{Caruana}, {Bunker}, {Wilkins}, {Stanway},
  {Lorenzoni}, {Jarvis}  \& {Ebert}}{{Caruana} et~al.}{2014}]{Caruna2014}
{Caruana} J.,  {Bunker} A.~J.,  {Wilkins} S.~M.,  {Stanway} E.~R.,  {Lorenzoni}
  S.,  {Jarvis} M.~J.,   {Ebert} H.,  2014, \mn@doi [\mnras]
  {10.1093/mnras/stu1341}, \href
  {https://ui.adsabs.harvard.edu/abs/2014MNRAS.443.2831C} {443, 2831}

\bibitem[\protect\citeauthoryear{Chirakkara, Federrath, Trivedi  \&
  Banerjee}{Chirakkara et~al.}{2021}]{chirakkara2021efficient}
Chirakkara R.~A.,  Federrath C.,  Trivedi P.,   Banerjee R.,  2021, Efficient
  highly-subsonic turbulent dynamo and growth of primordial magnetic fields
  (\mn@eprint {arXiv} {2101.08256})

\bibitem[\protect\citeauthoryear{{Chluba}, {Paoletti}, {Finelli}  \&
  {Rubi{\~n}o-Mart{\'\i}n}}{{Chluba} et~al.}{2015}]{Chluba2015}
{Chluba} J.,  {Paoletti} D.,  {Finelli} F.,   {Rubi{\~n}o-Mart{\'\i}n} J.~A.,
  2015, \mn@doi [\mnras] {10.1093/mnras/stv1096}, \href
  {https://ui.adsabs.harvard.edu/abs/2015MNRAS.451.2244C} {451, 2244}

\bibitem[\protect\citeauthoryear{{Chongchitnan} \& {Meiksin}}{{Chongchitnan} \&
  {Meiksin}}{2014}]{Meiksin2014}
{Chongchitnan} S.,  {Meiksin} A.,  2014, \mn@doi [\mnras]
  {10.1093/mnras/stt2169}, \href
  {https://ui.adsabs.harvard.edu/abs/2014MNRAS.437.3639C} {437, 3639}

\bibitem[\protect\citeauthoryear{{Chy{\.z}y}, {We{\.z}gowiec}, {Beck}  \&
  {Bomans}}{{Chy{\.z}y} et~al.}{2011}]{Chyzy2011}
{Chy{\.z}y} K.~T.,  {We{\.z}gowiec} M.,  {Beck} R.,   {Bomans} D.~J.,  2011,
  \mn@doi [\aap] {10.1051/0004-6361/201015393}, \href
  {https://ui.adsabs.harvard.edu/abs/2011A&A...529A..94C} {529, A94}

\bibitem[\protect\citeauthoryear{{Commer{\c{c}}on}, {Debout}  \&
  {Teyssier}}{{Commer{\c{c}}on} et~al.}{2014}]{Commercon2014}
{Commer{\c{c}}on} B.,  {Debout} V.,   {Teyssier} R.,  2014, \mn@doi [\aap]
  {10.1051/0004-6361/201322858}, \href
  {https://ui.adsabs.harvard.edu/abs/2014A&A...563A..11C} {563, A11}

\bibitem[\protect\citeauthoryear{{Crain} et~al.,}{{Crain}
  et~al.}{2015}]{Crain2015}
{Crain} R.~A.,  et~al., 2015, \mn@doi [\mnras] {10.1093/mnras/stv725}, \href
  {https://ui.adsabs.harvard.edu/abs/2015MNRAS.450.1937C} {450, 1937}

\bibitem[\protect\citeauthoryear{{Crutcher}}{{Crutcher}}{2012}]{Crutcher2012}
{Crutcher} R.~M.,  2012, \mn@doi [\araa] {10.1146/annurev-astro-081811-125514},
  \href {https://ui.adsabs.harvard.edu/abs/2012ARA&A..50...29C} {50, 29}

\bibitem[\protect\citeauthoryear{{D'Aloisio}, {McQuinn}, {Davies}  \&
  {Furlanetto}}{{D'Aloisio} et~al.}{2018}]{Daloisio2018}
{D'Aloisio} A.,  {McQuinn} M.,  {Davies} F.~B.,   {Furlanetto} S.~R.,  2018,
  \mn@doi [\mnras] {10.1093/mnras/stx2341}, \href
  {https://ui.adsabs.harvard.edu/abs/2018MNRAS.473..560D} {473, 560}

\bibitem[\protect\citeauthoryear{{Davis} \& {Greenstein}}{{Davis} \&
  {Greenstein}}{1951}]{Davis1951}
{Davis} Leverett J.,  {Greenstein} J.~L.,  1951, \mn@doi [\apj]
  {10.1086/145464}, \href
  {https://ui.adsabs.harvard.edu/abs/1951ApJ...114..206D} {114, 206}

\bibitem[\protect\citeauthoryear{{Dawoodbhoy} et~al.,}{{Dawoodbhoy}
  et~al.}{2018}]{Dawoodbhoy2018}
{Dawoodbhoy} T.,  et~al., 2018, \mn@doi [\mnras] {10.1093/mnras/sty1945}, \href
  {http://adsabs.harvard.edu/abs/2018MNRAS.480.1740D} {480, 1740}

\bibitem[\protect\citeauthoryear{{Dedner}, {Kemm}, {Kr{\"o}ner}, {Munz},
  {Schnitzer}  \& {Wesenberg}}{{Dedner} et~al.}{2002}]{Dedner2002}
{Dedner} A.,  {Kemm} F.,  {Kr{\"o}ner} D.,  {Munz} C.~D.,  {Schnitzer} T.,
  {Wesenberg} M.,  2002, \mn@doi [Journal of Computational Physics]
  {10.1006/jcph.2001.6961}, \href
  {https://ui.adsabs.harvard.edu/abs/2002JCoPh.175..645D} {175, 645}

\bibitem[\protect\citeauthoryear{{Dekel} \& {Silk}}{{Dekel} \&
  {Silk}}{1986}]{Dekel1986}
{Dekel} A.,  {Silk} J.,  1986, \mn@doi [\apj] {10.1086/164050}, \href
  {https://ui.adsabs.harvard.edu/abs/1986ApJ...303...39D} {303, 39}

\bibitem[\protect\citeauthoryear{{Diemer}}{{Diemer}}{2018}]{Diemer2018}
{Diemer} B.,  2018, \mn@doi [\apjs] {10.3847/1538-4365/aaee8c}, \href
  {https://ui.adsabs.harvard.edu/abs/2018ApJS..239...35D} {239, 35}

\bibitem[\protect\citeauthoryear{{Dolag} \& {Stasyszyn}}{{Dolag} \&
  {Stasyszyn}}{2009}]{Dolag2009}
{Dolag} K.,  {Stasyszyn} F.,  2009, \mn@doi [\mnras]
  {10.1111/j.1365-2966.2009.15181.x}, \href
  {https://ui.adsabs.harvard.edu/abs/2009MNRAS.398.1678D} {398, 1678}

\bibitem[\protect\citeauthoryear{{Dolag}, {Kachelriess}, {Ostapchenko}  \&
  {Tom{\`a}s}}{{Dolag} et~al.}{2011}]{Dolag2011}
{Dolag} K.,  {Kachelriess} M.,  {Ostapchenko} S.,   {Tom{\`a}s} R.,  2011,
  \mn@doi [\apjl] {10.1088/2041-8205/727/1/L4}, \href
  {https://ui.adsabs.harvard.edu/abs/2011ApJ...727L...4D} {727, L4}

\bibitem[\protect\citeauthoryear{{Donnert}, {Vazza}, {Br{\"u}ggen}  \&
  {ZuHone}}{{Donnert} et~al.}{2018}]{Donnert2018}
{Donnert} J.,  {Vazza} F.,  {Br{\"u}ggen} M.,   {ZuHone} J.,  2018, \mn@doi
  [\ssr] {10.1007/s11214-018-0556-8}, \href
  {https://ui.adsabs.harvard.edu/abs/2018SSRv..214..122D} {214, 122}

\bibitem[\protect\citeauthoryear{{Doumler} \& {Knebe}}{{Doumler} \&
  {Knebe}}{2010}]{Doumler2010}
{Doumler} T.,  {Knebe} A.,  2010, \mn@doi [\mnras]
  {10.1111/j.1365-2966.2009.16144.x}, \href
  {https://ui.adsabs.harvard.edu/abs/2010MNRAS.403..453D} {403, 453}

\bibitem[\protect\citeauthoryear{{Dubois} \& {Teyssier}}{{Dubois} \&
  {Teyssier}}{2008}]{Dubois2008}
{Dubois} Y.,  {Teyssier} R.,  2008, \mn@doi [\aap]
  {10.1051/0004-6361:200809513}, \href
  {https://ui.adsabs.harvard.edu/abs/2008A&A...482L..13D} {482, L13}

\bibitem[\protect\citeauthoryear{{Duffin} \& {Pudritz}}{{Duffin} \&
  {Pudritz}}{2009}]{Duffin2009}
{Duffin} D.~F.,  {Pudritz} R.~E.,  2009, \mn@doi [\apjl]
  {10.1088/0004-637X/706/1/L46}, \href
  {https://ui.adsabs.harvard.edu/abs/2009ApJ...706L..46D} {706, L46}

\bibitem[\protect\citeauthoryear{{Durrive} \& {Langer}}{{Durrive} \&
  {Langer}}{2015}]{Durrive2015}
{Durrive} J.~B.,  {Langer} M.,  2015, \mn@doi [\mnras] {10.1093/mnras/stv1578},
  \href {https://ui.adsabs.harvard.edu/abs/2015MNRAS.453..345D} {453, 345}

\bibitem[\protect\citeauthoryear{{Durrive}, {Tashiro}, {Langer}  \&
  {Sugiyama}}{{Durrive} et~al.}{2017}]{Durrive2017}
{Durrive} J.-B.,  {Tashiro} H.,  {Langer} M.,   {Sugiyama} N.,  2017, \mn@doi
  [\mnras] {10.1093/mnras/stx2007}, \href
  {https://ui.adsabs.harvard.edu/abs/2017MNRAS.472.1649D} {472, 1649}

\bibitem[\protect\citeauthoryear{{Efstathiou}}{{Efstathiou}}{1992}]{Efstathiou1992}
{Efstathiou} G.,  1992, \mn@doi [\mnras] {10.1093/mnras/256.1.43P}, \href
  {https://ui.adsabs.harvard.edu/abs/1992MNRAS.256P..43E} {256, 43P}

\bibitem[\protect\citeauthoryear{{Eisenstein} \& {Hu}}{{Eisenstein} \&
  {Hu}}{1998}]{Eisenstein1999}
{Eisenstein} D.~J.,  {Hu} W.,  1998, \mn@doi [\apj] {10.1086/305424}, \href
  {https://ui.adsabs.harvard.edu/abs/1998ApJ...496..605E} {496, 605}

\bibitem[\protect\citeauthoryear{{Eldridge}, {Izzard}  \& {Tout}}{{Eldridge}
  et~al.}{2008}]{Eldridge2008}
{Eldridge} J.~J.,  {Izzard} R.~G.,   {Tout} C.~A.,  2008, \mn@doi [\mnras]
  {10.1111/j.1365-2966.2007.12738.x}, \href
  {https://ui.adsabs.harvard.edu/abs/2008MNRAS.384.1109E} {384, 1109}

\bibitem[\protect\citeauthoryear{{Evans} \& {Hawley}}{{Evans} \&
  {Hawley}}{1988}]{Evans1988}
{Evans} C.~R.,  {Hawley} J.~F.,  1988, \mn@doi [\apj] {10.1086/166684}, \href
  {https://ui.adsabs.harvard.edu/abs/1988ApJ...332..659E} {332, 659}

\bibitem[\protect\citeauthoryear{{Fan} et~al.,}{{Fan} et~al.}{2006}]{Fan2006}
{Fan} X.,  et~al., 2006, \mn@doi [\aj] {10.1086/504836}, \href
  {https://ui.adsabs.harvard.edu/abs/2006AJ....132..117F} {132, 117}

\bibitem[\protect\citeauthoryear{{Federrath} \& {Klessen}}{{Federrath} \&
  {Klessen}}{2012}]{Federrath2012}
{Federrath} C.,  {Klessen} R.~S.,  2012, \mn@doi [\apj]
  {10.1088/0004-637X/761/2/156}, \href
  {https://ui.adsabs.harvard.edu/abs/2012ApJ...761..156F} {761, 156}

\bibitem[\protect\citeauthoryear{{Federrath}, {Schober}, {Bovino}  \&
  {Schleicher}}{{Federrath} et~al.}{2014}]{Federrath2014}
{Federrath} C.,  {Schober} J.,  {Bovino} S.,   {Schleicher} D. R.~G.,  2014,
  \mn@doi [\apjl] {10.1088/2041-8205/797/2/L19}, \href
  {https://ui.adsabs.harvard.edu/abs/2014ApJ...797L..19F} {797, L19}

\bibitem[\protect\citeauthoryear{{Feretti}, {Giovannini}, {Govoni}  \&
  {Murgia}}{{Feretti} et~al.}{2012}]{Feretti2012}
{Feretti} L.,  {Giovannini} G.,  {Govoni} F.,   {Murgia} M.,  2012, \mn@doi
  [\aapr] {10.1007/s00159-012-0054-z}, \href
  {https://ui.adsabs.harvard.edu/abs/2012A&ARv..20...54F} {20, 54}

\bibitem[\protect\citeauthoryear{{Ferland}, {Korista}, {Verner}, {Ferguson},
  {Kingdon}  \& {Verner}}{{Ferland} et~al.}{1998}]{Ferland1998}
{Ferland} G.~J.,  {Korista} K.~T.,  {Verner} D.~A.,  {Ferguson} J.~W.,
  {Kingdon} J.~B.,   {Verner} E.~M.,  1998, \mn@doi [\pasp] {10.1086/316190},
  \href {https://ui.adsabs.harvard.edu/abs/1998PASP..110..761F} {110, 761}

\bibitem[\protect\citeauthoryear{{Fialkov} \& {Barkana}}{{Fialkov} \&
  {Barkana}}{2019}]{Fialkov2019}
{Fialkov} A.,  {Barkana} R.,  2019, \mn@doi [\mnras] {10.1093/mnras/stz873},
  \href {https://ui.adsabs.harvard.edu/abs/2019MNRAS.486.1763F} {486, 1763}

\bibitem[\protect\citeauthoryear{{Field}}{{Field}}{1958}]{Field1958}
{Field} G.~B.,  1958, \mn@doi [Proceedings of the IRE]
  {10.1109/JRPROC.1958.286741}, \href
  {https://ui.adsabs.harvard.edu/abs/1958PIRE...46..240F} {46, 240}

\bibitem[\protect\citeauthoryear{{Finkelstein} et~al.,}{{Finkelstein}
  et~al.}{2015}]{Finkelstein2015}
{Finkelstein} S.~L.,  et~al., 2015, \mn@doi [\apj]
  {10.1088/0004-637X/810/1/71}, \href
  {https://ui.adsabs.harvard.edu/abs/2015ApJ...810...71F} {810, 71}

\bibitem[\protect\citeauthoryear{{Finkelstein} et~al.,}{{Finkelstein}
  et~al.}{2019}]{Finkelstein2019}
{Finkelstein} S.~L.,  et~al., 2019, \mn@doi [\apj] {10.3847/1538-4357/ab1ea8},
  \href {https://ui.adsabs.harvard.edu/abs/2019ApJ...879...36F} {879, 36}

\bibitem[\protect\citeauthoryear{{Friedmann}}{{Friedmann}}{1922}]{Friedmann1922}
{Friedmann} A.,  1922, \mn@doi [Zeitschrift fur Physik] {10.1007/BF01332580},
  \href {https://ui.adsabs.harvard.edu/abs/1922ZPhy...10..377F} {10, 377}

\bibitem[\protect\citeauthoryear{{Fromang}, {Hennebelle}  \&
  {Teyssier}}{{Fromang} et~al.}{2006}]{Fromang2006}
{Fromang} S.,  {Hennebelle} P.,   {Teyssier} R.,  2006, \mn@doi [\aap]
  {10.1051/0004-6361:20065371}, \href
  {https://ui.adsabs.harvard.edu/abs/2006A&A...457..371F} {457, 371}

\bibitem[\protect\citeauthoryear{{Fukugita} \& {Kawasaki}}{{Fukugita} \&
  {Kawasaki}}{1994}]{Fukugita1994}
{Fukugita} M.,  {Kawasaki} M.,  1994, \mn@doi [\mnras]
  {10.1093/mnras/269.3.563}, \href
  {https://ui.adsabs.harvard.edu/abs/1994MNRAS.269..563F} {269, 563}

\bibitem[\protect\citeauthoryear{{Furlanetto}}{{Furlanetto}}{2006}]{Furlanetto2006}
{Furlanetto} S.~R.,  2006, \mn@doi [\mnras] {10.1111/j.1365-2966.2006.10725.x},
  \href {https://ui.adsabs.harvard.edu/abs/2006MNRAS.371..867F} {371, 867}

\bibitem[\protect\citeauthoryear{{Furlanetto}, {Oh}  \& {Briggs}}{{Furlanetto}
  et~al.}{2006}]{Furlanetto2006b}
{Furlanetto} S.~R.,  {Oh} S.~P.,   {Briggs} F.~H.,  2006, \mn@doi [\physrep]
  {10.1016/j.physrep.2006.08.002}, \href
  {https://ui.adsabs.harvard.edu/abs/2006PhR...433..181F} {433, 181}

\bibitem[\protect\citeauthoryear{{Garaldi}, {Pakmor}  \& {Springel}}{{Garaldi}
  et~al.}{2020}]{Garaldi2020}
{Garaldi} E.,  {Pakmor} R.,   {Springel} V.,  2020, arXiv e-prints, \href
  {https://ui.adsabs.harvard.edu/abs/2020arXiv201009729G} {p. arXiv:2010.09729}

\bibitem[\protect\citeauthoryear{{Gardner} et~al.,}{{Gardner}
  et~al.}{2006}]{Gardner2006}
{Gardner} J.~P.,  et~al., 2006, \mn@doi [\ssr] {10.1007/s11214-006-8315-7},
  \href {https://ui.adsabs.harvard.edu/abs/2006SSRv..123..485G} {123, 485}

\bibitem[\protect\citeauthoryear{{Garel}, {Blaizot}, {Rosdahl},
  {Michel-Dansac}, {Haehnelt}, {Katz}, {Kimm}  \& {Verhamme}}{{Garel}
  et~al.}{2021}]{Garel2021}
{Garel} T.,  {Blaizot} J.,  {Rosdahl} J.,  {Michel-Dansac} L.,  {Haehnelt}
  M.~G.,  {Katz} H.,  {Kimm} T.,   {Verhamme} A.,  2021, \mn@doi [\mnras]
  {10.1093/mnras/stab990}, \href
  {https://ui.adsabs.harvard.edu/abs/2021MNRAS.504.1902G} {504, 1902}

\bibitem[\protect\citeauthoryear{{Gnedin} \& {Abel}}{{Gnedin} \&
  {Abel}}{2001}]{Gnedin2001}
{Gnedin} N.~Y.,  {Abel} T.,  2001, \mn@doi [\na]
  {10.1016/S1384-1076(01)00068-9}, \href
  {https://ui.adsabs.harvard.edu/abs/2001NewA....6..437G} {6, 437}

\bibitem[\protect\citeauthoryear{{Gnedin} \& {Hollon}}{{Gnedin} \&
  {Hollon}}{2012}]{Gnedin2012}
{Gnedin} N.~Y.,  {Hollon} N.,  2012, \mn@doi [\apjs]
  {10.1088/0067-0049/202/2/13}, \href
  {https://ui.adsabs.harvard.edu/abs/2012ApJS..202...13G} {202, 13}

\bibitem[\protect\citeauthoryear{{Gnedin} \& {Kaurov}}{{Gnedin} \&
  {Kaurov}}{2014}]{Gnedin2014}
{Gnedin} N.~Y.,  {Kaurov} A.~A.,  2014, \mn@doi [\apj]
  {10.1088/0004-637X/793/1/30}, \href
  {https://ui.adsabs.harvard.edu/abs/2014ApJ...793...30G} {793, 30}

\bibitem[\protect\citeauthoryear{{Gnedin}, {Ferrara}  \& {Zweibel}}{{Gnedin}
  et~al.}{2000}]{Gnedin2000}
{Gnedin} N.~Y.,  {Ferrara} A.,   {Zweibel} E.~G.,  2000, \mn@doi [\apj]
  {10.1086/309272}, \href
  {https://ui.adsabs.harvard.edu/abs/2000ApJ...539..505G} {539, 505}

\bibitem[\protect\citeauthoryear{{Gopal} \& {Sethi}}{{Gopal} \&
  {Sethi}}{2003}]{Gopal2003}
{Gopal} R.,  {Sethi} S.~K.,  2003, \mn@doi [Journal of Astrophysics and
  Astronomy] {10.1007/BF02702312}, \href
  {https://ui.adsabs.harvard.edu/abs/2003JApA...24...51G} {24, 51}

\bibitem[\protect\citeauthoryear{{Grasso} \& {Rubinstein}}{{Grasso} \&
  {Rubinstein}}{1995}]{Grasso1995}
{Grasso} D.,  {Rubinstein} H.~R.,  1995, \mn@doi [Astroparticle Physics]
  {10.1016/0927-6505(94)00030-7}, \href
  {https://ui.adsabs.harvard.edu/abs/1995APh.....3...95G} {3, 95}

\bibitem[\protect\citeauthoryear{{Guillet} \& {Teyssier}}{{Guillet} \&
  {Teyssier}}{2011}]{Guillet2011}
{Guillet} T.,  {Teyssier} R.,  2011, \mn@doi [Journal of Computational Physics]
  {10.1016/j.jcp.2011.02.044}, \href
  {https://ui.adsabs.harvard.edu/abs/2011JCoPh.230.4756G} {230, 4756}

\bibitem[\protect\citeauthoryear{{Haardt} \& {Madau}}{{Haardt} \&
  {Madau}}{1996}]{Haardt1996}
{Haardt} F.,  {Madau} P.,  1996, \mn@doi [\apj] {10.1086/177035}, \href
  {https://ui.adsabs.harvard.edu/abs/1996ApJ...461...20H} {461, 20}

\bibitem[\protect\citeauthoryear{{Hahn} \& {Abel}}{{Hahn} \&
  {Abel}}{2011}]{Hahn2011}
{Hahn} O.,  {Abel} T.,  2011, \mn@doi [\mnras]
  {10.1111/j.1365-2966.2011.18820.x}, \href
  {https://ui.adsabs.harvard.edu/abs/2011MNRAS.415.2101H} {415, 2101}

\bibitem[\protect\citeauthoryear{{Harikane} et~al.,}{{Harikane}
  et~al.}{2018}]{Harikane2018}
{Harikane} Y.,  et~al., 2018, \mn@doi [\pasj] {10.1093/pasj/psx097}, \href
  {https://ui.adsabs.harvard.edu/abs/2018PASJ...70S..11H} {70, S11}

\bibitem[\protect\citeauthoryear{{Haugen}, {Brandenburg}  \& {Dobler}}{{Haugen}
  et~al.}{2004a}]{Haugen2004b}
{Haugen} N.~E.,  {Brandenburg} A.,   {Dobler} W.,  2004a, \mn@doi [\pre]
  {10.1103/PhysRevE.70.016308}, \href
  {https://ui.adsabs.harvard.edu/abs/2004PhRvE..70a6308H} {70, 016308}

\bibitem[\protect\citeauthoryear{{Haugen}, {Brandenburg}  \& {Mee}}{{Haugen}
  et~al.}{2004b}]{Haugen2004}
{Haugen} N. E.~L.,  {Brandenburg} A.,   {Mee} A.~J.,  2004b, \mn@doi [\mnras]
  {10.1111/j.1365-2966.2004.08127.x}, \href
  {https://ui.adsabs.harvard.edu/abs/2004MNRAS.353..947H} {353, 947}

\bibitem[\protect\citeauthoryear{{Hennebelle} \& {Chabrier}}{{Hennebelle} \&
  {Chabrier}}{2011}]{Hennebelle2011}
{Hennebelle} P.,  {Chabrier} G.,  2011, \mn@doi [\apjl]
  {10.1088/2041-8205/743/2/L29}, \href
  {https://ui.adsabs.harvard.edu/abs/2011ApJ...743L..29H} {743, L29}

\bibitem[\protect\citeauthoryear{{Hills}, {Kulkarni}, {Meerburg}  \&
  {Puchwein}}{{Hills} et~al.}{2018}]{Hills2018}
{Hills} R.,  {Kulkarni} G.,  {Meerburg} P.~D.,   {Puchwein} E.,  2018, \mn@doi
  [\nat] {10.1038/s41586-018-0796-5}, \href
  {https://ui.adsabs.harvard.edu/abs/2018Natur.564E..32H} {564, E32}

\bibitem[\protect\citeauthoryear{{Hinshaw} et~al.,}{{Hinshaw}
  et~al.}{2013}]{Hinshaw2013}
{Hinshaw} G.,  et~al., 2013, \mn@doi [\apjs] {10.1088/0067-0049/208/2/19},
  \href {https://ui.adsabs.harvard.edu/abs/2013ApJS..208...19H} {208, 19}

\bibitem[\protect\citeauthoryear{{Hirata}}{{Hirata}}{2006}]{Hirata2006}
{Hirata} C.~M.,  2006, \mn@doi [\mnras] {10.1111/j.1365-2966.2005.09949.x},
  \href {https://ui.adsabs.harvard.edu/abs/2006MNRAS.367..259H} {367, 259}

\bibitem[\protect\citeauthoryear{{Hopkins} \& {Raives}}{{Hopkins} \&
  {Raives}}{2016}]{Hopkins2016}
{Hopkins} P.~F.,  {Raives} M.~J.,  2016, \mn@doi [\mnras]
  {10.1093/mnras/stv2180}, \href
  {https://ui.adsabs.harvard.edu/abs/2016MNRAS.455...51H} {455, 51}

\bibitem[\protect\citeauthoryear{{Hopkins}, {Grudi{\'c}}, {Wetzel},
  {Kere{\v{s}}}, {Faucher-Gigu{\`e}re}, {Ma}, {Murray}  \& {Butcher}}{{Hopkins}
  et~al.}{2020}]{Hopkins2020}
{Hopkins} P.~F.,  {Grudi{\'c}} M.~Y.,  {Wetzel} A.,  {Kere{\v{s}}} D.,
  {Faucher-Gigu{\`e}re} C.-A.,  {Ma} X.,  {Murray} N.,   {Butcher} N.,  2020,
  \mn@doi [\mnras] {10.1093/mnras/stz3129}, \href
  {https://ui.adsabs.harvard.edu/abs/2020MNRAS.491.3702H} {491, 3702}

\bibitem[\protect\citeauthoryear{{Hui} \& {Gnedin}}{{Hui} \&
  {Gnedin}}{1997}]{Hui1997}
{Hui} L.,  {Gnedin} N.~Y.,  1997, \mn@doi [\mnras] {10.1093/mnras/292.1.27},
  \href {https://ui.adsabs.harvard.edu/abs/1997MNRAS.292...27H} {292, 27}

\bibitem[\protect\citeauthoryear{{Iliev}, {Mellema}, {Pen}, {Merz}, {Shapiro}
  \& {Alvarez}}{{Iliev} et~al.}{2006}]{Iliev2006}
{Iliev} I.~T.,  {Mellema} G.,  {Pen} U.~L.,  {Merz} H.,  {Shapiro} P.~R.,
  {Alvarez} M.~A.,  2006, \mn@doi [\mnras] {10.1111/j.1365-2966.2006.10502.x},
  \href {https://ui.adsabs.harvard.edu/abs/2006MNRAS.369.1625I} {369, 1625}

\bibitem[\protect\citeauthoryear{{Iliev}, {Mellema}, {Ahn}, {Shapiro}, {Mao}
  \& {Pen}}{{Iliev} et~al.}{2014}]{Iliev2014}
{Iliev} I.~T.,  {Mellema} G.,  {Ahn} K.,  {Shapiro} P.~R.,  {Mao} Y.,   {Pen}
  U.-L.,  2014, \mn@doi [\mnras] {10.1093/mnras/stt2497}, \href
  {https://ui.adsabs.harvard.edu/abs/2014MNRAS.439..725I} {439, 725}

\bibitem[\protect\citeauthoryear{{Ishigaki}, {Kawamata}, {Ouchi}, {Oguri},
  {Shimasaku}  \& {Ono}}{{Ishigaki} et~al.}{2018}]{Ishigaki2018}
{Ishigaki} M.,  {Kawamata} R.,  {Ouchi} M.,  {Oguri} M.,  {Shimasaku} K.,
  {Ono} Y.,  2018, \mn@doi [\apj] {10.3847/1538-4357/aaa544}, \href
  {https://ui.adsabs.harvard.edu/abs/2018ApJ...854...73I} {854, 73}

\bibitem[\protect\citeauthoryear{{Jedamzik} \& {Saveliev}}{{Jedamzik} \&
  {Saveliev}}{2019}]{Jedamzik2019}
{Jedamzik} K.,  {Saveliev} A.,  2019, \mn@doi [\prl]
  {10.1103/PhysRevLett.123.021301}, \href
  {https://ui.adsabs.harvard.edu/abs/2019PhRvL.123b1301J} {123, 021301}

\bibitem[\protect\citeauthoryear{{Jedamzik}, {Katalini{\'c}}  \&
  {Olinto}}{{Jedamzik} et~al.}{1998}]{Jedamzik1998}
{Jedamzik} K.,  {Katalini{\'c}} V.,   {Olinto} A.~V.,  1998, \mn@doi [\prd]
  {10.1103/PhysRevD.57.3264}, \href
  {https://ui.adsabs.harvard.edu/abs/1998PhRvD..57.3264J} {57, 3264}

\bibitem[\protect\citeauthoryear{{J\"{u}lich Supercomputing
  Centre}}{{J\"{u}lich Supercomputing Centre}}{2019}]{JUWELS}
{J\"{u}lich Supercomputing Centre} 2019, \mn@doi [Journal of large-scale
  research facilities] {10.17815/jlsrf-5-171}, 5

\bibitem[\protect\citeauthoryear{{Kannan}, {Vogelsberger}, {Marinacci},
  {McKinnon}, {Pakmor}  \& {Springel}}{{Kannan} et~al.}{2019}]{Kannan2019}
{Kannan} R.,  {Vogelsberger} M.,  {Marinacci} F.,  {McKinnon} R.,  {Pakmor} R.,
    {Springel} V.,  2019, \mn@doi [\mnras] {10.1093/mnras/stz287}, \href
  {https://ui.adsabs.harvard.edu/abs/2019MNRAS.485..117K} {485, 117}

\bibitem[\protect\citeauthoryear{{Katz}, {Kimm}, {Sijacki}  \&
  {Haehnelt}}{{Katz} et~al.}{2017}]{Katz2017}
{Katz} H.,  {Kimm} T.,  {Sijacki} D.,   {Haehnelt} M.~G.,  2017, \mn@doi
  [\mnras] {10.1093/mnras/stx608}, \href
  {https://ui.adsabs.harvard.edu/abs/2017MNRAS.468.4831K} {468, 4831}

\bibitem[\protect\citeauthoryear{{Katz}, {Kimm}, {Haehnelt}, {Sijacki},
  {Rosdahl}  \& {Blaizot}}{{Katz} et~al.}{2018}]{Katz2018}
{Katz} H.,  {Kimm} T.,  {Haehnelt} M.,  {Sijacki} D.,  {Rosdahl} J.,
  {Blaizot} J.,  2018, \mn@doi [\mnras] {10.1093/mnras/sty1225}, \href
  {https://ui.adsabs.harvard.edu/abs/2018MNRAS.478.4986K} {478, 4986}

\bibitem[\protect\citeauthoryear{{Katz}, {Martin-Alvarez}, {Devriendt}, {Slyz}
  \& {Kimm}}{{Katz} et~al.}{2019}]{Katz2019c}
{Katz} H.,  {Martin-Alvarez} S.,  {Devriendt} J.,  {Slyz} A.,   {Kimm} T.,
  2019, \mn@doi [\mnras] {10.1093/mnras/stz055}, \href
  {https://ui.adsabs.harvard.edu/abs/2019MNRAS.484.2620K} {484, 2620}

\bibitem[\protect\citeauthoryear{{Katz} et~al.,}{{Katz}
  et~al.}{2020a}]{Katz2020b}
{Katz} H.,  et~al., 2020a, arXiv e-prints, \href
  {https://ui.adsabs.harvard.edu/abs/2020arXiv200501734K} {p. arXiv:2005.01734}

\bibitem[\protect\citeauthoryear{{Katz} et~al.,}{{Katz}
  et~al.}{2020b}]{Katz2020}
{Katz} H.,  et~al., 2020b, \mn@doi [\mnras] {10.1093/mnras/staa639}, \href
  {https://ui.adsabs.harvard.edu/abs/2020MNRAS.494.2200K} {494, 2200}

\bibitem[\protect\citeauthoryear{{Kawamata}, {Ishigaki}, {Shimasaku}, {Oguri},
  {Ouchi}  \& {Tanigawa}}{{Kawamata} et~al.}{2018}]{Kawamata2018}
{Kawamata} R.,  {Ishigaki} M.,  {Shimasaku} K.,  {Oguri} M.,  {Ouchi} M.,
  {Tanigawa} S.,  2018, \mn@doi [\apj] {10.3847/1538-4357/aaa6cf}, \href
  {https://ui.adsabs.harvard.edu/abs/2018ApJ...855....4K} {855, 4}

\bibitem[\protect\citeauthoryear{{Keating}, {Weinberger}, {Kulkarni},
  {Haehnelt}, {Chardin}  \& {Aubert}}{{Keating} et~al.}{2020}]{Keating2020}
{Keating} L.~C.,  {Weinberger} L.~H.,  {Kulkarni} G.,  {Haehnelt} M.~G.,
  {Chardin} J.,   {Aubert} D.,  2020, \mn@doi [\mnras] {10.1093/mnras/stz3083},
  \href {https://ui.adsabs.harvard.edu/abs/2020MNRAS.491.1736K} {491, 1736}

\bibitem[\protect\citeauthoryear{{Kim}, {Olinto}  \& {Rosner}}{{Kim}
  et~al.}{1996}]{Kim1996}
{Kim} E.-J.,  {Olinto} A.~V.,   {Rosner} R.,  1996, \mn@doi [\apj]
  {10.1086/177667}, \href
  {https://ui.adsabs.harvard.edu/abs/1996ApJ...468...28K} {468, 28}

\bibitem[\protect\citeauthoryear{{Kimm}, {Cen}, {Devriendt}, {Dubois}  \&
  {Slyz}}{{Kimm} et~al.}{2015}]{Kimm2015}
{Kimm} T.,  {Cen} R.,  {Devriendt} J.,  {Dubois} Y.,   {Slyz} A.,  2015,
  \mn@doi [\mnras] {10.1093/mnras/stv1211}, \href
  {https://ui.adsabs.harvard.edu/abs/2015MNRAS.451.2900K} {451, 2900}

\bibitem[\protect\citeauthoryear{{Kimm}, {Katz}, {Haehnelt}, {Rosdahl},
  {Devriendt}  \& {Slyz}}{{Kimm} et~al.}{2017}]{Kimm2017}
{Kimm} T.,  {Katz} H.,  {Haehnelt} M.,  {Rosdahl} J.,  {Devriendt} J.,   {Slyz}
  A.,  2017, \mn@doi [\mnras] {10.1093/mnras/stx052}, \href
  {https://ui.adsabs.harvard.edu/abs/2017MNRAS.466.4826K} {466, 4826}

\bibitem[\protect\citeauthoryear{{K{\"o}rtgen}, {Banerjee}, {Pudritz}  \&
  {Schmidt}}{{K{\"o}rtgen} et~al.}{2019}]{Kortgen2019}
{K{\"o}rtgen} B.,  {Banerjee} R.,  {Pudritz} R.~E.,   {Schmidt} W.,  2019,
  \mn@doi [\mnras] {10.1093/mnras/stz2491}, \href
  {https://ui.adsabs.harvard.edu/abs/2019MNRAS.489.5004K} {489, 5004}

\bibitem[\protect\citeauthoryear{{Kroupa}}{{Kroupa}}{2001}]{Kroupa2001}
{Kroupa} P.,  2001, \mn@doi [\mnras] {10.1046/j.1365-8711.2001.04022.x}, \href
  {https://ui.adsabs.harvard.edu/abs/2001MNRAS.322..231K} {322, 231}

\bibitem[\protect\citeauthoryear{{Krumholz} \& {Federrath}}{{Krumholz} \&
  {Federrath}}{2019}]{Krumholz2019}
{Krumholz} M.~R.,  {Federrath} C.,  2019, \mn@doi [Frontiers in Astronomy and
  Space Sciences] {10.3389/fspas.2019.00007}, \href
  {https://ui.adsabs.harvard.edu/abs/2019FrASS...6....7K} {6, 7}

\bibitem[\protect\citeauthoryear{{Kuhlen} \& {Faucher-Gigu{\`e}re}}{{Kuhlen} \&
  {Faucher-Gigu{\`e}re}}{2012}]{Kuhlen2012}
{Kuhlen} M.,  {Faucher-Gigu{\`e}re} C.-A.,  2012, \mn@doi [\mnras]
  {10.1111/j.1365-2966.2012.20924.x}, \href
  {https://ui.adsabs.harvard.edu/abs/2012MNRAS.423..862K} {423, 862}

\bibitem[\protect\citeauthoryear{{Kulkarni}, {Keating}, {Haehnelt}, {Bosman},
  {Puchwein}, {Chardin}  \& {Aubert}}{{Kulkarni} et~al.}{2019}]{Kulkarni2019}
{Kulkarni} G.,  {Keating} L.~C.,  {Haehnelt} M.~G.,  {Bosman} S. E.~I.,
  {Puchwein} E.,  {Chardin} J.,   {Aubert} D.,  2019, \mn@doi [\mnras]
  {10.1093/mnrasl/slz025}, \href
  {https://ui.adsabs.harvard.edu/abs/2019MNRAS.485L..24K} {485, L24}

\bibitem[\protect\citeauthoryear{{Lee}, {Cen}, {Gott}  \& {Trac}}{{Lee}
  et~al.}{2008}]{Lee2008}
{Lee} K.-G.,  {Cen} R.,  {Gott} J.~Richard I.,   {Trac} H.,  2008, \mn@doi
  [\apj] {10.1086/525520}, \href
  {https://ui.adsabs.harvard.edu/abs/2008ApJ...675....8L} {675, 8}

\bibitem[\protect\citeauthoryear{{Levermore}}{{Levermore}}{1984}]{Levermore1984}
{Levermore} C.~D.,  1984, \mn@doi [\jqsrt] {10.1016/0022-4073(84)90112-2},
  \href {https://ui.adsabs.harvard.edu/abs/1984JQSRT..31..149L} {31, 149}

\bibitem[\protect\citeauthoryear{{Lewis}, {Challinor}  \& {Lasenby}}{{Lewis}
  et~al.}{2000}]{CAMB}
{Lewis} A.,  {Challinor} A.,   {Lasenby} A.,  2000, \mn@doi [\apj]
  {10.1086/309179}, \href
  {https://ui.adsabs.harvard.edu/abs/2000ApJ...538..473L} {538, 473}

\bibitem[\protect\citeauthoryear{{Lewis} et~al.,}{{Lewis}
  et~al.}{2020}]{Lewis2020}
{Lewis} J. S.~W.,  et~al., 2020, \mn@doi [\mnras] {10.1093/mnras/staa1748},
  \href {https://ui.adsabs.harvard.edu/abs/2020MNRAS.496.4342L} {496, 4342}

\bibitem[\protect\citeauthoryear{{Livermore}, {Finkelstein}  \&
  {Lotz}}{{Livermore} et~al.}{2017}]{Livermore2017}
{Livermore} R.~C.,  {Finkelstein} S.~L.,   {Lotz} J.~M.,  2017, \mn@doi [\apj]
  {10.3847/1538-4357/835/2/113}, \href
  {https://ui.adsabs.harvard.edu/abs/2017ApJ...835..113L} {835, 113}

\bibitem[\protect\citeauthoryear{{Lotz} et~al.,}{{Lotz}
  et~al.}{2017}]{Lotz2017}
{Lotz} J.~M.,  et~al., 2017, \mn@doi [\apj] {10.3847/1538-4357/837/1/97}, \href
  {https://ui.adsabs.harvard.edu/abs/2017ApJ...837...97L} {837, 97}

\bibitem[\protect\citeauthoryear{{Ma}, {Hopkins}, {Kasen}, {Quataert},
  {Faucher-Gigu{\`e}re}, {Kere{\v{s}}}, {Murray}  \& {Strom}}{{Ma}
  et~al.}{2016}]{Ma2016}
{Ma} X.,  {Hopkins} P.~F.,  {Kasen} D.,  {Quataert} E.,  {Faucher-Gigu{\`e}re}
  C.-A.,  {Kere{\v{s}}} D.,  {Murray} N.,   {Strom} A.,  2016, \mn@doi [\mnras]
  {10.1093/mnras/stw941}, \href
  {https://ui.adsabs.harvard.edu/abs/2016MNRAS.459.3614M} {459, 3614}

\bibitem[\protect\citeauthoryear{{Ma} et~al.,}{{Ma} et~al.}{2018}]{Ma2018}
{Ma} X.,  et~al., 2018, \mn@doi [\mnras] {10.1093/mnras/sty684}, \href
  {https://ui.adsabs.harvard.edu/abs/2018MNRAS.477..219M} {477, 219}

\bibitem[\protect\citeauthoryear{{Ma}, {Quataert}, {Wetzel}, {Hopkins},
  {Faucher-Gigu{\`e}re}  \& {Kere{\v{s}}}}{{Ma} et~al.}{2020}]{Ma2020}
{Ma} X.,  {Quataert} E.,  {Wetzel} A.,  {Hopkins} P.~F.,  {Faucher-Gigu{\`e}re}
  C.-A.,   {Kere{\v{s}}} D.,  2020, \mn@doi [\mnras] {10.1093/mnras/staa2404},
  \href {https://ui.adsabs.harvard.edu/abs/2020MNRAS.498.2001M} {498, 2001}

\bibitem[\protect\citeauthoryear{{Machida}, {Tomisaka}, {Matsumoto}  \&
  {Inutsuka}}{{Machida} et~al.}{2008}]{Machida2008}
{Machida} M.~N.,  {Tomisaka} K.,  {Matsumoto} T.,   {Inutsuka} S.-i.,  2008,
  \mn@doi [\apj] {10.1086/529133}, \href
  {https://ui.adsabs.harvard.edu/abs/2008ApJ...677..327M} {677, 327}

\bibitem[\protect\citeauthoryear{{Madau} \& {Dickinson}}{{Madau} \&
  {Dickinson}}{2014}]{Madau2014}
{Madau} P.,  {Dickinson} M.,  2014, \mn@doi [\araa]
  {10.1146/annurev-astro-081811-125615}, \href
  {https://ui.adsabs.harvard.edu/abs/2014ARA&A..52..415M} {52, 415}

\bibitem[\protect\citeauthoryear{{Madau}, {Haardt}  \& {Rees}}{{Madau}
  et~al.}{1999}]{Madau1999}
{Madau} P.,  {Haardt} F.,   {Rees} M.~J.,  1999, \mn@doi [\apj]
  {10.1086/306975}, \href
  {https://ui.adsabs.harvard.edu/abs/1999ApJ...514..648M} {514, 648}

\bibitem[\protect\citeauthoryear{{Marchand}, {Commer{\c{c}}on}  \&
  {Chabrier}}{{Marchand} et~al.}{2018}]{Marchand2018}
{Marchand} P.,  {Commer{\c{c}}on} B.,   {Chabrier} G.,  2018, \mn@doi [\aap]
  {10.1051/0004-6361/201832907}, \href
  {https://ui.adsabs.harvard.edu/abs/2018A&A...619A..37M} {619, A37}

\bibitem[\protect\citeauthoryear{{Marinacci} \& {Vogelsberger}}{{Marinacci} \&
  {Vogelsberger}}{2016}]{Marinacci2016}
{Marinacci} F.,  {Vogelsberger} M.,  2016, \mn@doi [\mnras]
  {10.1093/mnrasl/slv176}, \href
  {https://ui.adsabs.harvard.edu/abs/2016MNRAS.456L..69M} {456, L69}

\bibitem[\protect\citeauthoryear{{Marinacci} et~al.,}{{Marinacci}
  et~al.}{2018}]{Marinacci2018}
{Marinacci} F.,  et~al., 2018, \mn@doi [\mnras] {10.1093/mnras/sty2206}, \href
  {https://ui.adsabs.harvard.edu/abs/2018MNRAS.480.5113M} {480, 5113}

\bibitem[\protect\citeauthoryear{{Martin-Alvarez}, {Devriendt}, {Slyz}  \&
  {Teyssier}}{{Martin-Alvarez} et~al.}{2018}]{Sergio2018}
{Martin-Alvarez} S.,  {Devriendt} J.,  {Slyz} A.,   {Teyssier} R.,  2018,
  \mn@doi [\mnras] {10.1093/mnras/sty1623}, \href
  {https://ui.adsabs.harvard.edu/abs/2018MNRAS.479.3343M} {479, 3343}

\bibitem[\protect\citeauthoryear{{Martin-Alvarez}, {Katz}, {Sijacki},
  {Devriendt}  \& {Slyz}}{{Martin-Alvarez} et~al.}{2020a}]{Sergio2020b}
{Martin-Alvarez} S.,  {Katz} H.,  {Sijacki} D.,  {Devriendt} J.,   {Slyz} A.,
  2020a, arXiv e-prints, \href
  {https://ui.adsabs.harvard.edu/abs/2020arXiv201111648M} {p. arXiv:2011.11648}

\bibitem[\protect\citeauthoryear{{Martin-Alvarez}, {Slyz}, {Devriendt}  \&
  {G{\'o}mez-Guijarro}}{{Martin-Alvarez} et~al.}{2020b}]{Sergio2020}
{Martin-Alvarez} S.,  {Slyz} A.,  {Devriendt} J.,   {G{\'o}mez-Guijarro} C.,
  2020b, \mn@doi [\mnras] {10.1093/mnras/staa1438}, \href
  {https://ui.adsabs.harvard.edu/abs/2020MNRAS.495.4475M} {495, 4475}

\bibitem[\protect\citeauthoryear{{McGreer}, {Mesinger}  \&
  {D'Odorico}}{{McGreer} et~al.}{2015}]{McGreer2015}
{McGreer} I.~D.,  {Mesinger} A.,   {D'Odorico} V.,  2015, \mn@doi [\mnras]
  {10.1093/mnras/stu2449}, \href
  {https://ui.adsabs.harvard.edu/abs/2015MNRAS.447..499M} {447, 499}

\bibitem[\protect\citeauthoryear{{McQuinn}, {Hernquist}, {Zaldarriaga}  \&
  {Dutta}}{{McQuinn} et~al.}{2007}]{McQuinn2007}
{McQuinn} M.,  {Hernquist} L.,  {Zaldarriaga} M.,   {Dutta} S.,  2007, \mn@doi
  [\mnras] {10.1111/j.1365-2966.2007.12085.x}, \href
  {https://ui.adsabs.harvard.edu/abs/2007MNRAS.381...75M} {381, 75}

\bibitem[\protect\citeauthoryear{{McQuinn}, {Lidz}, {Zaldarriaga}, {Hernquist}
  \& {Dutta}}{{McQuinn} et~al.}{2008}]{McQuinn2008}
{McQuinn} M.,  {Lidz} A.,  {Zaldarriaga} M.,  {Hernquist} L.,   {Dutta} S.,
  2008, \mn@doi [\mnras] {10.1111/j.1365-2966.2008.13271.x}, \href
  {https://ui.adsabs.harvard.edu/abs/2008MNRAS.388.1101M} {388, 1101}

\bibitem[\protect\citeauthoryear{{Meiksin} \& {Madau}}{{Meiksin} \&
  {Madau}}{2020}]{Meiksin2020}
{Meiksin} A.,  {Madau} P.,  2020, arXiv e-prints, \href
  {https://ui.adsabs.harvard.edu/abs/2020arXiv200615108M} {p. arXiv:2006.15108}

\bibitem[\protect\citeauthoryear{{Michel-Dansac}, {Blaizot}, {Garel},
  {Verhamme}, {Kimm}  \& {Trebitsch}}{{Michel-Dansac} et~al.}{2020}]{Leo2020}
{Michel-Dansac} L.,  {Blaizot} J.,  {Garel} T.,  {Verhamme} A.,  {Kimm} T.,
  {Trebitsch} M.,  2020, \mn@doi [\aap] {10.1051/0004-6361/201834961}, \href
  {https://ui.adsabs.harvard.edu/abs/2020A&A...635A.154M} {635, A154}

\bibitem[\protect\citeauthoryear{{Mineo}, {Gilfanov}  \& {Sunyaev}}{{Mineo}
  et~al.}{2012}]{Mineo2012}
{Mineo} S.,  {Gilfanov} M.,   {Sunyaev} R.,  2012, \mn@doi [\mnras]
  {10.1111/j.1365-2966.2011.19862.x}, \href
  {https://ui.adsabs.harvard.edu/abs/2012MNRAS.419.2095M} {419, 2095}

\bibitem[\protect\citeauthoryear{{Minoda}, {Tashiro}  \& {Takahashi}}{{Minoda}
  et~al.}{2019}]{Minoda2019}
{Minoda} T.,  {Tashiro} H.,   {Takahashi} T.,  2019, \mn@doi [\mnras]
  {10.1093/mnras/stz1860}, \href
  {https://ui.adsabs.harvard.edu/abs/2019MNRAS.488.2001M} {488, 2001}

\bibitem[\protect\citeauthoryear{{Mirocha}}{{Mirocha}}{2014}]{Mirocha2014}
{Mirocha} J.,  2014, \mn@doi [\mnras] {10.1093/mnras/stu1193}, \href
  {https://ui.adsabs.harvard.edu/abs/2014MNRAS.443.1211M} {443, 1211}

\bibitem[\protect\citeauthoryear{{Mirocha}}{{Mirocha}}{2020}]{Ares2020}
{Mirocha} J.,  2020, {ARES: Accelerated Reionization Era Simulations}
  (\mn@eprint {ascl} {2011.010})

\bibitem[\protect\citeauthoryear{{Mirocha}, {Harker}  \& {Burns}}{{Mirocha}
  et~al.}{2013}]{Mirocha2013}
{Mirocha} J.,  {Harker} G. J.~A.,   {Burns} J.~O.,  2013, \mn@doi [\apj]
  {10.1088/0004-637X/777/2/118}, \href
  {https://ui.adsabs.harvard.edu/abs/2013ApJ...777..118M} {777, 118}

\bibitem[\protect\citeauthoryear{{Mittal} \& {Kulkarni}}{{Mittal} \&
  {Kulkarni}}{2020}]{Mittal2020}
{Mittal} S.,  {Kulkarni} G.,  2020, \mn@doi [\mnras] {10.1093/mnras/staa3811},
  \href {https://ui.adsabs.harvard.edu/abs/2020MNRAS.tmp.3593M} {}

\bibitem[\protect\citeauthoryear{{Miyoshi} \& {Kusano}}{{Miyoshi} \&
  {Kusano}}{2005}]{Miyoshi2005}
{Miyoshi} T.,  {Kusano} K.,  2005, \mn@doi [Journal of Computational Physics]
  {10.1016/j.jcp.2005.02.017}, \href
  {https://ui.adsabs.harvard.edu/abs/2005JCoPh.208..315M} {208, 315}

\bibitem[\protect\citeauthoryear{{Mo}, {Mao}  \& {White}}{{Mo}
  et~al.}{1998}]{Mo1998}
{Mo} H.~J.,  {Mao} S.,   {White} S. D.~M.,  1998, \mn@doi [\mnras]
  {10.1046/j.1365-8711.1998.01227.x}, \href
  {https://ui.adsabs.harvard.edu/abs/1998MNRAS.295..319M} {295, 319}

\bibitem[\protect\citeauthoryear{{Mortlock} et~al.,}{{Mortlock}
  et~al.}{2011}]{Mortlock2011}
{Mortlock} D.~J.,  et~al., 2011, \mn@doi [\nat] {10.1038/nature10159}, \href
  {https://ui.adsabs.harvard.edu/abs/2011Natur.474..616M} {474, 616}

\bibitem[\protect\citeauthoryear{{Moster}, {Naab}  \& {White}}{{Moster}
  et~al.}{2018}]{Moster2018}
{Moster} B.~P.,  {Naab} T.,   {White} S. D.~M.,  2018, \mn@doi [\mnras]
  {10.1093/mnras/sty655}, \href
  {https://ui.adsabs.harvard.edu/abs/2018MNRAS.477.1822M} {477, 1822}

\bibitem[\protect\citeauthoryear{{Mu{\~n}oz} \& {Loeb}}{{Mu{\~n}oz} \&
  {Loeb}}{2018}]{Munoz2018}
{Mu{\~n}oz} J.~B.,  {Loeb} A.,  2018, \mn@doi [\nat]
  {10.1038/s41586-018-0151-x}, \href
  {https://ui.adsabs.harvard.edu/abs/2018Natur.557..684M} {557, 684}

\bibitem[\protect\citeauthoryear{{Nakajima}, {Ellis}, {Robertson}, {Tang}  \&
  {Stark}}{{Nakajima} et~al.}{2020}]{Nakajima2020}
{Nakajima} K.,  {Ellis} R.~S.,  {Robertson} B.~E.,  {Tang} M.,   {Stark} D.~P.,
   2020, \mn@doi [\apj] {10.3847/1538-4357/ab6604}, \href
  {https://ui.adsabs.harvard.edu/abs/2020ApJ...889..161N} {889, 161}

\bibitem[\protect\citeauthoryear{{Naoz}, {Yoshida}  \& {Gnedin}}{{Naoz}
  et~al.}{2012}]{Naoz2012}
{Naoz} S.,  {Yoshida} N.,   {Gnedin} N.~Y.,  2012, \mn@doi [\apj]
  {10.1088/0004-637X/747/2/128}, \href
  {https://ui.adsabs.harvard.edu/abs/2012ApJ...747..128N} {747, 128}

\bibitem[\protect\citeauthoryear{{Naoz}, {Yoshida}  \& {Gnedin}}{{Naoz}
  et~al.}{2013}]{Naoz2013}
{Naoz} S.,  {Yoshida} N.,   {Gnedin} N.~Y.,  2013, \mn@doi [\apj]
  {10.1088/0004-637X/763/1/27}, \href
  {https://ui.adsabs.harvard.edu/abs/2013ApJ...763...27N} {763, 27}

\bibitem[\protect\citeauthoryear{{Natwariya}}{{Natwariya}}{2020}]{Natwariya2020}
{Natwariya} P.~K.,  2020, arXiv e-prints, \href
  {https://ui.adsabs.harvard.edu/abs/2020arXiv200709938N} {p. arXiv:2007.09938}

\bibitem[\protect\citeauthoryear{{Neronov} \& {Vovk}}{{Neronov} \&
  {Vovk}}{2010}]{Neronov2010}
{Neronov} A.,  {Vovk} I.,  2010, \mn@doi [Science] {10.1126/science.1184192},
  \href {https://ui.adsabs.harvard.edu/abs/2010Sci...328...73N} {328, 73}

\bibitem[\protect\citeauthoryear{{O'Shea}, {Wise}, {Xu}  \& {Norman}}{{O'Shea}
  et~al.}{2015}]{Oshea2015}
{O'Shea} B.~W.,  {Wise} J.~H.,  {Xu} H.,   {Norman} M.~L.,  2015, \mn@doi
  [\apjl] {10.1088/2041-8205/807/1/L12}, \href
  {https://ui.adsabs.harvard.edu/abs/2015ApJ...807L..12O} {807, L12}

\bibitem[\protect\citeauthoryear{{Okamoto}, {Gao}  \& {Theuns}}{{Okamoto}
  et~al.}{2008}]{Okamoto2008}
{Okamoto} T.,  {Gao} L.,   {Theuns} T.,  2008, \mn@doi [\mnras]
  {10.1111/j.1365-2966.2008.13830.x}, \href
  {https://ui.adsabs.harvard.edu/abs/2008MNRAS.390..920O} {390, 920}

\bibitem[\protect\citeauthoryear{{Ono} et~al.,}{{Ono} et~al.}{2012}]{Ono2012}
{Ono} Y.,  et~al., 2012, \mn@doi [\apj] {10.1088/0004-637X/744/2/83}, \href
  {https://ui.adsabs.harvard.edu/abs/2012ApJ...744...83O} {744, 83}

\bibitem[\protect\citeauthoryear{{Ota} et~al.,}{{Ota} et~al.}{2008}]{Ota2008}
{Ota} K.,  et~al., 2008, \mn@doi [\apj] {10.1086/529006}, \href
  {https://ui.adsabs.harvard.edu/abs/2008ApJ...677...12O} {677, 12}

\bibitem[\protect\citeauthoryear{{Ouchi} et~al.,}{{Ouchi}
  et~al.}{2010}]{Ouchi2010}
{Ouchi} M.,  et~al., 2010, \mn@doi [\apj] {10.1088/0004-637X/723/1/869}, \href
  {https://ui.adsabs.harvard.edu/abs/2010ApJ...723..869O} {723, 869}

\bibitem[\protect\citeauthoryear{{Padoan} \& {Nordlund}}{{Padoan} \&
  {Nordlund}}{2011}]{Padoan2011}
{Padoan} P.,  {Nordlund} {\r{A}}.,  2011, \mn@doi [\apj]
  {10.1088/0004-637X/730/1/40}, \href
  {https://ui.adsabs.harvard.edu/abs/2011ApJ...730...40P} {730, 40}

\bibitem[\protect\citeauthoryear{{Pandey} \& {Sethi}}{{Pandey} \&
  {Sethi}}{2013}]{Pandey2013}
{Pandey} K.~L.,  {Sethi} S.~K.,  2013, \mn@doi [\apj]
  {10.1088/0004-637X/762/1/15}, \href
  {https://ui.adsabs.harvard.edu/abs/2013ApJ...762...15P} {762, 15}

\bibitem[\protect\citeauthoryear{{Pandey}, {Choudhury}, {Sethi}  \&
  {Ferrara}}{{Pandey} et~al.}{2015}]{Pandey2015}
{Pandey} K.~L.,  {Choudhury} T.~R.,  {Sethi} S.~K.,   {Ferrara} A.,  2015,
  \mn@doi [\mnras] {10.1093/mnras/stv1055}, \href
  {https://ui.adsabs.harvard.edu/abs/2015MNRAS.451.1692P} {451, 1692}

\bibitem[\protect\citeauthoryear{{Pawlik} \& {Schaye}}{{Pawlik} \&
  {Schaye}}{2008}]{Pawlik2008}
{Pawlik} A.~H.,  {Schaye} J.,  2008, \mn@doi [\mnras]
  {10.1111/j.1365-2966.2008.13601.x}, \href
  {https://ui.adsabs.harvard.edu/abs/2008MNRAS.389..651P} {389, 651}

\bibitem[\protect\citeauthoryear{{Pawlik}, {Schaye}  \& {van
  Scherpenzeel}}{{Pawlik} et~al.}{2009}]{Pawlik2009}
{Pawlik} A.~H.,  {Schaye} J.,   {van Scherpenzeel} E.,  2009, \mn@doi [\mnras]
  {10.1111/j.1365-2966.2009.14486.x}, \href
  {https://ui.adsabs.harvard.edu/abs/2009MNRAS.394.1812P} {394, 1812}

\bibitem[\protect\citeauthoryear{{Pentericci} et~al.,}{{Pentericci}
  et~al.}{2014}]{Pentericci2014}
{Pentericci} L.,  et~al., 2014, \mn@doi [\apj] {10.1088/0004-637X/793/2/113},
  \href {https://ui.adsabs.harvard.edu/abs/2014ApJ...793..113P} {793, 113}

\bibitem[\protect\citeauthoryear{{Pillepich} et~al.,}{{Pillepich}
  et~al.}{2018}]{IllustrisTNG}
{Pillepich} A.,  et~al., 2018, \mn@doi [\mnras] {10.1093/mnras/stx2656}, \href
  {https://ui.adsabs.harvard.edu/abs/2018MNRAS.473.4077P} {473, 4077}

\bibitem[\protect\citeauthoryear{{Planck Collaboration} et~al.,}{{Planck
  Collaboration} et~al.}{2014}]{Planck2014}
{Planck Collaboration} et~al., 2014, \mn@doi [\aap]
  {10.1051/0004-6361/201321529}, \href
  {https://ui.adsabs.harvard.edu/abs/2014A&A...571A...1P} {571, A1}

\bibitem[\protect\citeauthoryear{{Planck Collaboration} et~al.,}{{Planck
  Collaboration} et~al.}{2016}]{PlanckPMF2016}
{Planck Collaboration} et~al., 2016, \mn@doi [\aap]
  {10.1051/0004-6361/201525821}, \href
  {https://ui.adsabs.harvard.edu/abs/2016A&A...594A..19P} {594, A19}

\bibitem[\protect\citeauthoryear{{Planck Collaboration} et~al.,}{{Planck
  Collaboration} et~al.}{2018}]{Planck2018}
{Planck Collaboration} et~al., 2018, arXiv e-prints, \href
  {https://ui.adsabs.harvard.edu/abs/2018arXiv180706209P} {p. arXiv:1807.06209}

\bibitem[\protect\citeauthoryear{{Pontzen} \& {Governato}}{{Pontzen} \&
  {Governato}}{2012}]{Pontzen2012}
{Pontzen} A.,  {Governato} F.,  2012, \mn@doi [\mnras]
  {10.1111/j.1365-2966.2012.20571.x}, \href
  {https://ui.adsabs.harvard.edu/abs/2012MNRAS.421.3464P} {421, 3464}

\bibitem[\protect\citeauthoryear{{Powell}, {Roe}, {Linde}, {Gombosi}  \& {De
  Zeeuw}}{{Powell} et~al.}{1999}]{Powell1999}
{Powell} K.~G.,  {Roe} P.~L.,  {Linde} T.~J.,  {Gombosi} T.~I.,   {De Zeeuw}
  D.~L.,  1999, \mn@doi [Journal of Computational Physics]
  {10.1006/jcph.1999.6299}, \href
  {https://ui.adsabs.harvard.edu/abs/1999JCoPh.154..284P} {154, 284}

\bibitem[\protect\citeauthoryear{{Press} \& {Schechter}}{{Press} \&
  {Schechter}}{1974}]{Press1974}
{Press} W.~H.,  {Schechter} P.,  1974, \mn@doi [\apj] {10.1086/152650}, \href
  {https://ui.adsabs.harvard.edu/abs/1974ApJ...187..425P} {187, 425}

\bibitem[\protect\citeauthoryear{{Pritchard} \& {Loeb}}{{Pritchard} \&
  {Loeb}}{2012}]{Pritchard2012}
{Pritchard} J.~R.,  {Loeb} A.,  2012, \mn@doi [Reports on Progress in Physics]
  {10.1088/0034-4885/75/8/086901}, \href
  {https://ui.adsabs.harvard.edu/abs/2012RPPh...75h6901P} {75, 086901}

\bibitem[\protect\citeauthoryear{{Ratra}}{{Ratra}}{1992}]{Ratra1992}
{Ratra} B.,  1992, \mn@doi [\apjl] {10.1086/186384}, \href
  {https://ui.adsabs.harvard.edu/abs/1992ApJ...391L...1R} {391, L1}

\bibitem[\protect\citeauthoryear{{Read}, {Iorio}, {Agertz}  \&
  {Fraternali}}{{Read} et~al.}{2017}]{Read2017}
{Read} J.~I.,  {Iorio} G.,  {Agertz} O.,   {Fraternali} F.,  2017, \mn@doi
  [\mnras] {10.1093/mnras/stx147}, \href
  {https://ui.adsabs.harvard.edu/abs/2017MNRAS.467.2019R} {467, 2019}

\bibitem[\protect\citeauthoryear{{Rees}}{{Rees}}{1986}]{Rees1986}
{Rees} M.~J.,  1986, \mn@doi [\mnras] {10.1093/mnras/218.1.25P}, \href
  {https://ui.adsabs.harvard.edu/abs/1986MNRAS.218P..25R} {218, 25P}

\bibitem[\protect\citeauthoryear{{Rieder} \& {Teyssier}}{{Rieder} \&
  {Teyssier}}{2017}]{Rieder2017}
{Rieder} M.,  {Teyssier} R.,  2017, \mn@doi [\mnras] {10.1093/mnras/stx1670},
  \href {https://ui.adsabs.harvard.edu/abs/2017MNRAS.471.2674R} {471, 2674}

\bibitem[\protect\citeauthoryear{{Robertson} et~al.,}{{Robertson}
  et~al.}{2013}]{Robertson2013}
{Robertson} B.~E.,  et~al., 2013, \mn@doi [\apj] {10.1088/0004-637X/768/1/71},
  \href {https://ui.adsabs.harvard.edu/abs/2013ApJ...768...71R} {768, 71}

\bibitem[\protect\citeauthoryear{Roe}{Roe}{1986}]{Roe1986}
Roe P.~L.,  1986, \mn@doi [Annual Review of Fluid Mechanics]
  {10.1146/annurev.fl.18.010186.002005}, 18, 337

\bibitem[\protect\citeauthoryear{{Rosdahl} \& {Teyssier}}{{Rosdahl} \&
  {Teyssier}}{2015}]{Rosdahl2015}
{Rosdahl} J.,  {Teyssier} R.,  2015, \mn@doi [\mnras] {10.1093/mnras/stv567},
  \href {https://ui.adsabs.harvard.edu/abs/2015MNRAS.449.4380R} {449, 4380}

\bibitem[\protect\citeauthoryear{{Rosdahl}, {Blaizot}, {Aubert}, {Stranex}  \&
  {Teyssier}}{{Rosdahl} et~al.}{2013}]{Rosdahl2013}
{Rosdahl} J.,  {Blaizot} J.,  {Aubert} D.,  {Stranex} T.,   {Teyssier} R.,
  2013, \mn@doi [\mnras] {10.1093/mnras/stt1722}, \href
  {https://ui.adsabs.harvard.edu/abs/2013MNRAS.436.2188R} {436, 2188}

\bibitem[\protect\citeauthoryear{{Rosdahl}, {Schaye}, {Dubois}, {Kimm}  \&
  {Teyssier}}{{Rosdahl} et~al.}{2017}]{Rosdahl2017}
{Rosdahl} J.,  {Schaye} J.,  {Dubois} Y.,  {Kimm} T.,   {Teyssier} R.,  2017,
  \mn@doi [\mnras] {10.1093/mnras/stw3034}, \href
  {https://ui.adsabs.harvard.edu/abs/2017MNRAS.466...11R} {466, 11}

\bibitem[\protect\citeauthoryear{{Rosdahl} et~al.,}{{Rosdahl}
  et~al.}{2018}]{Rosdahl2018}
{Rosdahl} J.,  et~al., 2018, \mn@doi [\mnras] {10.1093/mnras/sty1655}, \href
  {https://ui.adsabs.harvard.edu/abs/2018MNRAS.479..994R} {479, 994}

\bibitem[\protect\citeauthoryear{{Rosen} \& {Bregman}}{{Rosen} \&
  {Bregman}}{1995}]{Rosen1995}
{Rosen} A.,  {Bregman} J.~N.,  1995, \mn@doi [\apj] {10.1086/175303}, \href
  {https://ui.adsabs.harvard.edu/abs/1995ApJ...440..634R} {440, 634}

\bibitem[\protect\citeauthoryear{{Saga}, {Tashiro}  \& {Yokoyama}}{{Saga}
  et~al.}{2018}]{Saga2018}
{Saga} S.,  {Tashiro} H.,   {Yokoyama} S.,  2018, \mn@doi [\mnras]
  {10.1093/mnrasl/slx195}, \href
  {https://ui.adsabs.harvard.edu/abs/2018MNRAS.474L..52S} {474, L52}

\bibitem[\protect\citeauthoryear{{Sanati}, {Revaz}, {Schober}, {Kunze}  \&
  {Jablonka}}{{Sanati} et~al.}{2020}]{Sanati2020}
{Sanati} M.,  {Revaz} Y.,  {Schober} J.,  {Kunze} K.~E.,   {Jablonka} P.,
  2020, arXiv e-prints, \href
  {https://ui.adsabs.harvard.edu/abs/2020arXiv200505401S} {p. arXiv:2005.05401}

\bibitem[\protect\citeauthoryear{{Schenker}, {Ellis}, {Konidaris}  \&
  {Stark}}{{Schenker} et~al.}{2014}]{Schenker2014}
{Schenker} M.~A.,  {Ellis} R.~S.,  {Konidaris} N.~P.,   {Stark} D.~P.,  2014,
  \mn@doi [\apj] {10.1088/0004-637X/795/1/20}, \href
  {https://ui.adsabs.harvard.edu/abs/2014ApJ...795...20S} {795, 20}

\bibitem[\protect\citeauthoryear{{Schleicher}, {Banerjee}  \&
  {Klessen}}{{Schleicher} et~al.}{2009}]{Schleichler2009}
{Schleicher} D. R.~G.,  {Banerjee} R.,   {Klessen} R.~S.,  2009, \mn@doi [\apj]
  {10.1088/0004-637X/692/1/236}, \href
  {https://ui.adsabs.harvard.edu/abs/2009ApJ...692..236S} {692, 236}

\bibitem[\protect\citeauthoryear{{Schmidt}}{{Schmidt}}{1959}]{Schmidt}
{Schmidt} M.,  1959, \mn@doi [\apj] {10.1086/146614}, \href
  {https://ui.adsabs.harvard.edu/abs/1959ApJ...129..243S} {129, 243}

\bibitem[\protect\citeauthoryear{{Schober}, {Schleicher}, {Federrath},
  {Glover}, {Klessen}  \& {Banerjee}}{{Schober} et~al.}{2012}]{Schober2012}
{Schober} J.,  {Schleicher} D.,  {Federrath} C.,  {Glover} S.,  {Klessen}
  R.~S.,   {Banerjee} R.,  2012, \mn@doi [\apj] {10.1088/0004-637X/754/2/99},
  \href {https://ui.adsabs.harvard.edu/abs/2012ApJ...754...99S} {754, 99}

\bibitem[\protect\citeauthoryear{{Schober}, {Schleicher}, {Federrath}, {Bovino}
   \& {Klessen}}{{Schober} et~al.}{2015}]{Schober2015}
{Schober} J.,  {Schleicher} D.~R.~G.,  {Federrath} C.,  {Bovino} S.,
  {Klessen} R.~S.,  2015, \mn@doi [\pre] {10.1103/PhysRevE.92.023010}, \href
  {https://ui.adsabs.harvard.edu/abs/2015PhRvE..92b3010S} {92, 023010}

\bibitem[\protect\citeauthoryear{{Schroeder}, {Mesinger}  \&
  {Haiman}}{{Schroeder} et~al.}{2013}]{Schroeder2013}
{Schroeder} J.,  {Mesinger} A.,   {Haiman} Z.,  2013, \mn@doi [\mnras]
  {10.1093/mnras/sts253}, \href
  {https://ui.adsabs.harvard.edu/abs/2013MNRAS.428.3058S} {428, 3058}

\bibitem[\protect\citeauthoryear{{Sethi} \& {Subramanian}}{{Sethi} \&
  {Subramanian}}{2005}]{Sethi2005}
{Sethi} S.~K.,  {Subramanian} K.,  2005, \mn@doi [\mnras]
  {10.1111/j.1365-2966.2004.08520.x}, \href
  {https://ui.adsabs.harvard.edu/abs/2005MNRAS.356..778S} {356, 778}

\bibitem[\protect\citeauthoryear{{Sethi} \& {Subramanian}}{{Sethi} \&
  {Subramanian}}{2009}]{Sethi2009}
{Sethi} S.~K.,  {Subramanian} K.,  2009, \mn@doi [\jcap]
  {10.1088/1475-7516/2009/11/021}, \href
  {https://ui.adsabs.harvard.edu/abs/2009JCAP...11..021S} {2009, 021}

\bibitem[\protect\citeauthoryear{{Shaw} \& {Lewis}}{{Shaw} \&
  {Lewis}}{2012}]{Shaw2012}
{Shaw} J.~R.,  {Lewis} A.,  2012, \mn@doi [\prd] {10.1103/PhysRevD.86.043510},
  \href {https://ui.adsabs.harvard.edu/abs/2012PhRvD..86d3510S} {86, 043510}

\bibitem[\protect\citeauthoryear{{Shibuya}, {Ouchi}  \& {Harikane}}{{Shibuya}
  et~al.}{2015}]{Shibuya2015}
{Shibuya} T.,  {Ouchi} M.,   {Harikane} Y.,  2015, \mn@doi [\apjs]
  {10.1088/0067-0049/219/2/15}, \href
  {https://ui.adsabs.harvard.edu/abs/2015ApJS..219...15S} {219, 15}

\bibitem[\protect\citeauthoryear{{Shull} \& {van Steenberg}}{{Shull} \& {van
  Steenberg}}{1985}]{Shull1985}
{Shull} J.~M.,  {van Steenberg} M.~E.,  1985, \mn@doi [\apj] {10.1086/163605},
  \href {https://ui.adsabs.harvard.edu/abs/1985ApJ...298..268S} {298, 268}

\bibitem[\protect\citeauthoryear{{Sobacchi} \& {Mesinger}}{{Sobacchi} \&
  {Mesinger}}{2015}]{Sobacchi2015}
{Sobacchi} E.,  {Mesinger} A.,  2015, \mn@doi [\mnras] {10.1093/mnras/stv1751},
  \href {https://ui.adsabs.harvard.edu/abs/2015MNRAS.453.1843S} {453, 1843}

\bibitem[\protect\citeauthoryear{{Sparke}}{{Sparke}}{1982}]{Sparke1982}
{Sparke} L.~S.,  1982, \mn@doi [\apj] {10.1086/160237}, \href
  {https://ui.adsabs.harvard.edu/abs/1982ApJ...260..104S} {260, 104}

\bibitem[\protect\citeauthoryear{{Stanway}, {Eldridge}  \& {Becker}}{{Stanway}
  et~al.}{2016}]{Stanway2016}
{Stanway} E.~R.,  {Eldridge} J.~J.,   {Becker} G.~D.,  2016, \mn@doi [\mnras]
  {10.1093/mnras/stv2661}, \href
  {https://ui.adsabs.harvard.edu/abs/2016MNRAS.456..485S} {456, 485}

\bibitem[\protect\citeauthoryear{{Subramanian}}{{Subramanian}}{2010}]{Subramanian2010}
{Subramanian} K.,  2010, \mn@doi [Astronomische Nachrichten]
  {10.1002/asna.200911312}, \href
  {https://ui.adsabs.harvard.edu/abs/2010AN....331..110S} {331, 110}

\bibitem[\protect\citeauthoryear{{Subramanian}}{{Subramanian}}{2016}]{Subramanian2016}
{Subramanian} K.,  2016, \mn@doi [Reports on Progress in Physics]
  {10.1088/0034-4885/79/7/076901}, \href
  {https://ui.adsabs.harvard.edu/abs/2016RPPh...79g6901S} {79, 076901}

\bibitem[\protect\citeauthoryear{{Subramanian} \& {Barrow}}{{Subramanian} \&
  {Barrow}}{1998}]{Subramanian1998}
{Subramanian} K.,  {Barrow} J.~D.,  1998, \mn@doi [\prd]
  {10.1103/PhysRevD.58.083502}, \href
  {https://ui.adsabs.harvard.edu/abs/1998PhRvD..58h3502S} {58, 083502}

\bibitem[\protect\citeauthoryear{{Subramanian}, {Narasimha}  \&
  {Chitre}}{{Subramanian} et~al.}{1994}]{Subramanian1994}
{Subramanian} K.,  {Narasimha} D.,   {Chitre} S.~M.,  1994, \mn@doi [\mnras]
  {10.1093/mnras/271.1.L15}, \href
  {https://ui.adsabs.harvard.edu/abs/1994MNRAS.271L..15S} {271, L15}

\bibitem[\protect\citeauthoryear{{Tabatabaei}, {Krause}, {Fletcher}  \&
  {Beck}}{{Tabatabaei} et~al.}{2008}]{Tabatabaei2008}
{Tabatabaei} F.~S.,  {Krause} M.,  {Fletcher} A.,   {Beck} R.,  2008, \mn@doi
  [\aap] {10.1051/0004-6361:200810590}, \href
  {https://ui.adsabs.harvard.edu/abs/2008A&A...490.1005T} {490, 1005}

\bibitem[\protect\citeauthoryear{{Tashiro} \& {Sugiyama}}{{Tashiro} \&
  {Sugiyama}}{2006}]{Tashiro2006}
{Tashiro} H.,  {Sugiyama} N.,  2006, \mn@doi [\mnras]
  {10.1111/j.1365-2966.2006.10901.x}, \href
  {https://ui.adsabs.harvard.edu/abs/2006MNRAS.372.1060T} {372, 1060}

\bibitem[\protect\citeauthoryear{{Tasinato}}{{Tasinato}}{2015}]{Tasinato2015}
{Tasinato} G.,  2015, \mn@doi [\jcap] {10.1088/1475-7516/2015/03/040}, \href
  {https://ui.adsabs.harvard.edu/abs/2015JCAP...03..040T} {2015, 040}

\bibitem[\protect\citeauthoryear{{Tavecchio}, {Ghisellini}, {Bonnoli}  \&
  {Foschini}}{{Tavecchio} et~al.}{2011}]{Tavecchio2011}
{Tavecchio} F.,  {Ghisellini} G.,  {Bonnoli} G.,   {Foschini} L.,  2011,
  \mn@doi [\mnras] {10.1111/j.1365-2966.2011.18657.x}, \href
  {https://ui.adsabs.harvard.edu/abs/2011MNRAS.414.3566T} {414, 3566}

\bibitem[\protect\citeauthoryear{{Teyssier}}{{Teyssier}}{2002}]{Teyssier2002}
{Teyssier} R.,  2002, \mn@doi [\aap] {10.1051/0004-6361:20011817}, \href
  {https://ui.adsabs.harvard.edu/abs/2002A&A...385..337T} {385, 337}

\bibitem[\protect\citeauthoryear{{Teyssier}, {Fromang}  \& {Dormy}}{{Teyssier}
  et~al.}{2006}]{Teyssier2006}
{Teyssier} R.,  {Fromang} S.,   {Dormy} E.,  2006, \mn@doi [Journal of
  Computational Physics] {10.1016/j.jcp.2006.01.042}, \href
  {https://ui.adsabs.harvard.edu/abs/2006JCoPh.218...44T} {218, 44}

\bibitem[\protect\citeauthoryear{{Tilvi} et~al.,}{{Tilvi}
  et~al.}{2014}]{Tilvi2014}
{Tilvi} V.,  et~al., 2014, \mn@doi [\apj] {10.1088/0004-637X/794/1/5}, \href
  {https://ui.adsabs.harvard.edu/abs/2014ApJ...794....5T} {794, 5}

\bibitem[\protect\citeauthoryear{Toro}{Toro}{2009}]{Toro2009}
Toro E.~F.,  2009, Riemann Solvers and Numerical Methods for Fluid Dynamics, 3
  edn.
Springer

\bibitem[\protect\citeauthoryear{{Totani}, {Kawai}, {Kosugi}, {Aoki}, {Yamada},
  {Iye}, {Ohta}  \& {Hattori}}{{Totani} et~al.}{2006}]{Totani2006}
{Totani} T.,  {Kawai} N.,  {Kosugi} G.,  {Aoki} K.,  {Yamada} T.,  {Iye} M.,
  {Ohta} K.,   {Hattori} T.,  2006, \mn@doi [\pasj] {10.1093/pasj/58.3.485},
  \href {https://ui.adsabs.harvard.edu/abs/2006PASJ...58..485T} {58, 485}

\bibitem[\protect\citeauthoryear{{T{\'o}th}}{{T{\'o}th}}{2000}]{Toth2000}
{T{\'o}th} G.,  2000, \mn@doi [Journal of Computational Physics]
  {10.1006/jcph.2000.6519}, \href
  {https://ui.adsabs.harvard.edu/abs/2000JCoPh.161..605T} {161, 605}

\bibitem[\protect\citeauthoryear{{Trac}, {Cen}  \& {Loeb}}{{Trac}
  et~al.}{2008}]{Trac2008}
{Trac} H.,  {Cen} R.,   {Loeb} A.,  2008, \mn@doi [\apjl] {10.1086/595678},
  \href {https://ui.adsabs.harvard.edu/abs/2008ApJ...689L..81T} {689, L81}

\bibitem[\protect\citeauthoryear{{Trebitsch}, {Blaizot}, {Rosdahl}, {Devriendt}
   \& {Slyz}}{{Trebitsch} et~al.}{2017}]{Trebitsch2017}
{Trebitsch} M.,  {Blaizot} J.,  {Rosdahl} J.,  {Devriendt} J.,   {Slyz} A.,
  2017, \mn@doi [\mnras] {10.1093/mnras/stx1060}, \href
  {https://ui.adsabs.harvard.edu/abs/2017MNRAS.470..224T} {470, 224}

\bibitem[\protect\citeauthoryear{{Tricco}, {Price}  \& {Federrath}}{{Tricco}
  et~al.}{2016}]{Tricco2016}
{Tricco} T.~S.,  {Price} D.~J.,   {Federrath} C.,  2016, \mn@doi [\mnras]
  {10.1093/mnras/stw1280}, \href
  {https://ui.adsabs.harvard.edu/abs/2016MNRAS.461.1260T} {461, 1260}

\bibitem[\protect\citeauthoryear{{Tseliakhovich} \& {Hirata}}{{Tseliakhovich}
  \& {Hirata}}{2010}]{Tseliakhovich2010}
{Tseliakhovich} D.,  {Hirata} C.,  2010, \mn@doi [\prd]
  {10.1103/PhysRevD.82.083520}, \href
  {https://ui.adsabs.harvard.edu/abs/2010PhRvD..82h3520T} {82, 083520}

\bibitem[\protect\citeauthoryear{{Turner} \& {Widrow}}{{Turner} \&
  {Widrow}}{1988}]{Turner1988}
{Turner} M.~S.,  {Widrow} L.~M.,  1988, \mn@doi [\prd]
  {10.1103/PhysRevD.37.2743}, \href
  {https://ui.adsabs.harvard.edu/abs/1988PhRvD..37.2743T} {37, 2743}

\bibitem[\protect\citeauthoryear{{Tweed}, {Devriendt}, {Blaizot}, {Colombi}  \&
  {Slyz}}{{Tweed} et~al.}{2009}]{Tweed2009}
{Tweed} D.,  {Devriendt} J.,  {Blaizot} J.,  {Colombi} S.,   {Slyz} A.,  2009,
  \mn@doi [\aap] {10.1051/0004-6361/200911787}, \href
  {https://ui.adsabs.harvard.edu/abs/2009A&A...506..647T} {506, 647}

\bibitem[\protect\citeauthoryear{Tóth \& Roe}{Tóth \& Roe}{2002}]{Toth2002}
Tóth G.,  Roe P.,  2002, \mn@doi [Journal of Computational Physics]
  {https://doi.org/10.1006/jcph.2002.7120}, 180, 736

\bibitem[\protect\citeauthoryear{{Vazza}, {Br{\"u}ggen}, {Gheller},
  {Hackstein}, {Wittor}  \& {Hinz}}{{Vazza} et~al.}{2017}]{Vazza2017}
{Vazza} F.,  {Br{\"u}ggen} M.,  {Gheller} C.,  {Hackstein} S.,  {Wittor} D.,
  {Hinz} P.~M.,  2017, \mn@doi [Classical and Quantum Gravity]
  {10.1088/1361-6382/aa8e60}, \href
  {https://ui.adsabs.harvard.edu/abs/2017CQGra..34w4001V} {34, 234001}

\bibitem[\protect\citeauthoryear{{Vazza}, {Brunetti}, {Br{\"u}ggen}  \&
  {Bonafede}}{{Vazza} et~al.}{2018}]{Vazza2018}
{Vazza} F.,  {Brunetti} G.,  {Br{\"u}ggen} M.,   {Bonafede} A.,  2018, \mn@doi
  [\mnras] {10.1093/mnras/stx2830}, \href
  {https://ui.adsabs.harvard.edu/abs/2018MNRAS.474.1672V} {474, 1672}

\bibitem[\protect\citeauthoryear{{Vogelsberger}, {Genel}, {Sijacki}, {Torrey},
  {Springel}  \& {Hernquist}}{{Vogelsberger} et~al.}{2013}]{Vogelsberger2013}
{Vogelsberger} M.,  {Genel} S.,  {Sijacki} D.,  {Torrey} P.,  {Springel} V.,
  {Hernquist} L.,  2013, \mn@doi [\mnras] {10.1093/mnras/stt1789}, \href
  {https://ui.adsabs.harvard.edu/abs/2013MNRAS.436.3031V} {436, 3031}

\bibitem[\protect\citeauthoryear{{Wasserman}}{{Wasserman}}{1978}]{Wasserman1978}
{Wasserman} I.,  1978, \mn@doi [\apj] {10.1086/156381}, \href
  {https://ui.adsabs.harvard.edu/abs/1978ApJ...224..337W} {224, 337}

\bibitem[\protect\citeauthoryear{{Wise} \& {Abel}}{{Wise} \&
  {Abel}}{2011}]{Wise2011}
{Wise} J.~H.,  {Abel} T.,  2011, \mn@doi [\mnras]
  {10.1111/j.1365-2966.2011.18646.x}, \href
  {https://ui.adsabs.harvard.edu/abs/2011MNRAS.414.3458W} {414, 3458}

\bibitem[\protect\citeauthoryear{{Wise}, {Turk}, {Norman}  \& {Abel}}{{Wise}
  et~al.}{2012}]{Wise2012}
{Wise} J.~H.,  {Turk} M.~J.,  {Norman} M.~L.,   {Abel} T.,  2012, \mn@doi
  [\apj] {10.1088/0004-637X/745/1/50}, \href
  {https://ui.adsabs.harvard.edu/abs/2012ApJ...745...50W} {745, 50}

\bibitem[\protect\citeauthoryear{{Wise}, {Demchenko}, {Halicek}, {Norman},
  {Turk}, {Abel}  \& {Smith}}{{Wise} et~al.}{2014}]{Wise2014}
{Wise} J.~H.,  {Demchenko} V.~G.,  {Halicek} M.~T.,  {Norman} M.~L.,  {Turk}
  M.~J.,  {Abel} T.,   {Smith} B.~D.,  2014, \mn@doi [\mnras]
  {10.1093/mnras/stu979}, \href
  {https://ui.adsabs.harvard.edu/abs/2014MNRAS.442.2560W} {442, 2560}

\bibitem[\protect\citeauthoryear{{Wouthuysen}}{{Wouthuysen}}{1952}]{Wouthuysen1952}
{Wouthuysen} S.~A.,  1952, \mn@doi [\aj] {10.1086/106661}, \href
  {https://ui.adsabs.harvard.edu/abs/1952AJ.....57R..31W} {57, 31}

\bibitem[\protect\citeauthoryear{{Wyithe} \& {Bolton}}{{Wyithe} \&
  {Bolton}}{2011}]{Wyithe2011}
{Wyithe} J. S.~B.,  {Bolton} J.~S.,  2011, \mn@doi [\mnras]
  {10.1111/j.1365-2966.2010.18030.x}, \href
  {https://ui.adsabs.harvard.edu/abs/2011MNRAS.412.1926W} {412, 1926}

\bibitem[\protect\citeauthoryear{{Xu}, {Wise}, {Norman}, {Ahn}  \&
  {O'Shea}}{{Xu} et~al.}{2016}]{Xu2016}
{Xu} H.,  {Wise} J.~H.,  {Norman} M.~L.,  {Ahn} K.,   {O'Shea} B.~W.,  2016,
  \mn@doi [\apj] {10.3847/1538-4357/833/1/84}, \href
  {https://ui.adsabs.harvard.edu/abs/2016ApJ...833...84X} {833, 84}

\bibitem[\protect\citeauthoryear{{Yoo}, {Kimm}  \& {Rosdahl}}{{Yoo}
  et~al.}{2020}]{Yoo2020}
{Yoo} T.,  {Kimm} T.,   {Rosdahl} J.,  2020, \mn@doi [\mnras]
  {10.1093/mnras/staa3187}, \href
  {https://ui.adsabs.harvard.edu/abs/2020MNRAS.499.5175Y} {499, 5175}

\bibitem[\protect\citeauthoryear{{Zygelman}}{{Zygelman}}{2005}]{Zygelman2005}
{Zygelman} B.,  2005, \mn@doi [\apj] {10.1086/427682}, \href
  {https://ui.adsabs.harvard.edu/abs/2005ApJ...622.1356Z} {622, 1356}

\bibitem[\protect\citeauthoryear{{{\v{D}}urov{\v{c}}{\'\i}kov{\'a}}, {Katz},
  {Bosman}, {Davies}, {Devriendt}  \&
  {Slyz}}{{{\v{D}}urov{\v{c}}{\'\i}kov{\'a}} et~al.}{2020}]{Durovcikova2020}
{{\v{D}}urov{\v{c}}{\'\i}kov{\'a}} D.,  {Katz} H.,  {Bosman} S. E.~I.,
  {Davies} F.~B.,  {Devriendt} J.,   {Slyz} A.,  2020, \mn@doi [\mnras]
  {10.1093/mnras/staa505}, \href
  {https://ui.adsabs.harvard.edu/abs/2020MNRAS.493.4256D} {493, 4256}

\makeatother
\end{thebibliography}







\bsp	
\label{lastpage}
\end{document}